\documentclass[a4paper,fleqn,usenatbib]{mnras}
\usepackage{newtxtext,newtxmath}

\usepackage{ae,aecompl}
\usepackage{graphicx}	
\usepackage{amsmath}	
\usepackage{amssymb}	
\usepackage{mathtools}
\usepackage{hyperref}
\usepackage{lineno}
\title[DEVILS: Motivation, Design and Targets]{Deep Extragalactic VIsible Legacy Survey (DEVILS): Motivation, Design and Target Catalogue}
\author[L. J. M. Davies et. al.]{L. J. M. Davies$^{1}$\thanks{E-mail:
 luke.j.davies@uwa.edu.au}, A. S. G. Robotham$^{1}$, S. P. Driver$^{1}$, C. P. Lagos$^{1,2}$, L. Cortese$^{1}$, \newauthor E. Mannering$^{1,3}$, C. Foster$^{4, 5}$,  C. Lidman$^{6,7}$, A. Hashemizadeh$^{1}$, S. Koushan$^{1}$, \newauthor S. O'Toole$^{3}$, I. K. Baldry$^{7}$, M. Bilicki$^{8, 9}$,  J. Bland-Hawthorn$^{4,5}$, M. N. Bremer$^{10}$, \newauthor M. J. I. Brown$^{11}$,  J. J. Bryant$^{3, 4}$, B. Catinella$^{1}$, S. M. Croom$^{4}$,  M. W. Grootes$^{12}$,  \newauthor  B. W. Holwerda$^{13}$, M. J. Jarvis$^{14, 15}$, N. Maddox$^{16}$, M. Meyer$^{1}$, A. J. Moffett$^{17}$, \newauthor  S. Phillipps$^{10}$, E. N. Taylor$^{18}$,  R. A. Windhorst$^{19}$ and C. Wolf$^{6}$  \\
\\ 
$^{1}$ ICRAR, The University of Western Australia, 35 Stirling Highway, Crawley, WA 6009, Australia\\
$^{2}$ Australian Research Council Centre of Excellence for All-sky Astrophysics (CAASTRO), 44 Rosehill Street Redfern, NSW 2016, Australia \\
$^{3}$ Australian Astronomical Observatory, 105 Delhi Rd, North Ryde, NSW 2113, Australia\\
$^{4}$ Sydney Institute for Astronomy, School of Physics, A28, The University of Sydney, NSW 2006, Australia\\
$^{5}$ Australian Research Council Centre of Excellence for All Sky Astrophysics in 3 Dimensions (ASTRO 3D)\\
$^{6}$ Research School of Astronomy and Astrophysics, Australian National University, Canberra, ACT 2611, Australia\\
\\
Remaining affiliations listed at the end of this paper
}

\date{Accepted 08/06/18. Received 03/06/18; in original form 28/04/18}

\pubyear{2018}

\begin{document}
\label{firstpage}
\pagerange{\pageref{firstpage}--\pageref{lastpage}}
\maketitle

\begin{abstract}

The Deep Extragalactic VIsible Legacy Survey (DEVILS) is a large spectroscopic campaign at the Anglo-Australian Telescope (AAT) aimed at bridging the near and distant Universe by producing the highest completeness survey of galaxies and groups at intermediate redshifts ($0.3<z<1.0$). Our sample consists of $\sim$60,000 galaxies to Y$<$21.2\,mag, over $\sim$6\,deg$^{2}$ in three well-studied deep extragalactic fields (Cosmic Origins Survey field, COSMOS, Extended Chandra Deep Field South, ECDFS and the X-ray Multi-Mirror Mission Large-Scale Structure region, XMM-LSS - all Large Synoptic Survey Telescope deep-drill fields). This paper presents the broad experimental design of DEVILS. Our target sample has been selected from deep Visible and Infrared Survey Telescope for Astronomy (VISTA) Y-band imaging (VISTA Deep Extragalactic Observations, VIDEO and UltraVISTA), with photometry measured by \textsc{ProFound}. Photometric star/galaxy separation is done on the basis of NIR colours, and has been validated by visual inspection. To maximise our observing efficiency for faint targets we employ a redshift feedback strategy, which continually updates our target lists, feeding back the results from the previous night's observations. We also present an overview of the initial spectroscopic observations undertaken in late 2017 and early 2018.

\end{abstract}


\section{Introduction}

\begin{figure*}
\begin{center}
\includegraphics[scale=0.6]{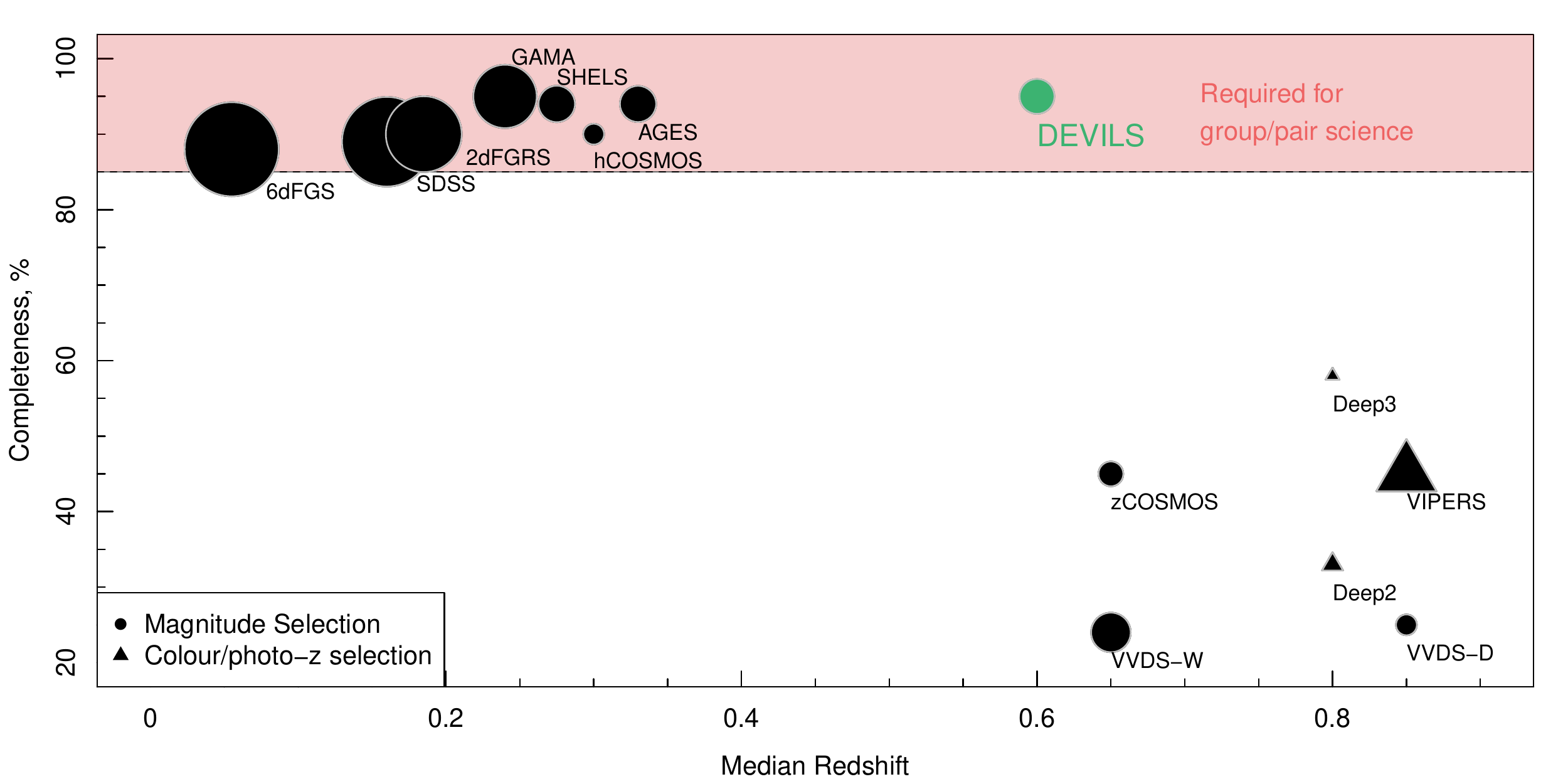}
\caption{DEVILS comparison to other existing large spectroscopic surveys in terms of completeness and median redshift of the target sample. Surveys are split by those that use a simple single-band selection function (circle) and those which use a more complex colour and/or photometric redshift selection (triangles). Point size is representative of log10[number of sources]. In order to explore the effect of sub-Mpc structure on the evolution of galaxies and to probe the evolution of group-scale haloes high completeness to magnitude-limited samples are required. Surveys with a completeness of $\lesssim$85\% miss a significant fraction of group galaxies and therefore can only parameterise the most massive haloes (see Section \ref{sec:Depth} and Figure \ref{fig:Halos}), and are also very incomplete to interacting pairs. For example, with uniform random sampling, to zeroth order, a survey which is $50\%$ complete at a given stellar mass will only identify $25\%$ of major merger pairs at that mass, while a similar $95\%$ complete sample will identify $90\%$ of pairs. Low redshift surveys, such as SDSS, 2dFGRS and GAMA probe to these high levels of completeness in the relatively local Universe. However, until DEVILS, there has been no survey undertaken with this level of completeness at intermediate redshift.}
\label{fig:Comp}
\end{center}
\end{figure*}

Over the past two decades large low redshift ($z<0.3$) galaxy evolution-focused surveys such as the Sloan Digital Sky Survey \citep[SDSS, $e.g.$][]{Abazajian09}, the Two Degree Field Galaxy Redshift Survey \citep[2dFGRS, ][]{Colless01}, and Galaxy And Mass Assembly Survey \citep[GAMA, ][]{Driver11, Liske15} have unequivocally changed our view of the local Universe. These surveys have parameterised structures on physical scales covering $\sim$5 orders of magnitude and characterised many of the astrophysical processes occurring at the current epoch. They have transformed our understanding of large-scale structure on scales of $>$1\,Mpc \citep[$e.g.$][]{Peacock01}, the baryon-dark matter interface on scales of a few kpc to 1\,Mpc \citep[ $e.g.$][]{Robotham11,Yang07}, the internal growth of galaxy structure of 1 to a few\,kpc scales \citep[$e.g.$][]{Lange15,Belli17}, the low stellar mass Universe \citep[$e.g.$][]{Baldry10}, and the effect of both large-scale environment \citep[$e.g.$][]{Peng10} and local environment \citep{Patton11,Davies15b,Davies16a} on galaxy evolution. However, by design these surveys have only targeted the relatively local Universe ($z<0.3$). While they provide a wealth of information about galaxies at the current epoch, they cannot measure the astrophysical processes that led to their formation. It is not the processes occurring today which shaped the $z\sim0$ Universe, but the factors which drove galaxy evolution and structure formation over the preceding 10 billion years.

\begin{figure*}
\begin{center}
\includegraphics[scale=0.47]{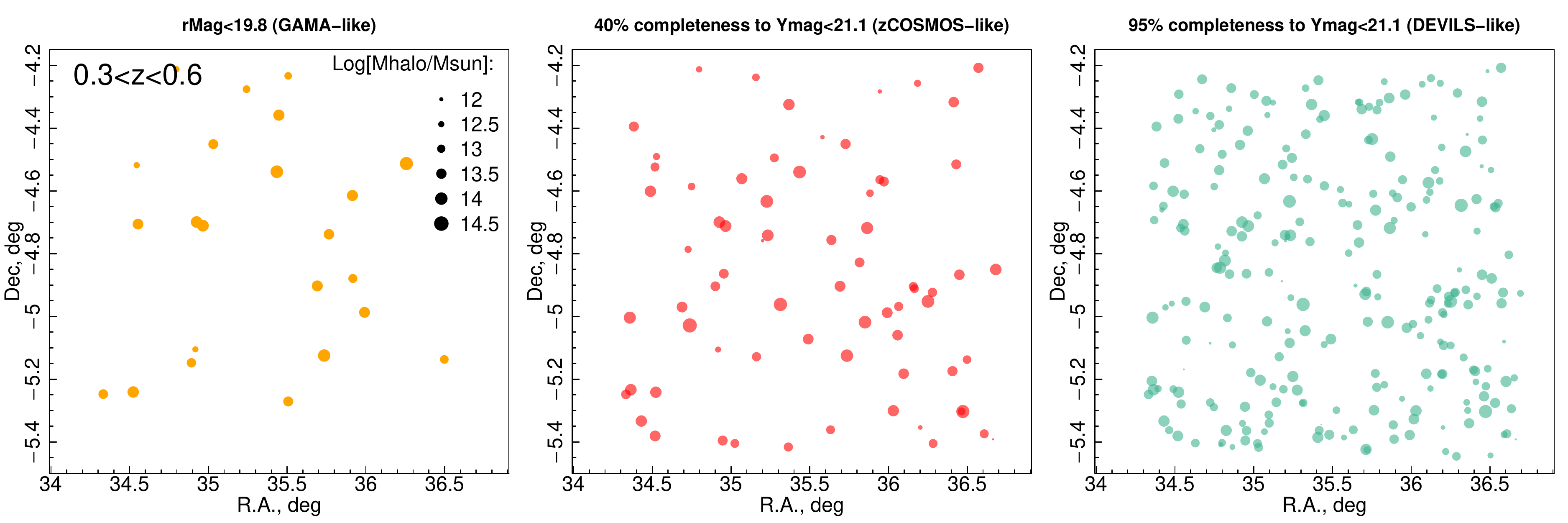}
\caption{The predicted distribution of multiplicity N$>$3 halos identifiable in the D02 region for our DEVILS TAO simulations (see Section \ref{sec:Depth}). The size of the point represents halo mass on a log scale given in the legend. Left: distribution of halos that would be identified at $.3<z<0.6$ using a GAMA-like r$<$19.8 limit. Middle: distribution of halos that would be identified at $0.3<z<0.6$ using a DEVILS-like Y$<$21.1\,mag limit, but with only 40\% completeness (comparable to other surveys at this epoch, such as zCOSMOS-bight and VIPERS). We assume uniform incompleteness and remove 60\% of Y$<$21.1\,mag galaxies. Right: distribution of halos that would be identified at $0.3<z<0.6$ using a DEVILS-like Y$<$21.1\,mag limit and with 95\% completeness limit, highlighting that low completeness surveys miss a large fraction of low mass halos.}
\label{fig:Halos}
\end{center}
\end{figure*}

Deep but small-area spectroscopic surveys such as zCOSMOS-deep \citep{Lilly07}, the Very Large Telescope (VLT) VIsible Multi-Object Spectrograph (VIMOS) Deep Survey \citep[VVDS, ][]{LeFevre13} and VIMOS Ultra-Deep Survey \citep[VUDS, ][]{LeFevre15}, have explored earlier epochs ($z>1$), probing the initial stages of galaxy evolution. However, it is the relatively under-sampled epoch at intermediate redshifts ($0.3<z<1.0$), where both galaxies and their host haloes undergo significant coeval evolution, specifically in terms of the environmental effects on galaxies.  At this epoch many of the $z\sim0$ environmental trends observed in surveys such as GAMA and SDSS were shaped \citep[$e.g.$][]{Darvish16}. It is here that roughly half of all stars were formed \citep{Madau14, Driver18}, galaxies underwent significant mass, size, morphology and angular momentum evolution \citep[$e.g.$][]{Lotz11,vanderWel12,Lange15, Codis15}, and our current cold dark matter model $\Lambda$CDM predicts a strong and testable evolution of the halo mass function \citep[$e.g.$][]{Murray13,Elahi18}. Surprisingly, this key epoch in the formation of the fundamental relationships we observe today has been left comparatively unexplored. 

\begin{figure*}
\begin{center}
\includegraphics[scale=0.28]{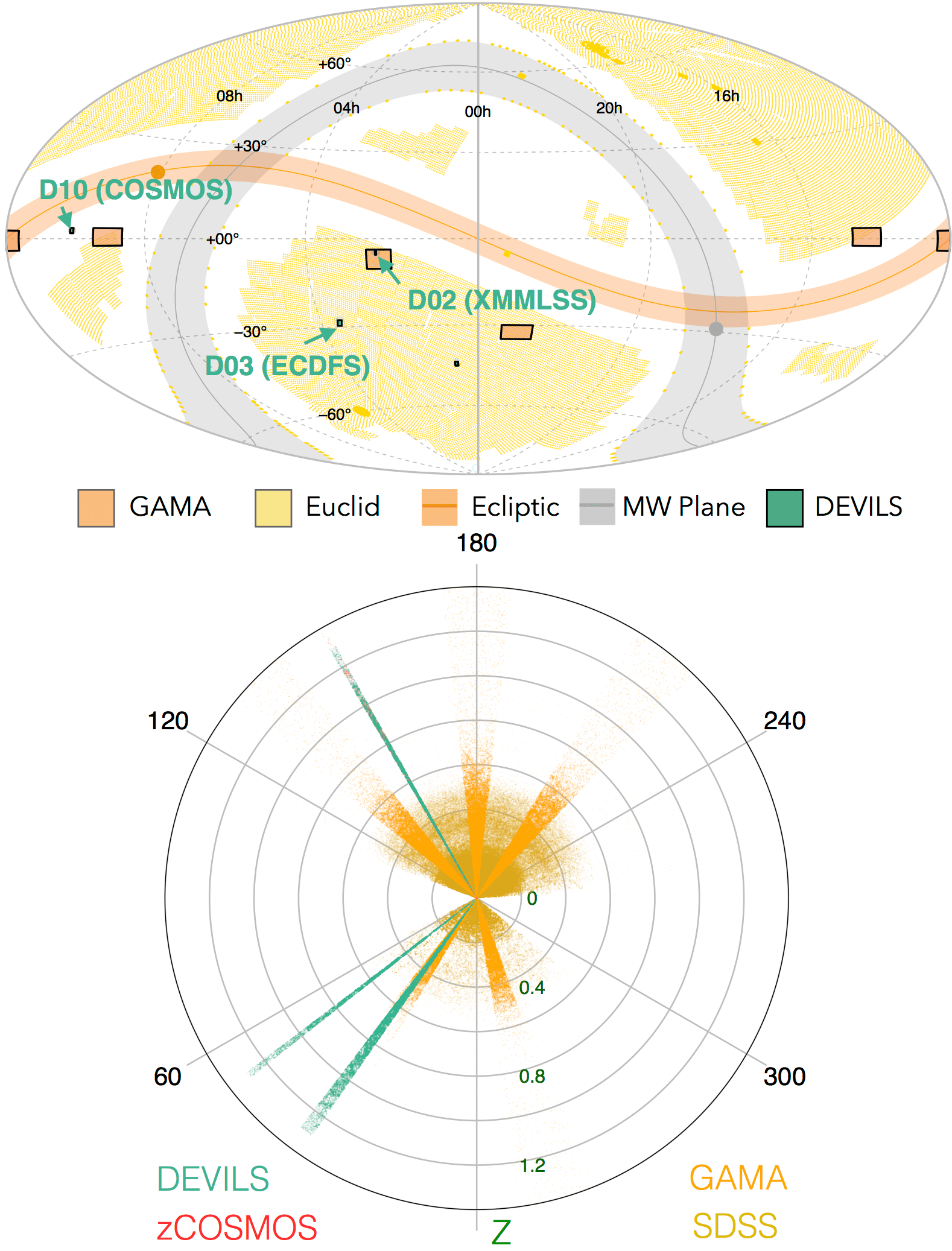}
\caption{DEVILS field positions with respect to the GAMA regions and current Euclid footprint (top) produced by \textsc{AstroMap} (\url{astromap.icrar.org}) and predicted light cones distribution of DEVILS sources from our Theoretical Astrophysical Observatory (TAO) simulations (see Section \ref{sec:Depth}) in comparison to SDSS, GAMA and zCOSMOS (note zCOSMOS sits under the D10 distribution at $\sim150$\,deg, but extends to slightly higher redshift).}
\label{fig:Fields}
\end{center}
\end{figure*}

\vspace{1mm}

To probe the processes that shaped the local Universe we require a consistent parameterisation of both galaxies and the larger scale dark matter distribution in which they reside. This can only be achieved through the identification of structures on sub-Mpc (group) scales. On these scales, dark matter haloes virialize and merge, and gas collapses to form galaxies. Thus, this regime is paramount to our understanding of baryon physics, and the interplay between dark matter and directly-observable galaxy components. To study sub-Mpc scales spectroscopic completeness is key, as even the most high fidelity photometric redshifts are not precise enough to identify these structures. However, at intermediate redshifts there is a paucity of fully sampled and complete spectroscopic surveys.

\vspace{1mm}

Until recently, the state of the art survey which probed this epoch \citep[zCOSMOS-bright,][]{Lilly07} was encumbered by its very small area ($\sim$1deg$^{2}$), sparse sampling, and complex footprint (due to slit-mask spectroscopy). This ultimately leads to low completeness \citep[only $\sim50\%$ to $i<22$, see][]{Davies15a}. Other surveys at this epoch have focused on the sparse sampling of colour-selected populations over large volumes, such as the VIMOS Public Extragalactic Redshift Survey \citep[VIPERS,][]{Garilli14} and DEEP2/3 \citep{Cooper12, Newman13}. More recently the VLT's Large Early Galaxy Census \citep[LEGA-C][]{vanderWel16} have focused on the sparse sampling of K-band selected galaxies at 0.6$<$z$<$1.0, but reaching high signal-to-noise continuum spectra for the detailed study the ages, metallicities and velocity dispersions of galaxies at this epoch. The designs of these surveys, while matched to their specific science goals, are not tuned to studying the evolution of galaxy groups, mergers, sub-Mpc structure and the influence of environment on galaxy evolution (see Figure \ref{fig:Comp} and Figure \ref{fig:Halos}). These are highly significant regimes which contribute to the formation of the fundamental relations observed today. To date, a holistic view of galaxy evolution over cosmic history relies on stitching together census-class surveys of the nearby Universe with these sparsely-sampled pencil-beam surveys of the very distant Universe. 

Recently, there have been a number of surveys that apply the high spectroscopic completeness approach to slightly fainter magnitudes than GAMA but in smaller area fields, such as the Active Galactic Nuclei (AGN) and Galaxy Evolution Survey \citep[AGES, covering 7.7\,deg$^{2}$ to $\sim$94\% completeness at i$<$20\,mag,][]{Kochanek12}, the Smithsonian Hectospec Lensing Survey \citep[SHELS, covering 8\,deg$^{2}$ to $\sim$94\% completeness at r$<$20.2\,mag,][]{Geller16}, and hCOSMOS \citep[covering 1\,deg$^{2}$ to $>$90\% completeness at r$<$20.6\,mag,][]{Damjanov18}.

To continue this trend out to higher redshifts and to overcome the issues associated with sparely sampled surveys, we are undertaking the Deep Extragalactic VIsible Legacy Survey (DEVILS\footnote{\url{https://devilsurvey.org/}}) - a magnitude-limited (Y$<$21.2\,mag, $\sim$1.5 magnitudes fainter than AGES), high completeness ($>95\%$) spectroscopic survey of three well-established legacy fields: XMM-Newton Large-Scale Structure field, XMM-LSS, Extended Chandra Deep Field-South, ECDFS, and Cosmological Evolution Survey field, COSMOS (see Figure \ref{fig:Fields}). DEVILS is designed to detect down to the stellar masses of M$^{*}_{z=0}$ galaxies to $z$\,=\,1 \citep[$10^{10.8}$\,M$_{\odot}$ - the typical galaxy in the local Universe in terms of mass-density budget,][]{Wright17}, major merger pairs of M$^{*}_{z=0}$ galaxies to $z$\,=\,0.8, and groups down to $10^{13}$\,M$_{\odot}$ to $z$\,=\,0.7 (Figure \ref{fig:Halos}). 

In this paper we present the DEVILS survey design - including key science, field selection and auxiliary data  (Section \ref{sec:Design}), generation of the DEVILS target list - including source finding, masking, photometric separation of potential stars and galaxies and visual classification (Section \ref{sec:Targets}),  an overview of our final input catalogue and observing strategy (Section \ref{sec:Input}), and early results from the 2017/B observations (Section \ref{sec:Obser}).

\section{Key Science Overview and Motivation}
\label{sec:KeyScience}
In this section we describe the key science objectives of DEVILS in the context of target selection and survey strategy. While the legacy impact of DEVILS will be open-ended, here we highlight the two key science goals of the project. This section is designed to provide an introduction to the core DEVILS science and motivate the subsequent survey design.

\subsection{The Late-time Evolution of the Halo Mass Function}

The distribution of dark matter haloes \citep[Halo Mass Function, HMF, $e.g.$][]{Press74,White78} and its evolution is one of the strongest predictions of the $\Lambda$CDM cosmological model. With the advent of precision cosmology from the Wilkinson Microwave Anisotropy Probe \citep[WMAP,][]{Bennett13} and $Planck$ \citep{Planck16}, the theoretical prediction of the HMF and its evolution places a robust constraint on the distribution of matter at a given epoch. Combined, the $\Lambda$CDM model and concordance cosmology now predict the complete development of structures on $>$5\,Mpc scales from the surface of last scattering (CMB, $z\sim1100$), to the current epoch, with zero free parameters (on smaller scales baryon physics is required to explain the growth of structure). Comparison of the HMF at $z<0.1$ over three orders of magnitude (GAMA - Robotham et al in prep), demonstrates a remarkable affirmation of the $z=0$ $\Lambda$CDM paradigm. The next critical challenge for $\Lambda$CDM is to test the predicted strong evolution of the HMF from $z=1$ to $z=0.1$. This evolution arises from the late time assembly of clusters and massive groups \citep[$e.g.$][]{Vikhlinin09}.

One of the primary ways to directly parameterise dark matter to low halo masses (group scale) is through the construction of group catalogues \citep[$i.e.$][]{Yang07,Robotham11}. Other approaches (CMB, redshift space distortions, strong/weak lensing, X-ray luminosities) can statistically recover the total dark matter content and provide some information of its distribution, but cannot recover individual halo masses except for the most massive, and rare, clusters. However, through groups we can parameterise the dark matter mass (and average density) of individual haloes down to $\sim$10$^{12}$\,M$_{\odot}$, as the group member's motions directly map to halo mass \citep[M$_{\mathrm{halo}}$\,$\propto$\,$\sigma^2r$\,or\,$r^3$, ][with typical errors in halo mass of $\sim$0.7dex for low multiplicity groups, N=4-5]{Eke06, Robotham11}. At this limit approximately 30\% of the dark matter is bound (and hence constrained, Robotham et al in prep), with the remainder either entirely unbound or in lower mass haloes. Nevertheless, group catalogues provide an extremely powerful mechanism to directly uncover a considerable fraction of the underlying dark matter {\it distribution} within a specific volume, by tracing out the gravitational potential defined by filamentary structure \citep[$e.g.$][]{Alpaslan14, Darvish15, Viola15, Kraljic18}.

\begin{figure}
\begin{center}
\includegraphics[scale=0.6]{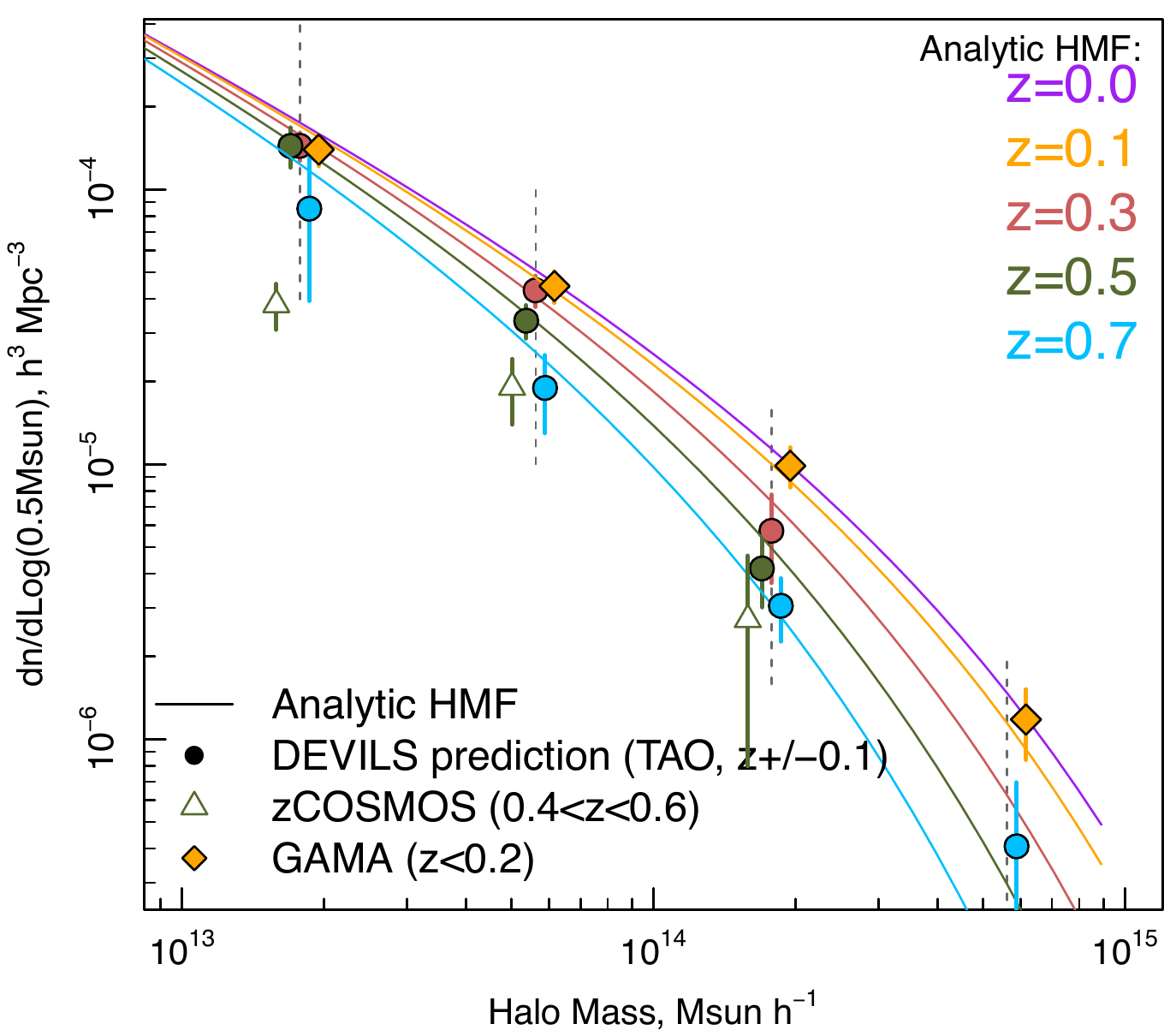}
\caption{Evolution of the analytical form of the halo mass function (HMF) from $\Lambda$CDM and DEVILS observational predictions from our TAO simulations (see Section \ref{sec:Depth}). Solid lines display the analytic form of the HMF from \citet{Murray13} at a range of redshifts. We assume that a dark matter halo has its mass parameterised if we detect $>3$ group members at our Y-mag<21.2 limit and show the resultant predicted DEVILS HMF at each epoch. We also display the HMF measured using the zCOSMOS groups sample of \citet{Knobel12} at $0.4<z<0.6$. If present, DEVILS will measure the late-time evolution of the high mass end of the HMF predicted by $\Lambda$CDM.}
\label{fig:HMF}
\end{center}
\end{figure}

GAMA has led the way in constraining the distribution of haloes to low masses in the local Universe \citep[M$_{\mathrm{Halo}}\sim10^{12}$\,M$_{\odot}$  at $z<0.2$][]{Robotham11}. Observationally however, we know little about the evolution of group-scale haloes at higher redshifts, and have previously been restricted by the use of photometric redshifts \citep[$e.g.$][]{Leauthaud12, Hatfield16}. By conducting a survey specifically optimised for the identification of group-scale dark matter haloes to $z=0.7$ (see Figure \ref{fig:Halos}), DEVILS will confirm the evolution, or lack thereof, in the massive-end of the HMF; extending below the massive cluster regime probed by X-ray cluster surveys \citep[$e.g.$][]{Vikhlinin09}. This will provide a stringent test of one of the clearest predictions of $\Lambda$CDM (Figure \ref{fig:HMF}). In addition, recently proposed alternative approaches to identify $\Lambda$CDM tension using cluster/group velocity dispersions alone \citep{Caldwell16}, can be directly tested via the construction of group catalogues in deep spectroscopic surveys (such as DEVILS).

\subsection{The Effect of Environment in Regulating Galaxy Evolution over the Last 8\,Gyr}

One of the most fundamental questions in astrophysics is what processes have shaped the formation and evolution of galaxies we observe today? To answer this question we typically use two complementary methods; either probing the factors which affect the growth of individual galaxies as a function of stellar mass, environment, epoch, etc, or by measuring the global evolution of the ensemble galaxy population via distributions such as the stellar mass function \citep[SMF $e.g.$ see][]{Marchesini09,Behroozi13,Muzzin13,Tomczak14,Driver18} or cosmic star-formation history \citep[CSFH $e.g.$ see][]{Bell05,Hopkins06,Schaye10, Madau14, Davies16b, Driver18}. It is the former of these methods which ultimately probes the processes which shape the latter. As such, we would ideally like to target both the underlying astrophysics $and$ the fundamental relations which they produce. To date, we have exquisitely probed these global relations in the local Universe \citep[$e.g.$][]{Bell03, Panter04, Elbaz07, Peng10, Baldry12, Chang15, Moffett16, Davies16b, Wright17}, and also have some understanding of how the distributions vary with time \citep[$e.g.$][]{Behroozi13, Vulcani13, Madau14, Lee15} and environment \citep[$e.g.$][]{Yang09,Bolzonella10, McNaught-Roberts14,Eardley15,Davidzon16,Tomczak17}. However, the astrophysics involved in $shaping$ these distributions at each epoch/environment are far from clear. 

Measuring the factors which govern the growth of galaxies is problematic. Most simplistically, there are two high level mechanisms that shape the UV-FIR properties by which we trace galaxy evolution: star formation \citep[SF, $e.g.$ see review of ][]{Kennicutt12} and mergers \citep[$e.g.$][, and see review of Conselice 2014]{Bundy04, Baugh06, Kartaltepe07, Bundy09,Jogee09,deRavel09,Lotz11,Robotham14}, which respectively form and redistribute the stellar material. However, a complex array of effects all have a significant impact on the evolution of galaxies, such as black hole growth \cite{Hopkins08,Kormendy13}, AGN feedback \citep{Kauffmann04,Fabian12}, gas-accretion \citep{Kauffmann06,Sancisi08}, starvation/strangulation \citep{Moore99}, atomic-to-molecular gas fraction \citep[$e.g.$][]{Wong02, Bigiel08, Popping14}, tidal stripping \citep{Gunn72, Moore99, Poggianti17,Brown17}, harassment \citep{Larson1980},  morphological transformations \citep{Conselice14, Eales15}, and very local \citep{Patton11, Scudder12, Scudder15, Davies15b} and larger-scale \citep{Dressler80, Giovanelli85, Peng10, Cortese11, Darvish16} environments. All of these factors largely drive the observed changes to star-formation and merger state. It is the varying contribution of these processes over the history of the Universe that lead to the relations we observe today.

\subsubsection{Star Formation, Mergers and the Growth of Stellar Mass} 

In the local Universe we have begun to explore how mergers and star formation are changing the distribution of stellar material at the current epoch. \cite{Robotham14} determined robust merger rates for GAMA, and estimated the relative contribution of in-situ star formation and mergers in shaping the $z\sim0$ SMF. They found that M$^{*}_{z=0}$ represents an important transition between merger- and SF-dominated growth, with sub-M$^{*}_{z=0}$ galaxies predominantly growing via star-formation and larger mass galaxies gaining their mass via mergers. However, we know very little about how these relative contributions evolve with redshift. For example, we know that both star formation and merger rates increase with look-back time \citep[$e.g.$][]{Lotz11, Lee15}, but does the relative stellar mass growth via these mechanisms evolve consistently? And can their relative contribution account for the evolution in the shape of the SMF?  Interestingly, \cite{Bell03} find that while the majority of stars formed at $z<1$ are produced in blue late-type galaxies, the bulk of stellar mass is accumulated in red early type galaxies, suggesting it is both star-formation and mergers at intermediate redshift that contribute strongly to formation of the observed $z\sim0$ mass function.

\begin{figure}
\begin{center}
\includegraphics[scale=0.42]{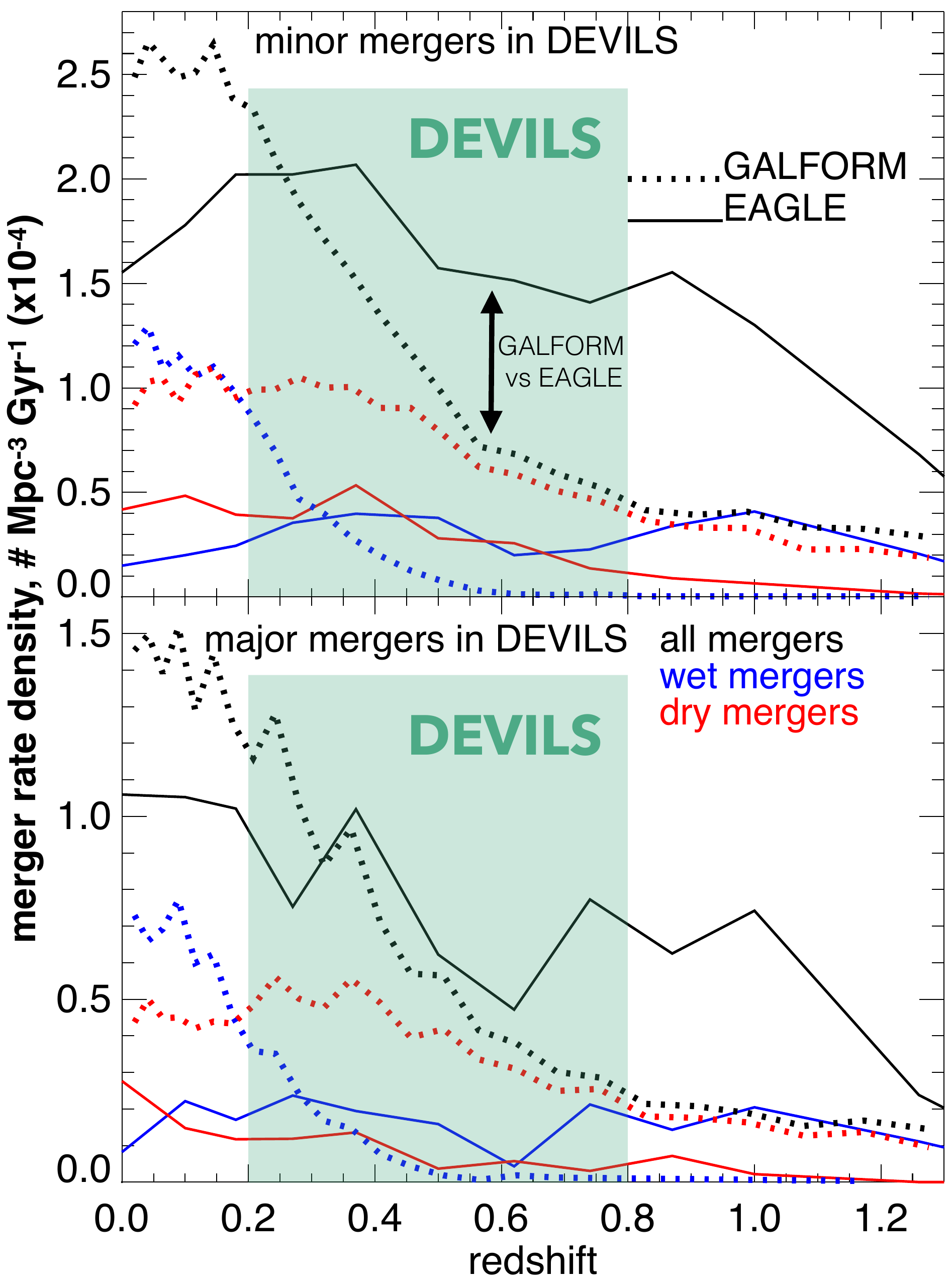}
\caption{GALFORM (dotted lines) and EAGLE (solid lines) predictions for minor (top) / major (bottom), and wet (blue) / dry (red) merger rates at a DEVILS-like stellar mass limit. Here wet/dry mergers are defined by Hydrogen gas content with (M$_{\mathrm{HI}}$+M$_{\mathrm{H2}}$)/M$_{\mathrm{star}}>0.5$ and $<0.1$ for wet and dry respectively ($i.e.$ intermediate gas fractions are not displayed). Over the DEVILS volume, GALFORM and EAGLE predict very different major-minor merger rates (factor of two) and major wet-dry merger fractions (factor of $\sim10$). The red and blue arrows are }
\label{fig:clau}
\end{center}
\end{figure}

\begin{figure*}
\begin{center}
\includegraphics[scale=0.205]{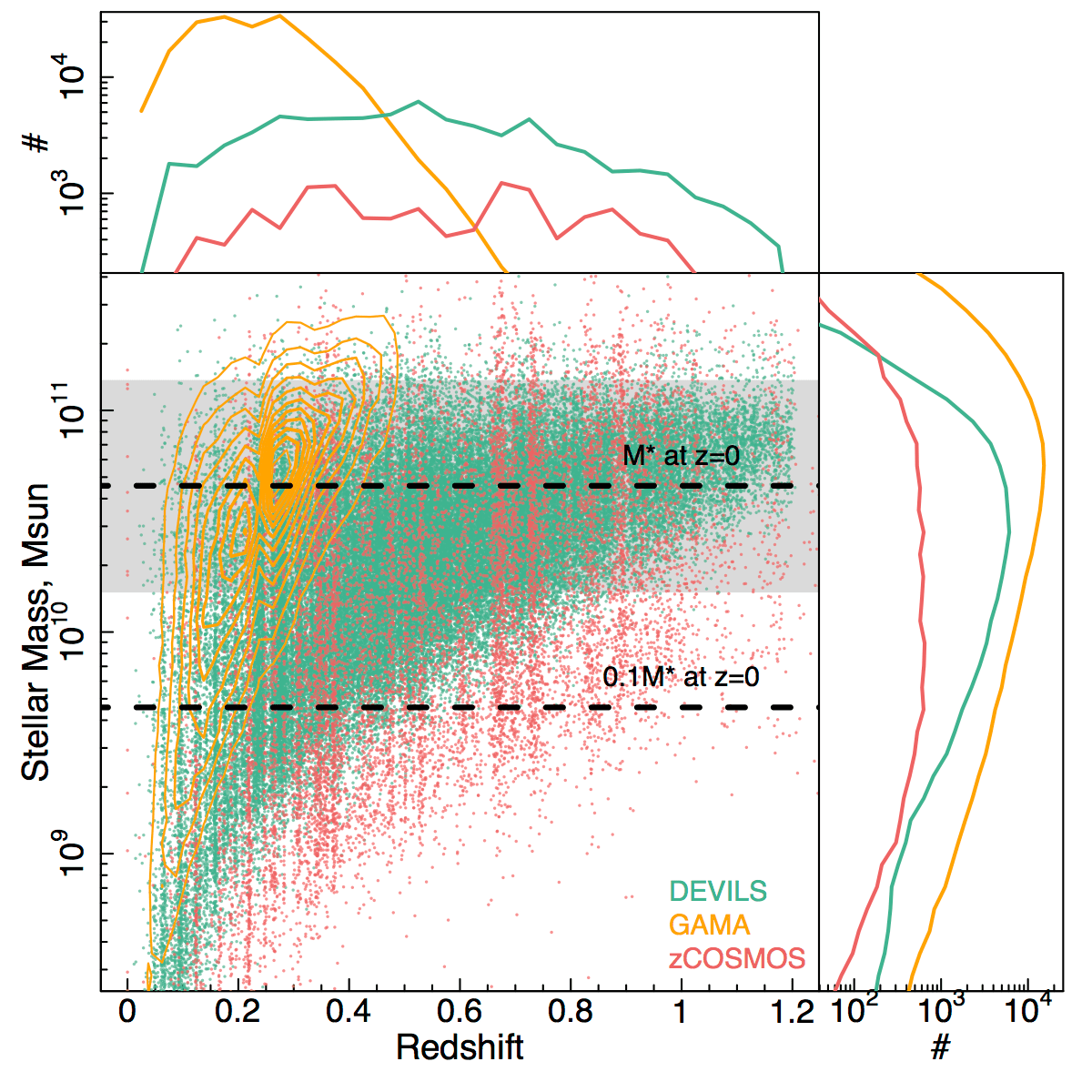}
\includegraphics[scale=0.57]{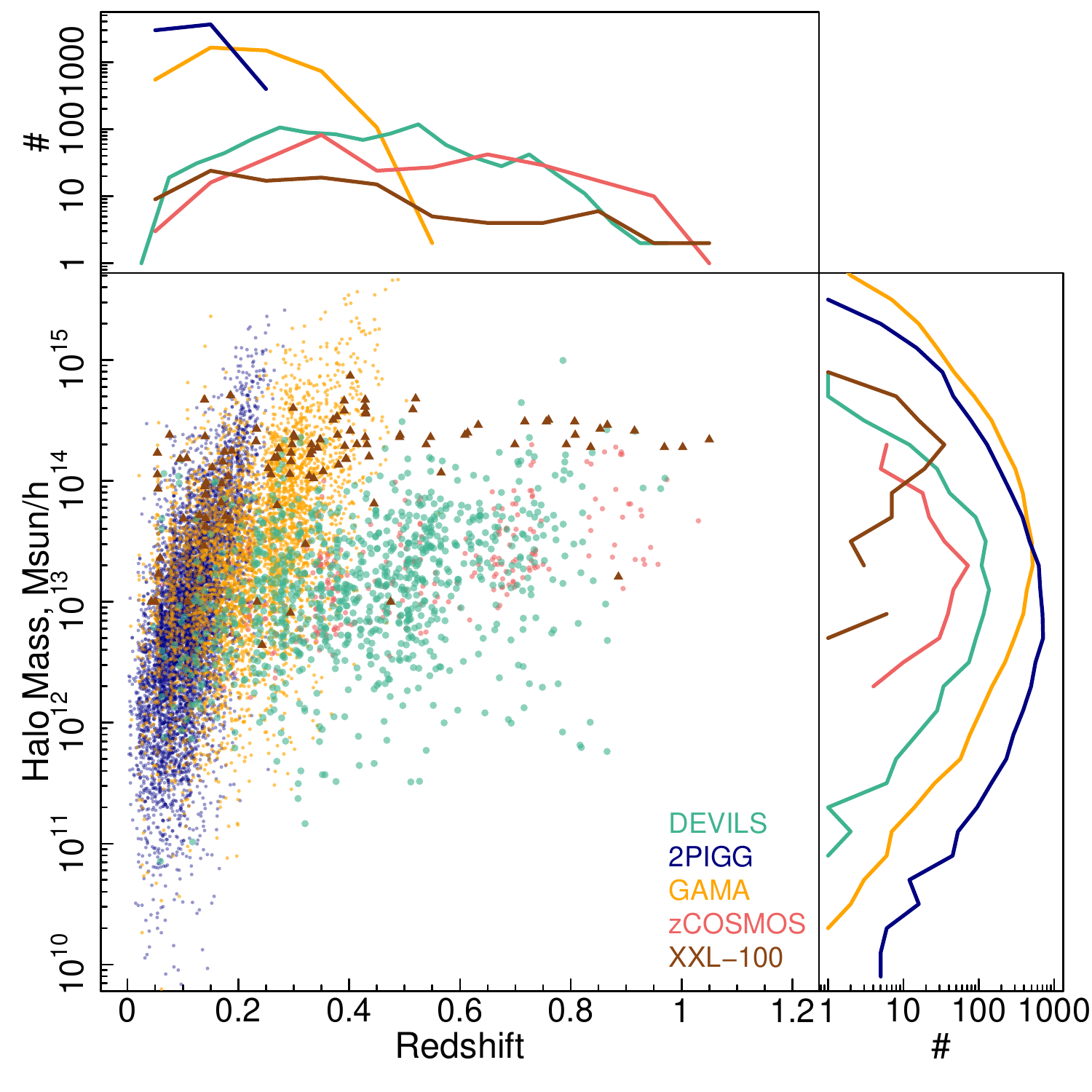}
\caption{DEVILS predicted stellar mass-redshift distribution (left) and halo mass-redshift distribution (right). For clarity, in the left hand panel, we only show density contours for the GAMA sample. Distributions are predicted using the Theoretic Astrophysical Observatory (TAO) for 6deg$^{2}$,  a Y-mag$<21.2$ selection and 95$\%$ completeness (we randomly remove $5\%$ of sources). For group halo masses, we assume a group with $>3$ members has its mass constrained. For comparison we show the same distributions for both GAMA and zCOSMOS (and for groups 2dFGRS-2PIGG and the XXL-100 bright cluster sample). DEVILS will detect M$^{*}_{z\sim0}$ galaxies out to $z\sim1$ (dashed line on the left panel), identify major merger ($<$3:1 mass ratio) pairs to M$^{*}_{z\sim0}$ galaxies out to $z\sim0.8$ (grey shaded band on left panel),  and parameterise 10$^{13}$M$_{\odot}$ haloes out to $z\sim0.7$  (right panel).}
\label{fig:Mz}
\end{center}
\end{figure*}

To investigate these processes requires, at least, measurements of both star formation and mergers over a broad redshift range and to a much higher velocity resolution than obtainable by photo-$z$s alone ($<100$\,km\,s$^{-1}$ compared to $\sim1000$\,km\,s$^{-1}$ for photo-$z$s). Previously a number of surveys have explored the interplay between star-formation and mergers at intermediate redshift, using precise photo-$z$s and high resolution imaging. Most notable of these is the Galaxy Evolution from Morphologies and SEDs (GEMS) survey \citep{Rix04}, which combined deep Hubble Space Telescope (HST) imaging with the precise ($\delta z$/(1+$z$) = 0.006 $\pm$0.020) photo-$z$s of the Classifying Objects by Medium-Band Observations in 17 Filters survey \citep[COMBO17,][]{Wolf01,Wolf04}. The GEMS+COMBO17 survey provides invaluable information about the contribution of ongoing major mergers at intermediate redshifts \citep[$e.g.$][]{Bell06,Robaina09}; $i.e.$ those identified via visual disturbance in HST imaging. However, without spectroscopic redshifts to identify close, but not visually disturbed, pair systems they may be missing a significant fraction of merger events. In addition, the lack of environmental resolution ($i.e.$ to the group-scale) provided by photo-$z$s inhibits these surveys from exploring the environmental trends of mergers and star-formation activity.  

Simulations add little constraint to merger rates with models predicting half an order of magnitude variation of different merger types: $i.e.$ major/minor - defined at a 3:1 stellar mass ratio, and wet/dry - based on HI gas content. Thus, to constrain these models further we require stronger observational constraints. For example, at a DEVILS-like stellar mass limit, the Evolution and Assembly of GaLaxies and their Environments \citep[EAGLE][]{Schaye15} and Galform \citep{Lacey16} models predict vastly different minor merger rates and wet/dry merger fractions at \textit{z}\,$>$\,0.3 \citep[Figure \ref{fig:clau} and see][for an example of mergers in EAGLE]{Lagos18}. 

To differentiate between the effect of interactions in competing galaxy formation models, we must robustly identify highly complete samples of merging galaxies at intermediate redshift. Within DEVILS we will have a sample with which to identify pre-merger pairs \citep[as in][]{Robotham14}, deep high resolution imaging to identify post-merger systems, and complementary HI data to explore the importance of wet/dry mergers (see Section \ref{sec:Fields}). Combined this will allow us to explore the relative contribution of star-formation and mergers in shaping the local galaxy population.

\subsubsection{The Environmental Dependence of Galaxy Evolution} 

Observationally, it has long been known that in the local Universe the cluster environment on scales of $>$1\,Mpc can leave strong imprints on galaxies. These environments can affect properties such as morphology \citep{Dressler80}, colour \citep{Kodama01}, stellar mass \citep{Ostriker75}, AGN fraction \citep{Kauffmann04}, star formation \citep{Peng10}, gas content \citep{Gunn72} and kinematic structure \citep[$e.g.$][]{Cappellari11,Fogarty14}. More recently, within GAMA and SDSS we have begun to explore the effect of more local environment ($<$\,Mpc, groups $\&$ pairs). Numerous studies have found that while different environments can display different luminosity/mass functions, when controlled for mass, it is local galaxy-galaxy interactions that leave the strongest imprint on galaxy properties \citep[$e.g.$][]{Alpaslan15,Scudder15,Grootes17}. Specifically the effect of $<$\,Mpc environment can drive significant (factor of $\sim4$) changes to star formation \citep[$e.g.$][]{Patton11, Scudder12,  Davies15b, Davies16a}. These changes are intimately linked to the underlying atomic and molecular gas distribution, which can be easily disrupted in over-dense environments via turbulence, ram-pressure stripping and strangulation \citep[$i.e.$][]{Nichols11,Nichols13}. 

At higher redshifts, the quiescent fraction in over-dense environments is found to evolve dramatically to $z\sim1$ \citep{Darvish16} and potentially even reverse at $z>1$ \citep{Elbaz07}. However, more recent results from $Herschel$ suggest this reversal may occur at much earlier times \citep[for example, see][]{Elbaz07,Popesso11,Popesso12,Ziparo14}. In terms of structure, the morphology-density relation in the most massive clusters appears in place by $z\sim1$, but in group-sized haloes evolves dramatically between $0<z<1$ \citep{Smith05}. In addition, determining the relative importance of large-scale astrophysical processes such as pre-processing in group environments at these redshifts is essential in understanding the observed environmental trends at $z\sim0$.

DEVILS will simultaneously trace the evolution of galaxies \textit{and} structure on 0.01-10\,Mpc scales ranging from mergers and pairs, to groups, clusters, filaments and voids. This will allow us to finely grid in stellar mass, morphology, halo mass, environment and epoch (see Section \ref{sec:Design}) to determine the origin of the $z\sim0$ fundamental relations observed by GAMA and SDSS. 

\section{Survey Design}
\label{sec:Design}

\subsection{Facility and Instrument Characteristics}
\label{sec:Facility}

DEVILS spectroscopic observations are currently being undertaken at the 3.9m Anglo-Australian telescope at the Siding Spring Observatory in New South Wales with the AAOmega fibre-fed spectrograph \citep[][]{Sharp06,Saunders04} in conjunction with the Two-degree Field \citep[2dF,][]{Lewis02} positioner. The 2dF positioner has been at the forefront of large galaxy redshift surveys in the local Universe (such as 2dFGRS and GAMA) and has also been used to great success in targeting large numbers of sources to higher redshift \citep[$e.g.$ 2dFLenS and WiggleZ,][]{Blake16, Drinkwater18} and more recently to faint magnitudes \citep[OzDES,][]{Childress17}. As such, it is an ideal facility to perform our deep intermediate redshift survey.  2dF allows the simultaneous observation of $\sim400$ targets with the AAOmega spectrograph, with 2$^{\prime\prime}$ diameter fibres. The 2dF top-end consists of an atmospheric dispersion compensator (ADC) and robot gantry that positions fibres to 0.3$^{\prime\prime}$ accuracy on sky.     

AAOmega observes in two spectral channels (blue and red), both equipped with a 2k$\times$4k E2V CCD detector and an AAO2 CCD controller. We observe with the 5700\AA\ dichroic allowing for simultaneous coverage from 3750\AA\ to 8850\AA. For the blue CCD we use the 580V grating with central wavelength of 4820\AA\ providing a $\sim1.03$\AA/pix dispersion, while for the red CCD we use the 385R gating with central wavelength of 7250\AA\ providing a $\sim1.56$\AA/pix dispersion. This results in a spectral resolution that varies from R$\sim$1000 (blue) to R$\sim$1600 (red). This spectral resolution and wavelength range was selected to enable detection of at least the [OII] (3727\AA) emission line and 4000\AA\ break over our full target redshift range and provide sufficient velocity resolution with which to identify close pairs \citep[$<$50\,km\,sec$^{-1}$, see][]{Robotham14}.

\subsection{Selection Band, Depth and Area}
\label{sec:Depth}

\subsubsection{DEVILS Simulations Using the Theoretical Astrophysical Observatory Tool}

In the following subsections we describe how we define the DEVILS selection band, depth and area. This is largely achieved using simulated DEVILS lightcones generated using the Theoretical Astrophysical Observatory \citep[TAO,][]{Bernyk16} tool\footnote{\url{https://tao.asvo.org.au/tao/}}. Here we briefly describe the TAO simulations we generate and any assumptions made. Note that TAO provides all of the functionality to go from these input parameters to fully simulated lightcones.  

We use Millennium N-body simulation with WMAP cosmology \citep{Springel05} coupled with the Semi-Analytic Galaxy Evolution (SAGE) model \citep{Croton16} to simulate the galaxy distribution in each of the DEVILS regions independently. When generating lightcones, we restrict our simulated galaxies to $z<1.2$ to limit computational time (the small number galaxies detected at $z>1.2$ will be negligible in the DEVILS sample). Spectral energy distribution (SED) models from \cite{Bruzual03} with a Chabrier initial mass function \citep{Chabrier03} are used to estimate intrinsic galaxy magnitudes from the SAGE ages, SFHs and metallicities. Using the relationship between colour excess, E(B-V), and instantaneous star formation rate TAO estimates the dust content of each galaxy. The galaxy SED is then extincted using a Calzetti extinction law \citep{Calzetti94} and observed-frame apparent magnitudes are calculated using the galaxy's luminosity distance, extincted SED and common filter sets. TAO provides magnitude measurements for the VISTA Y-band, which we use for our DEVILS selection. All light cones are then cut at Y$<$22.0\,mag to limit computational time.

\subsubsection{Selection Band}

To achieve the science goals stated previously, DEVILS must probe to faint magnitudes and high completeness, and obtain robust samples of individual galaxies, pairs of galaxies and groups.  The majority of the galaxy evolution science we are undertaking in DEVILS requires stellar mass-selected samples \citep[for example, see][and many other GAMA papers for a detailed discussion of the benefits of a stellar mass-based selection\footnote{\url{http://www.gama-survey.org/pubs/onads.php}}]{Taylor11}. In addition, for our halo science we require a relatively unbiased tracer of galaxies within haloes. As such, we would ideally like our target selection to be in terms of stellar mass. However, without prior distance information this can not be achieved. Thus, we select based on a single imaging band which is closely correlated with stellar mass. While colours could be used, these create complex selection functions which can bias any galaxy evolution studies and may lead to biases in halo identification, as environment is a strong driver of galaxy colour \citep[$e.g.$][]{Kodama01}. Various authors \citep[$e.g.$][]{Bell00,Taylor11} have shown that the individual near-infrared (NIR) bands are most correlated with stellar mass, as they are dominated by emission from old stars (which in turn dominates the mass of the galaxy). 

The most significant gains in directly relating a single observation band to stellar mass, come from ensuring that the survey selection band remains above the 4000\AA\ break in the rest-frame of the galaxy for the entire sample. This is true for all NIR bands to $z\sim1$. Secondly, we also wish to stay as close as possible to the spectral observing range of the AAT+AAOmega, such that our selection band is representative of the spectral emission we will receive. Last, we also wish to select in a band which has consistent deep imaging, from the same facility/instrument over a number of deep fields spread in right ascension (see Section \ref{sec:Fields}). We find that the Visible and Infrared Survey Telescope for Astronomy (VISTA) Y-band (centred at 1.02$\mu$m) uniquely meets all of these criteria.

\subsubsection{Depth}

The key science goals of DEVILS require us to trace the evolution of galaxies, galaxy interactions, and the most massive groups and clusters out to intermediate redshifts. In order to parameterise these we aim to select galaxies to: i) measure the evolution of typical M$^{*}_{z\sim0}$-like galaxies out to $z\sim1$, ii) identify all major merger ($<$3:1 mass ratio) pairs to M$^{*}_{z\sim0}$-like galaxies out to $z\sim0.8$, and iii) measure the mass of 10$^{13}$M$_{\odot}$ haloes out to $z\sim0.7$. To test the depth required to achieve these goals, we use our simulated DEVILS survey volumes from TAO. We apply varying Y-band magnitude limits (between 20.0$<$Y$<$22.0\,mag) and investigate the resultant DEVILS sample we would obtain if we reached $95\%$ spectroscopic completeness (we randomly remove $5\%$ of sources, but note that in practice this will not be random but related to some galaxy characteristic). In terms of haloes, we assume that a dark matter halo has its mass measured if we detect $>3$ group members \citep[with $\sim$0.7\,dex error][]{Robotham11}. We find that to reach our target goals in terms of galaxies and haloes outlined above, requires a Y$<21.2$\,mag selection (and see Figure \ref{fig:Halos}). The resulting predicted stellar mass and halo mass distributions as a function of redshift are shown in Figure \ref{fig:Mz}. In the left panel we also highlight M$^{*}_{z\sim0}$ at 10$^{10.8}$M$_{\odot}$ \citep{Wright17} and the region where major merger pairs to M$^{*}_{z\sim0}$ galaxies (based on a 3:1 mass ratio) would be found - highlighting that we can detect both galaxies in these types of merger systems to $z\sim0.7$.  We then compare these to the distribution from GAMA \citep[at low redshift,][]{Liske15} and zCOSMOS-bright \citep[as a comparable sample at a similar redshift to DEVILS]{Lilly07}. For groups we also show the 2dFGRS-2PIGG \citep{Eke06} distribution as a $z\sim0$ and the ultimate XMM extragalactic survey (XXL) 100 bright cluster sample \citep{Pacaud16} to $z\sim1$ for comparison.           

\subsubsection{Area}

Using our TAO simulations, we can also make a simplistic estimation regarding the sample size we require to subdivide our population based on stellar mass, halo mass and redshift in order to achieve our key science goals. To measure the evolution in the HMF, at a minimum we require, $\sim$two/three $z>0.3$ redshift bins each containing a few tens of groups over a $\sim0.5$\,dex range in halo mass. This will allow us to differentiate between the HMF at each epoch, assuming the analytic form predicted by $\Lambda$CDM to be correct; $i.e.$ in our TAO simulations with a few tens of groups in each 0.5\,dex range the combined errors on the measured HMF at z=0.3 and z=0.7 are smaller than the evolution in the analytic form of the HMF over this epoch (see Figure \ref{fig:HMF}). To constrain the SMF to M$^{*}_{z\sim0}$ at the high redshift end of our sample, requires $\sim$\,1,500 galaxies per $\Delta z$\,=\,0.2 at $z$\,=\,0.8-1 (much more at lower $z$). We note that these estimations are based on relatively simplistic assumptions, and are conservative, such that we should achieve our science goals.  To reach these numbers in our TAO simulations, we require a sample size of $\gtrsim50,000$ galaxies to Y$<$21.2\,mag. Taking the deep NIR number counts of \cite{Driver16} we predict that to observe a sample of this size requires a target area $\sim6$deg$^2$.

\begin{table*}
\begin{center}
\caption{DEVILS field positions, areas and target selection imaging.}
\label{tab:Fields}
\scriptsize
\begin{tabular}{c c c c c c c c c c c }
Field & Common & RA,Dec (cen) & RA,Dec (cen) & RA,Dec (min) & RA,Dec (max) & Area & Band & Depth & Survey  &  Limit   \\
 & Name & (sex) & (deg) & (deg)  & (deg) & (deg) &  & (mag) &  & (5$\sigma$ AB)  \\
\hline
\hline
D02 & XMM-LSS & 02:22:06.0, -04:42:00.0 & 35.53, -4.70 & 34.00, -5.20 & 37.05, -4.20 & 3.00 & VISTA-Y & 21.2 & VIDEO & 25.0 \\
D03 & ECDFS & 03:32:27.12, -28:00:00.0 & 53.113, -28.00 	 & 52.263, -28.50 & 53.963, -27.50 & 1.50 & VISTA-Y & 21.2 & VIDEO & 24.9 \\
D10 & COSMOS &10:00:09.6, 02:13:12.0 &150.04, 2.22 & 149.38, 1.65 & 150.70, 2.79 & 1.50 & VISTA-Y & 21.2 & UltraVISTA & $>$24.8 \\

\end{tabular}
\end{center}
\end{table*}

\begin{figure*}
\begin{center}
\includegraphics[scale=0.52]{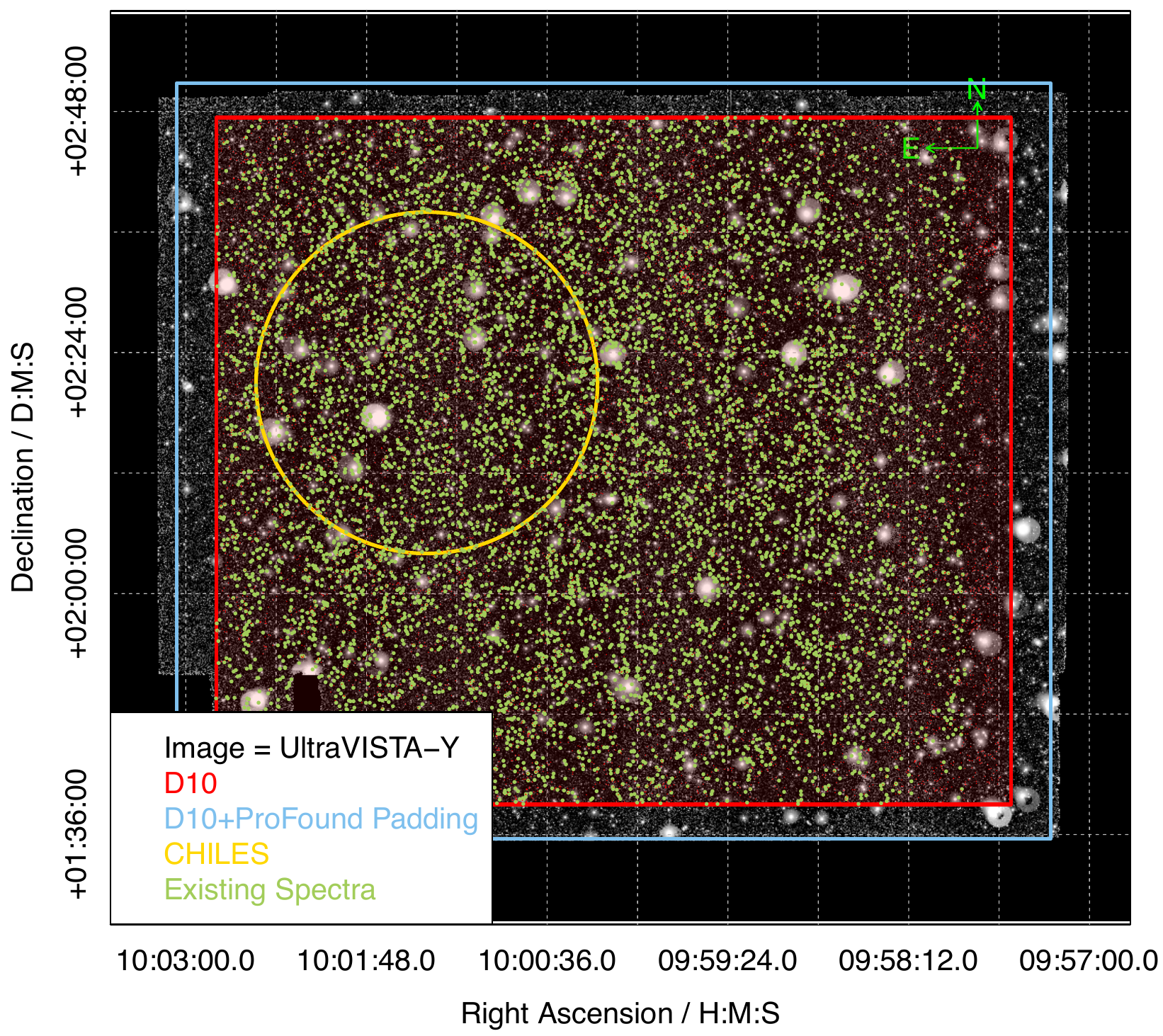}
\includegraphics[scale=0.52]{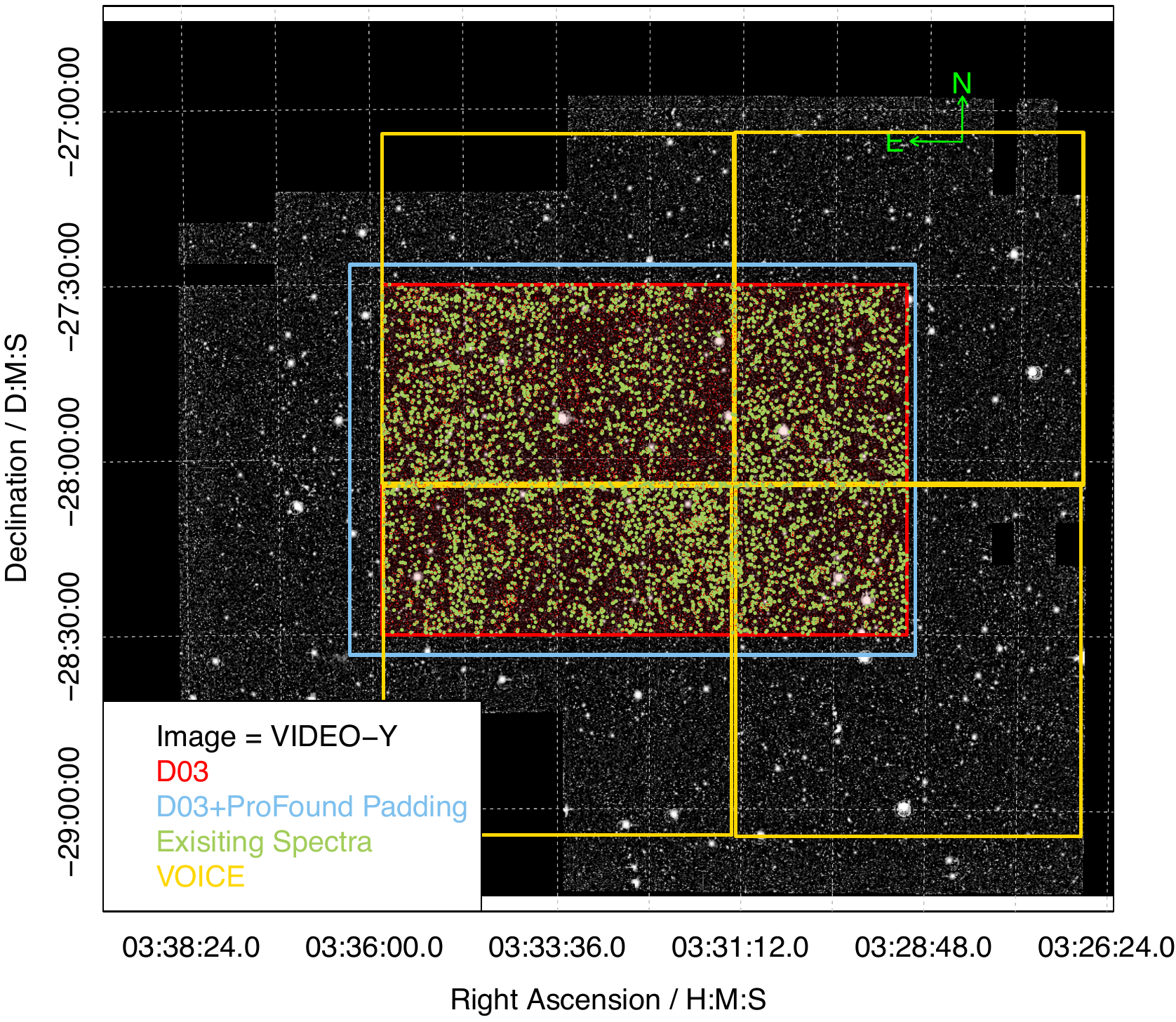}
\includegraphics[scale=0.52]{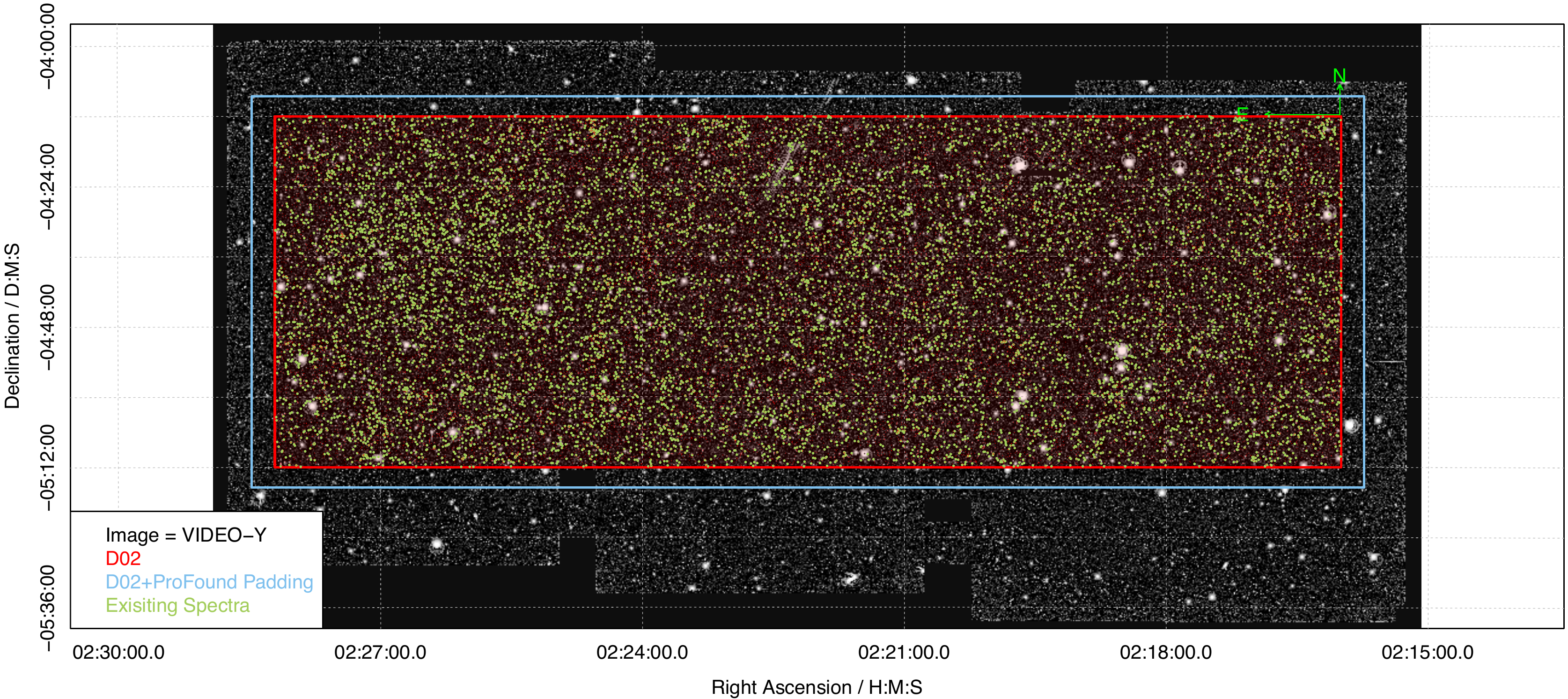}
\caption{DEVILS fields in relation to NIR VISTA Y-band imaging from VIDEO (D02, D03) and UltraVISTA (D10). Existing robust spectroscopic redshifts at Y-mag$<21.2$ are shown as faint green dots. The red box bounds the DEVILS field, while the blue box highlights the region over which we perform our \textsc{ProFound} source finding (as not to exclude sources which overlap the DEVILS field boundary). Gold regions display new, ongoing surveys which have driven our choice of region selection - the CHILES HI survey in D10 and VOICE VST imaging survey in D03.}
\label{fig:FieldPos}
\end{center}
\end{figure*}

\subsection{Field Selection and Existing Data}
\label{sec:Fields}

In order to maximise the scientific return of the DEVILS project, we aim to target well-studied fields with extensive multi-wavelength imaging and spectroscopy, which will also be targeted with upcoming next-generation facilities. Firstly, this will allow for a detailed parameterisation of the photometrically-derived properties for DEVILS sources, and secondly this will add lasting legacy to the DEVILS dataset. There are a number of such fields which are accessible to the AAT. We select three of the most well studied regions for our spectroscopic follow up: the X-ray Multi-Mirror Mission Large-Scale Structure region \citep[XMM-LSS, $e.g.$ see][]{Pierre04}, the Extended Chandra Deep Field-South region \citep[ECDFS, see][]{Virani06} and Cosmic Evolution Survey region \citep[COSMOS,][]{Scoville07}. Our priority condition for the selection of these fields is consistent deep Y-band imaging to produce our target selection covering $>6$\,deg$^2$. Each of these fields has deep VISTA imaging in the Y-band (XMM-LSS/ECDFS - VIDEO, and COSMOS - UltraVISTA) which covers a total area of $\sim10.5$\,deg$^2$. DEVILS field positions and areas are presented in Table \ref{tab:Fields}. From this point on we shall refer to the DEVILS 2h field (XMM-LSS) as D02, the 3h field (ECDFS) as D03 and 10h field (COSMOS) as D10.     

These three deep fields have also been targeted extensively with many multi-wavelength imaging and spectroscopic programs which are essential to the DEVILS survey. Deep imaging data will be used for SED fitting and the determination of galaxy properties \citep[see][for similar work in COSMOS]{Andrews17}, while existing spectra will be used within the DEVILS sample to maximise the scientific return of the program while minimising observational costs (see Section \ref{sec:existingSpec}). 

In this paper we briefly outline the previous key imaging and spectroscopic programs in the DEVILS regions, and used in our target selection. A more detailed analysis of the DEVILS multi-wavelength data and derived products will be presented in later work. We highlight that all three fields contain deep X-ray \citep[][Chen et al submitted]{Pierre06,Ranalli13,Cappelluti09}, UV \citep{Xu05, Zamojski07}, Optical \citep{deJong13,Taniguchi15,Vaccari16, Aihara17}, NIR \citep{McCracken12,Jarvis13}, MIR \citep{Lonsdale03,Surace04,Sanders07,Mauduit12,Lin16}, FIR \citep{Oliver12} and radio continuum \citep[][Hale et al in prep]{Tasse06, Tasse07, Norris06, Huynh12, Miller13, Aretxaga11, Smolcic14, Smolcic17, Schinnerer07} imaging. It is also worth noting that the deep X-ray data in these fields will provide complementary X-ray-derived group/cluster masses for the DEVILS sample; providing an independent test of our derived masses, and allowing a detailed exploration of the comparison between spectroscopically-derived group properties and those determined via X-ray emission. This will be explored further when the DEVILS sample is complete.    

These fields are also of strong interest for ongoing/upcoming large observational programs. Subaru-Hyper Suprime Camera \citep{Aihara17} is currently undertaking deep imaging programs over the DEVILS fields \citep[$e.g.$ see][in D10]{Tanaka17} and all three regions have been announced as Large Synoptic Survey Telescope (LSST) deep-drill fields (see Gawiser et al - LSST white paper\footnote{\url{https://project.lsst.org/sites/default/files/WP/Gawiser-ultradeep-extragalactic-01.pdf}}). Euclid \citep{Laureijs11} and the Wide Field Infrared Survey Telescope (WFIRST) will provide high resolution imaging in D02/D03 to complement the existing HST imaging in D10, and MeerKAT International GHz Tiered Extragalactic Exploration \citep[MIGHTEE]{Jarvis17} will produce deep 1.4\,GHz observations in all three fields to supplement existing Very Large Array (VLA) continuum observations in D10 \citep{Smolcic14, Smolcic17, Schinnerer07}. 

Finally, these fields are also the location of the next-generation deep HI surveys. The Jansky VLA's Cosmic HI Large Extragalactic Survey \cite[CHILES,][]{Fernandez13} is currently ongoing in D10 and will probe 21cm emission from galaxies to $z<0.45$, while MeerKAT programs MIGHTEE-HI in all three fields, and Looking At the Distant Universe with the MeerKAT Array \citep[LADUMA]{Holwerda11} in D03 will target HI to $z\sim0.58$ and $z\sim1.4$ respectively. Combined, MIGHTEE-HI and LADUMA will cover the full DEVILS area. The combination of the DEVILS sample with these HI surveys is a tantalising prospect both in terms of stacking DEVILS sources to extend further down the HI mass function \citep{Verheijen07, Lah07} and HI cosmic density \citep{Rhee16,Rhee18} at a given redshift, and using DEVILS-derived environmental metrics to explore the effect of environment on atomic gas content and distribution \citep[$e.g.$][]{Cortese11,Poggianti17,Brown17}.

\begin{figure*}
\begin{center}
\includegraphics[scale=0.4]{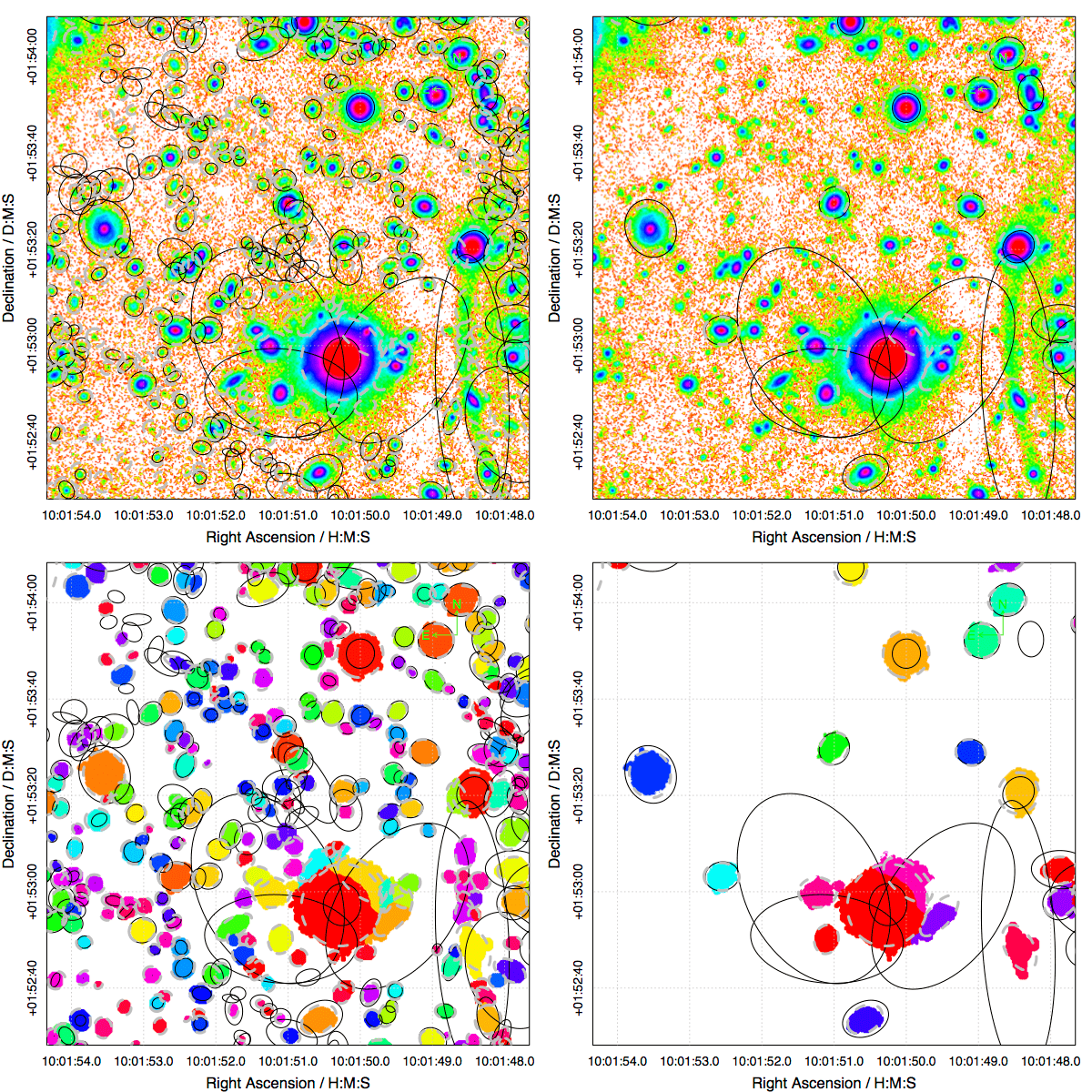}
\caption{\textsc{Profound} source detection and extraction in a complex subregion of D10. Top left: Input UltraVISTA image with rainbow colours used to display pixel flux, over-plotted as black ellipses are SExtractor-defined apertures from the UltraVISTA catalogues. Bottom left: \textsc{ProFound} segments identified in the same region (coloured by unique ID). We also over-plot in grey an ellipse defined by the \textsc{ProFound} segment; although this is not used in our analysis. The SExtractor-defined catalogue contains false detections ($i.e.$ around 10:01:53.5, 01:53:55 - top left), erroneously large apertures where multiple sources are identified as single source ($i.e.$ around the complex region at 10:01:50.4, 01:52:50), misses flux in the extended wings of some sources ($i.e.$ the source at 10:01:50, 01:53:45), and identifies the large stellar halo (bottom right) as a bright source. The right two panels display the same as the left but for a DEVILS-like Y$<$21.2\,mag selection. Clearly, many of these same issues persist even at relatively bright magnitudes.}
\label{fig:Profound}
\end{center}
\end{figure*}

\section{Target Selection}
\label{sec:Targets}

In the following section we provide a detailed description of the imaging, source extraction and photometry,  colour-based selection of potential stars and galaxies, masking and visual inspection. We refer the casual reader to Section \ref{sec:Input} for a brief summary of how our final sample was selected. 

\subsection{Imaging}

We use VISTA Y-band imaging for our input selection band from UltraVISTA (D10) and VIDEO (D02 and D03). The UltraVISTA data used in our sample were taken between 2009 and 2012 with VIRCAM Y, J, H, Ks and NB118 bands covering the central 1.5\,deg$^2$ of the COSMOS region (D10). The survey uses a `jitter' image technique to produce 0.15$^{\prime\prime}$/pix plate-scale image resolution. The observing strategy of UltraVISTA produced deep and ultra-deep (0.62\,deg$^{2}$) stripes covering the COSMOS region to 24.8, 24.5, 24.1, 23.8 mag in the deep stripes and 25.7, 25.4, 25.1, 24.9 mag for the ultra-deep stripes in Y, J, H and Ks respectively (5$\sigma$ 2$^{\prime\prime}$ apertures).  We note that even the deep stripes consist of data 3.5\,magnitudes fainter than our survey limit. Full details of the UltraVISTA survey and data characteristics can be found in \cite{McCracken12}.

VIDEO data are taken with VIRCAM Z, Y, J, H and Ks bands covering a total of 12deg$^{2}$ covering ECDFS (D03), Elais-S1 and XMM-LSS (D02). Data are stacked and re-sampled to obtain a 0.2$^{\prime\prime}$/pix image resolution.  We use the most recent (February 2017) VIDEO internal team data release which provide stacked images in all 5 bands and \textsc{SExtractor}-derived \citep{Bertin96} colour-optimised photometry catalogues. The 2017 data reaches 5$\sigma$ 2$^{\prime\prime}$ aperture detection limits of 25.43, 24.96, 24.54, 24.05, 23.67 (D02) and 24.76, 24.92, 24.52, 23.88,  23.57 (D03) in Z,Y,J,H,Ks respectively. Full details of the VIDEO survey and data characteristics can be found in \cite{Jarvis13}. Figure \ref{fig:FieldPos} displays the DEVILS regions (red box) in relation to the UltraVISTA and VIDEO imaging.

\subsection{\textsc{ProFound} Source Finding and Extraction}
\label{sec:ProFound}

To form a complete Y$<$21.2\,mag sample, we must derive robust total photometric measurements for all sources in the DEVILS regions. In addition, to perform our photometric star-galaxy separations, we also require Y, J, H and Ks colour-optimised photometry. 

Traditionally, widely used source detection software, such as \textsc{SExtractor}, would be used to identify target sources and extract photometry. However, within the GAMA survey \citep[see][]{Wright16} and our G10/COSMOS analysis \citep[see][]{Andrews17} we have found that \textsc{SExtractor} produces a non-negligible number of erroneous detections/measurements that require significant manual intervention (see previous references for in depth examples). To overcome some of these issues, we have developed a new source detection and extraction code \textsc{ProFound} \citep{Robotham18}. Briefly, \textsc{ProFound} identifies peak flux positions within the image, performs watershed deblending to identify source `segments', and then iteratively grows (dilates) these segments to measure total photometry. For full details of the \textsc{ProFound} package and comparison to other source detection algorithms see \cite{Robotham18}. 

For our input photometry we ran a bespoke wrapper to the base \textsc{ProFound} code which splits target images into a number of sub regions, identifies sources, extracts photometry, and then recombines the outputs to produce photometric measurements for the full survey region. This avoids some memory issues with deep imaging containing a large number of sources, and allows the code to be run in an easily parallelizable fashion. 

To perform source extraction with detailed deblending of flux, we would ideally like to identify all potential sources of flux in a given region; extending well below the proposed survey limit. While our sample only requires the identification and extraction of targets to Y$<$21.2\,mag, much fainter sources may lie close to these on sky, and thus require identification and extraction to remove their (potentially contaminating) flux. To maximise the removal of these confusing sources, and to provide a more robust identification of the segments required for a total flux measurement (extending to the low-surface brightness wings of sources), we produce an inverse variance weighted stacked image using the VISTA Y, J, H and Ks images. We first run \textsc{ProFound} independently over each band to determine a band-specific sky-RMS map. We then weight each image by its sky-RMS$^{-2}$ and combine to form a deep combined YJHKs image. This image does not conserve flux, but is only used to identify source segments to faint surface brightness limits (not to measure photometry). 

Using the stacked image we run \textsc{ProFound} with default parameters except:  tolerance=0.8 and skycut=1.5, to produce a segmentation map containing all segments to very faint magnitude limits ($\sim$25\,mag in the Y-band). These \textsc{ProFound} parameters were found to improve the segmentation in the deeper and `jittered'/resampled VISTA imaging of UltraVISTA and VIDEO, where the \textsc{ProFound} defaults were tuned to the shallower and 0.34$^{\prime\prime}$/pix resolution VISTA Kilo-degree Infrared Galaxy, VIKING. The tolerance parameter allows for less source fragmentation at higher resolutions, and the skycut parameter allows segments to extend to the lower surface brightness wings of sources in the deeper data \citep[see][for further details of these parameters]{Robotham18}. \textsc{ProFound} is run over the DEVILS regions with an $\sim$0.5\,deg padding to measure photometry for any sources which fall at the edge of the DEVILS region (bounded by the blue box in Figure \ref{fig:FieldPos}).

To derive our total NIR multi-band photometry, we then take the segmentation map defined using our stacked image and measure the total flux in each segment in the Y, J, H and Ks bands separately. However, there are a number of potential pitfalls to consider measuring multi-band photometry, such as matching the photometric aperture across bands, and accounting for point spread function (PSF) and seeing differences across the image/bands \citep[see discussions in][]{Driver16b,Wright16}. For our sample these effects are minimal. The VISTA data is pixel-matched in each field, and thus to first order, we can simply apply the stacked segments to the pixel data in each band individually. However, this does not account for small differences in PSF and seeing. To account for this, we re-calculate the sky value in each band independently and allow the segments to dilate to include extended flux if necessary with up to six dilations (iters=6), but not shrink \citep[see][for a detailed description of this process]{Robotham18}. Allowing this dilation can account for slightly varying point spread functions for different bands and across the image but does not significantly alter the segment used. 

In all bands we record the default \textsc{ProFound} output parameters, including the total flux, average surface brightness to a radius containing 90\% (R90 - where the radius is defined as the elliptical semi-major axis) and 50\% (R50) of the source flux, magnitude, segment statistics (such as axial ratio, R50, R90, number of pixels in segment, etc) and segment flags (such as the number of pixels which border another segment or the edge of the frame). Figure \ref{fig:Profound} displays the \textsc{ProFound} segments identified in a sub-region of the UltraVISTA data for the Y-band image to full depth (left) and to our Y$<$21.2\,mag limit (right). We also over plot SExtractor apertures from the publicly available UltraVISTA catalogues. Even at the relatively bright DEVILS limit, SExtractor produces erroneous apertures in crowded regions, leading to incorrect total photometry; \textsc{ProFound} does not suffer from these issues. However, note that in the main UltraVISTA papers they do not directly use these apertures for their scientific analysis, but opt for fixed-size colour-optimised apertures (see comparisons below).

\begin{table*}
\begin{center}
\caption{Zeropoint scaling applied to NIR data in DEVILS Fields. All offsets are additive.}
\label{tab:zeroPt}
\begin{tabular}{c c c c c}
Field & Imaging Survey & Band & zeropoint offset & Method \\
\hline
\hline

D02 & VIDEO & Y & 0.0943 & COSMOS2015 Y-J colour \\
D02 & VIDEO & J & 0.0512 & 2MASS J magnitude  \\
D02 & VIDEO & H & 0.0442 & 2MASS H magnitude  \\
D02 & VIDEO & Ks & 0.0498 & 2MASS K magnitude  \\
D03 & VIDEO & Y & 0.1000 & COSMOS2015 Y-J colour \\
D03 & VIDEO & J & 0.0510 & 2MASS J magnitude  \\
D03 & VIDEO & H & 0.0537 & 2MASS H magnitude  \\
D03 & VIDEO & Ks & 0.0421 & 2MASS Ks magnitude  \\
D10 & UltraVISTA & Y & 0.0474 & COSMOS2015 Y-J colour \\
D10 & UltraVISTA & J & 0.0489 & 2MASS J magnitude  \\
D10 & UltraVISTA & H & 0.0230 & 2MASS H magnitude  \\
D10 & UltraVISTA & Ks & 0.0270 & 2MASS Ks magnitude  \\

\end{tabular}
\end{center}
\end{table*}%

\begin{figure*}
\begin{center}
\includegraphics[scale=0.4]{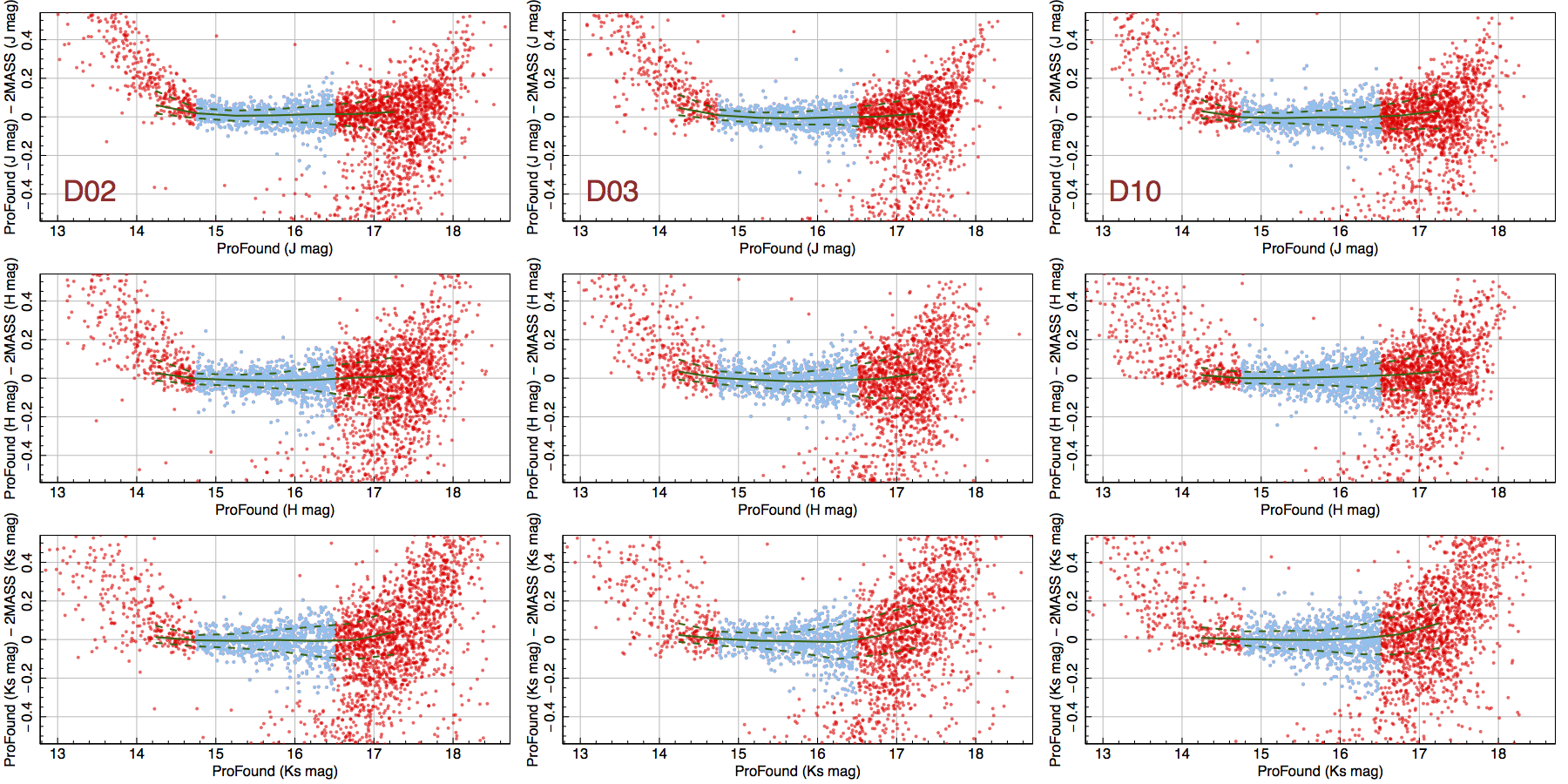}
\caption{ProFound magnitudes compared to 2MASS magnitudes in J, H and Ks after zeropoint scalings given in Table \ref{tab:zeroPt}  have been applied. Data used for zeropoint scaling is highlighted in blue. Running median and interquartile range are displayed as solid and dashed line respectively. The median interquartile range for the blue calibration points is $\sim$0.08, 0.1, and 0.1 in J, H and Ks respectively. The upward tail at the bright end is due to bright stars being saturated in VISTA, while the scatter/upward tail at the faint end is due to low-SN in the 2MASS data, and aperture/segment and seeing difference between VISTA+\textsc{ProFound} and 2MASS. We ignore both these regions in our zeropoint calibration}
\label{fig:2MASS}
\end{center}
\end{figure*}

\subsubsection{Colour-Optimised Photometry}
\label{sec:ColPhotom}

In addition to the total photometry outlined above, we also measure colour-optimised photometry in all bands using \textsc{ProFound}. For this colour-optimised photometry we ideally wish to measure flux in a fixed size aperture covering the central, high-surface-brightness region of the source, where colour gradients are minimal ($i.e.$ we only require that our measurement is consistent across all bands but does not require a total flux). When \textsc{ProFound} identifies sources it produces an initial (un-dilated) segmentation before dilating to obtain a total flux measurement. By design these `un-dilated' segments cover the central high surface brightness region of each source and are therefore ideal for colour-optimised photometry \citep[see][]{Robotham18}. We take the fixed size un-dilated segments derived from our stacked YJHK image and apply to each band individually. We once again recalculate the sky background independently for each band but do not allow any dilations (iters=0) to preserve the size/shape of each segment. Comparisons of our colour-optimised photometry with the VIDEO/UltraVISTA fixed-aperture photometry are presented in Section \ref{sec:photComp}.      

\begin{figure*}
\begin{center}
\includegraphics[scale=0.31]{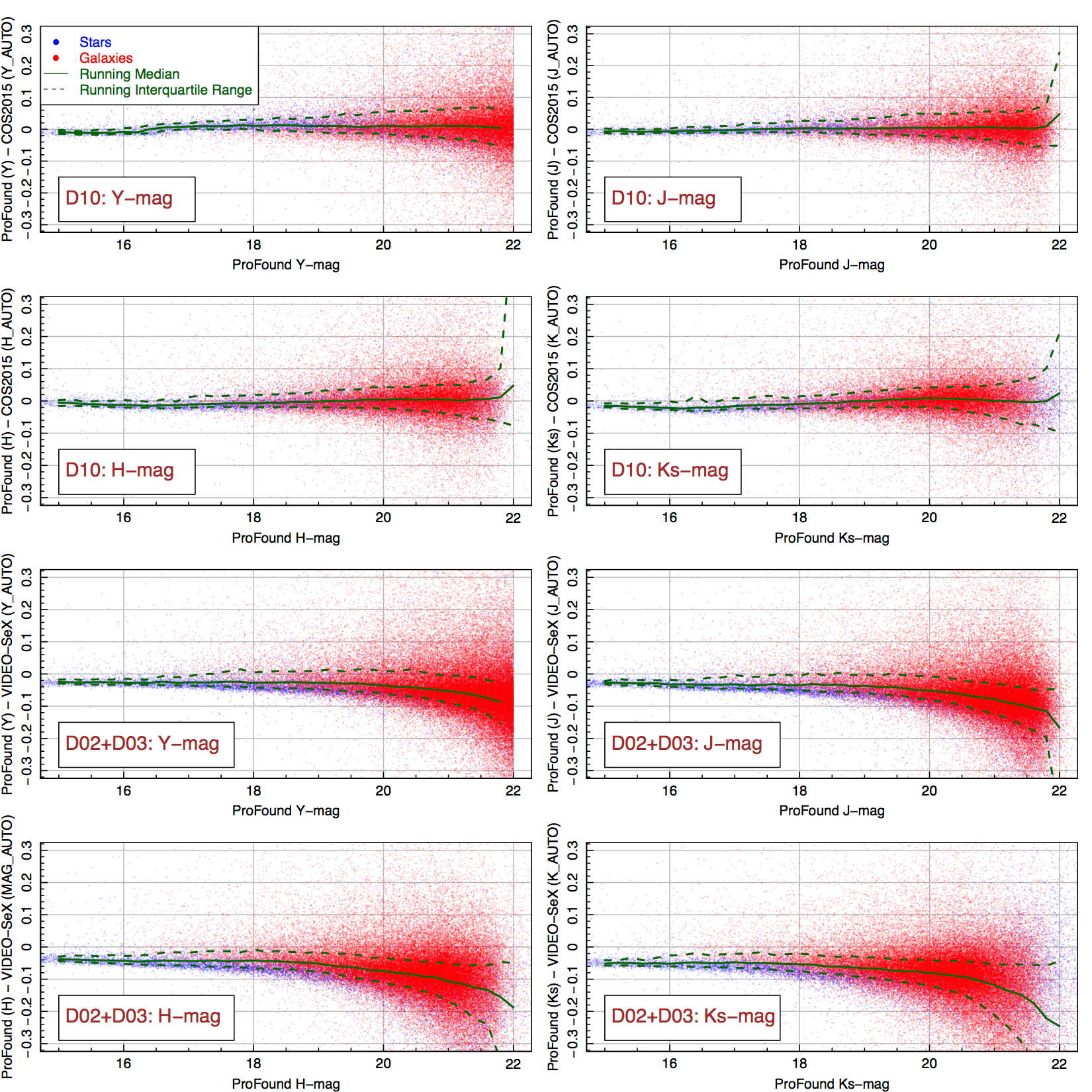}
\caption{ProFound magnitudes compared to VIDEO (\textsc{SExtractor} default runs) and UltraVISTA (COSMOS2015 catalogues) magnitudes in Y, J, H and Ks after zeropoint scalings given in Table \ref{tab:zeroPt} have been applied to both datasets. Sources are split into stars (blue) and galaxies (red) using the colour-based star-galaxy separator outlined in Section \ref{sec:StarGal}. Note that the magnitudes measured in both VIDEO and COSMOS2015 are not optimised for total magnitudes and thus the small normalisation offsets are expected. Our \textsc{ProFound} runs are extremely consistent with the COSMOS2015 measurements across the full magnitude range. In D02 \& D03 \textsc{ProFound} consistently measures more flux than \textsc{SExtractor}, with this offset increasing to fainter magnitudes. This is likely due to the fact that \textsc{ProFound} is measuring further into the low surface brightness wings of sources. }
\label{fig:MagOff}
\end{center}
\end{figure*}

\subsection{Photometric Re-Calibration}
\label{sec:2MASS}

In order to ensure accurate magnitude zeropoint scaling, both in terms of selection in our input band and for NIR colour-based star-galaxy separation, we perform our own independent zeropoint verification and scaling. We undertake this in a two step process, using observed Two Micron All Sky Survey \citep[2MASS]{Cohen03} magnitudes to calibrate the J, H, and Ks bands independently in each field, and then using the COSMOS2015 \citep{Laigle2016} Y-J colours to calibrate the Y-band.

2MASS provides a stable NIR flux calibration and is used in the VISTA data reduction performed by the Cambridge Astronomy Survey Unit (CASU). In each field we take all 2MASS sources and perform a 2$^{\prime\prime}$ positional match to our \textsc{ProFound} catalogue in J, H, and Ks. To directly compare default 2MASS (Vega) and VISTA-VIRCAM (AB) measurements we apply the magnitude system/colour scalings detailed on the CASU VISTA photometric calibration webpage\footnote{\url{http://casu.ast.cam.ac.uk/surveys-projects/vista/technical/photometric-properties}}. Both VIDEO and UltraVISTA surveys are processed using CASUVERS 1.3, and as observed with the same facility/instrument require the same colour terms:

\begin{equation*}
\begin{multlined}
J_{\mathrm{VISTA}}=J_{\mathrm{2MASS}}+0.937_{\mathrm{VEGA-AB}}-0.077(J-H)_{\mathrm{2MASS}}\\
\\
H_{\mathrm{VISTA}}=H_{\mathrm{2MASS}}+1.384_{\mathrm{VEGA-AB}}+0.032(J-H)_{\mathrm{2MASS}}\\
\\
Ks_{\mathrm{VISTA}}=Ks_{\mathrm{2MASS}}+1.839_{\mathrm{VEGA-AB}}+0.01(J-Ks)_{\mathrm{2MASS}}\\
\end{multlined}
\end{equation*}

\noindent where \textsc{vega-ab} terms are the conversion between Vega and AB magnitudes. We then calculate the median magnitude offset of \textsc{ProFound} - 2MASS at $14.75<$Mag$_{\mathrm{ProFound}}$$<16.5$ (given in Table \ref{tab:zeroPt}), applying the offsets to each band/field independently. Figure \ref{fig:2MASS} shows the resultant recalibrated comparison between \textsc{ProFound} and 2MASS. We see good agreement between 2MASS and \textsc{ProFound} in all bands and all zeropoint corrections applied in J, H and Ks are $<$0.06\,mag. We also note that there is very little evidence for zeropoint non-linearity in the data, which would manifest as a non-linear offset in our calibration region (the distribution of blue points at $14.75<$Mag$_{\mathrm{ProFound}}$$<16.5$ has little or no gradient).   

As 2MASS does not observe in the Y-band, we cannot directly compare to 2MASS to calibrate our Y-band zero-point. However, within the COSMOS2015 analysis of \cite{Laigle2016} they derive robust NIR colours for all sources in COSMOS, by calculating zeropoint offsets in comparison to a library of galaxy templates. As such, any systematic offset in Y-J colour between the COSMOS2015 catalogue and our recalibrated \textsc{ProFound} Y-J colours, is likely due to a zeropoint offset in Y. We select COSMOS2015 galaxies at Y$<$21.2\,mag and perform a 2$^{\prime\prime}$ match to our \textsc{ProFound} catalogue. Using the matched sources, we calculate the median Y-J colour in both COSMOS2015 and our catalogue. We then take the systematic offset between median values as our Y-band zeropoint scaling given in Table \ref{tab:zeroPt}. Note that, the VIDEO team has performed their own independent analysis and determined an $\sim$0.1\,mag offset in their Y-band zeropoint (provided via private correspondence from VIDEO P.I. M. Jarvis and noted in the VIDEO ESO data release document) - this is consistent with the offset we obtain here.

\subsection{Comparisons to Existing Photometry}
\label{sec:photComp}

Our primary aim in this process is to obtain consistency in colours and Y-band normalisation across all fields, both for our initial input Y-mag selection, and for our photometric star-galaxy separation. In Figure \ref{fig:MagOff} we show a direct comparison between our \textsc{ProFound} photometric measurements and those for the COSMOS2015 catalogues in D10 and our own independent \textsc{SExtractor} run on the VIDEO data using default \textsc{SExtractor} setting for D02 \& D03. We do not use the VIDEO team's internal catalogues as they use \textsc{SExtractor} parameters optimised for colour, not total, photometry ($>3\sigma$ for both detection and analysis thresholds leading to significant missed flux but robust colours). We also display sources which are colour selected (see Section \ref{sec:StarGal}) as stars and galaxies in blue and red respectively.  

All data points have the zeropoint scaling outlined in the previous section applied, and we display all sources to Y$<$22\,mag (0.8\,mag fainter than our DEVILS magnitude limit). For the D10 comparison (top four panels) we find excellent agreement in photometry between the COSMOS2015 catalogues and our \textsc{ProFound} photometry. All bands show no normalisation offset across the full magnitude range, and tight interquartile range to faint magnitudes ($\lesssim$0.2\,mag). The COSMOS2015 team also use a stacked (zYJHKs) image for their source detection and aperture definition. For the D02 \& D03 comparison to our \textsc{SExtractor} runs, we find offsets of $\sim$0.02-0.1\,mag (increasing to fainter magnitudes), with \textsc{ProFound} typically measuring more flux than \textsc{SExtractor}. This is likely to be caused by two effects: i) a systematic effect of \textsc{SExtractor} using Kron \citep{Kron80} magnitudes, which typically can miss 1-10\% of the total flux - equating to magnitude offsets up to 0.1\,mag, and ii) the combination of using a stacked detection image and \textsc{ProFound}'s methodology including the additional flux in the low-surface-brightness (LSB) wings of sources. This additional flux could potentially influence the faint galaxy number counts and in turn their contribution to the extragalactic background light \citep[EBL, see][]{Driver16}. This will be explored further in Section \ref{sec:NumberCounts}.            

\begin{figure*}
\begin{center}
\includegraphics[scale=0.55]{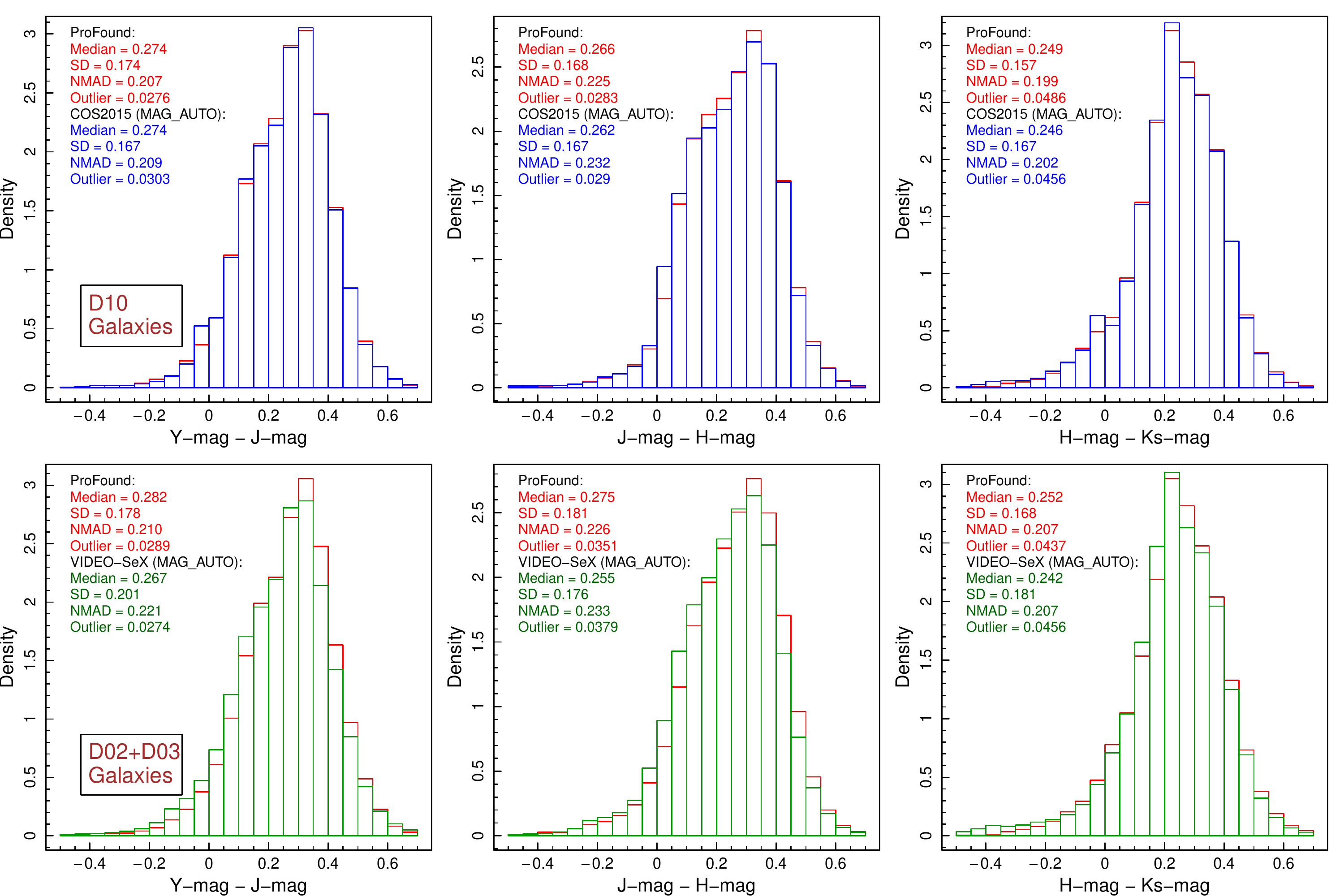}
\includegraphics[scale=0.55]{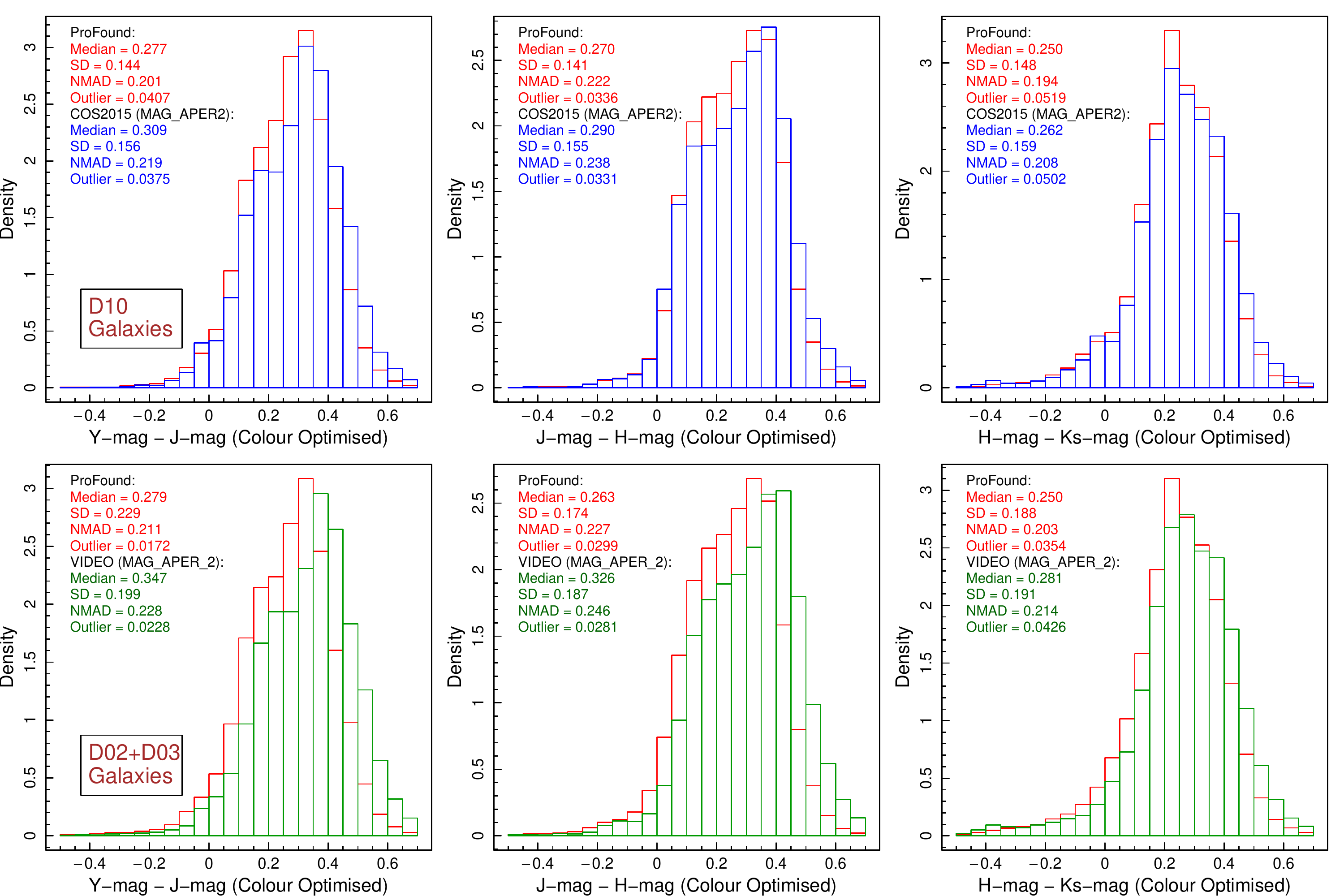}
\caption{Colour comparison between \textsc{ProFound} and VIDEO/COSMOS2015 after zeropoint scaling for total photometry measurements (top two rows) and colour-optimised photometry measurements (bottom two rows). For total photometry, we compare \textsc{ProFound} to the MAG\_AUTO parameters in the COSMOS2015 catalogue and our VIDEO \textsc{SExtractor} default runs on the D10 and D02/D03 data respectively. For colour-optimised photometry, we use the un-dilated segments from \textsc{ProFound} and fixed 2$^{\prime\prime}$ aperture measurements from both the COSMOS 2015 catalogue and VIDEO team's internal catalogue.}
\label{fig:ColOff}
\end{center}
\end{figure*}

In order to perform our photometric star-galaxy separation, we also require reliable galaxy colours. Both the COSMOS2015 and VIDEO team catalogues produce fixed aperture photometry specifically designed to provide robust colours. In Figure \ref{fig:ColOff} we display the NIR colour distributions for our \textsc{ProFound} total photometry measurements (top six panels) and our colour-optimised photometry measurements (bottom six panels) only for sources colour-selected as galaxies. For total photometry we compare to the COSMOS2015 and VIDEO points used in Figure \ref{fig:MagOff}, while for colour photometry we compare to each team's fixed 2$^{\prime\prime}$ aperture measurements. For each distribution we calculate the median, standard deviation, normalised median absolute deviation (NMAD or 1.4825$\times$MAD), and outlier rate. Outliers are defined as sources with absolute colour at $>2\times$ the standard deviation away from the median.

Firstly, taking the total photometry measurements we find that \textsc{ProFound} produces consistent median colours to both COSMOS2015 and our VIDEO-\textsc{SExtractor} runs to within 0.004\,mag (D10 - COSMOS2015) and 0.02\,mag (D02 \& D03 - VIDEO). We also find that our standard deviation, NMAD, and outlier rates are comparable or slightly lower than both COSMOS2015 and VIDEO-\textsc{SExtractor}. For the colour-optimised photometry we find slightly larger offsets in median colour of 0.02\,mag (D10 - COSMOS2015) and 0.07\,mag (D02 \& D03 - VIDEO). However, in most cases our standard deviation and NMAD values are consistently smaller than both COSMOS2015 and VIDEO, highlighting that  \textsc{ProFound} is producing accurate colour photometry for our star-galaxy separation. We also find that our \textsc{ProFound} colours are consistent across the UltraVISTA (D10) and VIDEO (D02 \& D03) imaging to $<0.08$\,mag with the largest offset in Y-J colour.

\begin{figure*}
\begin{center}
\includegraphics[scale=0.235]{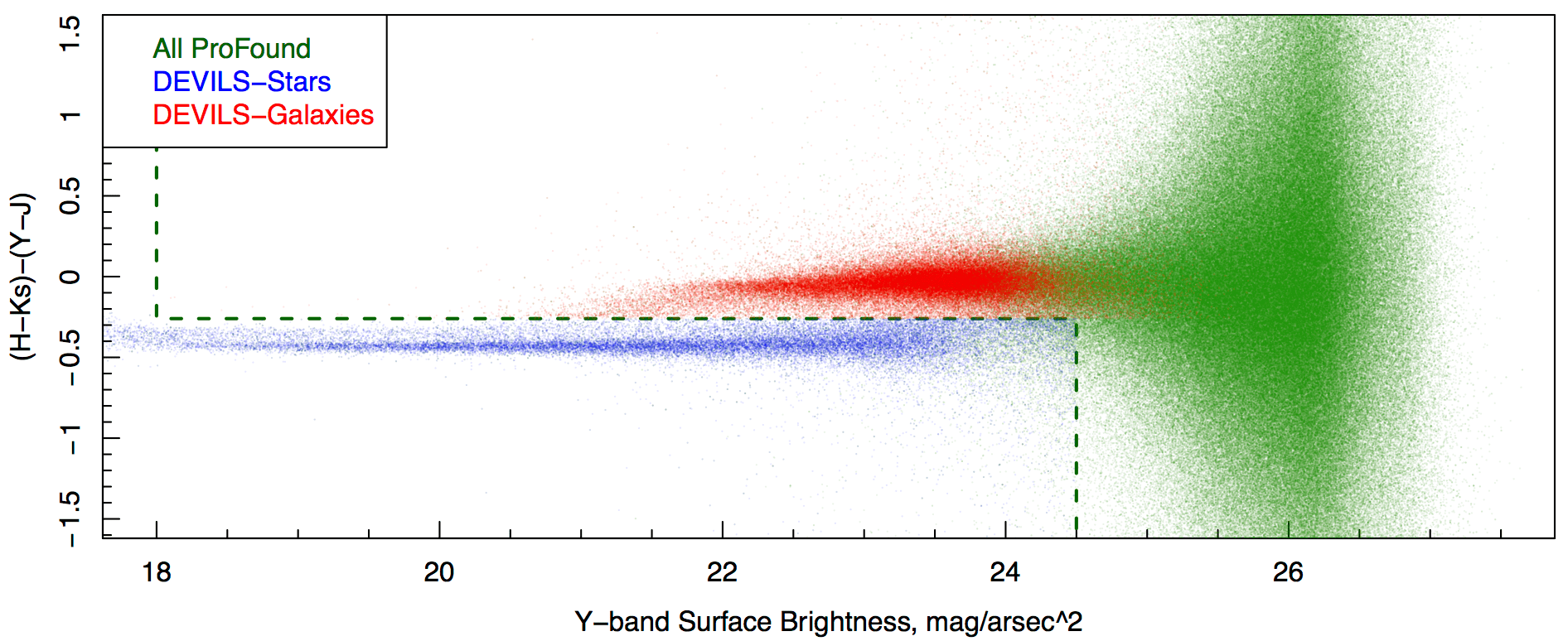}
\caption{Star-galaxy separation using NIR colours and surface brightness for all \textsc{ProFound}-detected sources in D02, D03 and D10 combined. All \textsc{ProFound} detections are shown in green, while galaxies/stars which meet our Y$<$21.2\,mag selection are shown in red and blue respectively. Our stellar selection window, as described in Section \ref{sec:StarGal}, is bounded by the green dashed box. }
\label{fig:StarGalall}
\end{center}
\end{figure*}

\subsection{Star-Galaxy Separation Using NIR Colours}
\label{sec:StarGal}

The outputs of \textsc{ProFound} provide us with a wealth of information with which to perform colour-based star-galaxy separation \citep[NIR colours, surface brightness, axial ratio, etc. See][]{Robotham18}. We note that sources selected in this manner are not spectroscopically confirmed stars and galaxies, and as such our selections only identify potential stars and galaxies. However, for clarity, from here on we shall refer to our colour selected samples as stars and galaxies.

\begin{figure*}
\begin{center}
\includegraphics[scale=0.22]{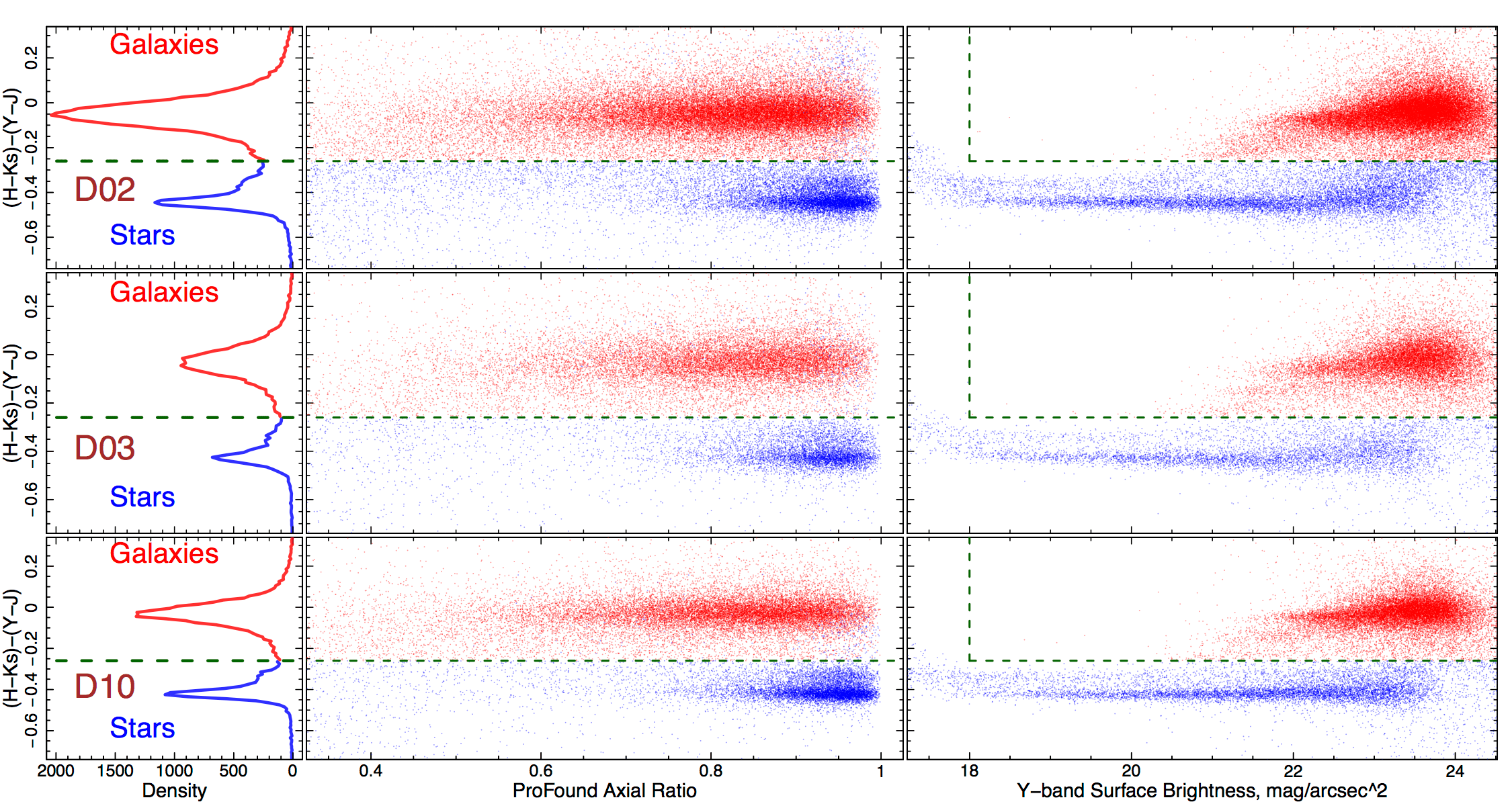}
\caption{Star-galaxy separation using NIR colours and surface brightness for DEVILS Y$<$21.2\,mag sources in D02, D03 and D10 (top to bottom respectively). Left: Histogram of (H-Ks)-(Y-J) colours in each field. We define the trough point between the two peaks as our colour separator between stars and galaxies. Middle: NIR Colour against axial ratio (Y-band) with galaxies and stars in red/blue respectively. Right: NIR Colour against surface brightness (Y-band). Red and blue points are identical to Figure \ref{fig:StarGalall}. In addition to our NIR colour separation, we also apply a surface brightness$>$18\,mag\,arcsec$^{-2}$ selection for galaxies to exclude saturated stars. Note that sources with stellar-colours and large axial ratio are largely identified as artefacts in out visual classifications}
\label{fig:StarGal}
\end{center}
\end{figure*}

Following a number of tests using various parameters we determined that the cleanest star-galaxy separation is produced when using a combination of NIR colours, with a bright surface brightness cut for saturated stars and a faint surface brightness cut to include galaxies with large photometric errors at the low surface brightness end. Note that these faint sources do not meet the DEVILS Y$<$21.2\,mag selection limit, and thus are not in our final sample. Figures \ref{fig:StarGalall} and \ref{fig:StarGal} display our star-galaxy separation procedure for all \textsc{ProFound} sources and just those that meet the DEVILS Y$<$21.2\,mag selection limit respectively. The former of these is only used for our number counts (see below).

To define our colour star-galaxy separation, we take our colour-based \textsc{ProFound} photometry (see Section \ref{sec:ColPhotom}), derive the distribution of (H-Ks)-(Y-J) colours for all sources which meet our magnitude selection (left column of Figure \ref{fig:StarGal}) and determine the trough point between the two peaks. The blue and red peaks identify stars and galaxies respectively. We note here, that this binary cut at the trough point will exclude a small number of galaxies. However, following this we do perform further visual classifications of all sources in the trough region to identify these sources (see \ref{sec:VISCLASS}). The distribution of these galaxies/stars in comparison to axial ratio and average Y-band surface brightness to the radius containing 90\% of the source flux, Y$\mu_{90}$, are shown in the middle and right columns of Figure \ref{fig:StarGal}. We do find that there are a small number of saturated stars which meet our NIR colour selection criteria, so we apply an additional, Y$\mu_{90}>18$\,mag\,arcsec$^{-2}$ cut for galaxies. In Figure \ref{fig:StarGalall}, we also find that when pushing well below the DEVILS limit galaxies with large error on their photometric measurements encroach on the stellar track. As such, we apply an additional Y$\mu_{90}<24.5$\,mag\,arcsec$^{-2}$ cut to our stellar selection when determining our number counts. In summary to select galaxies we use:

\begin{equation*}
(H-Ks)-(Y-J)>-0.26\mathrm{\,\,\,OR\,\,\,}Y\mu_{90}>24.5 \mathrm{\,mag\,arcsec}^{-2}
\end{equation*}
\begin{equation*}
Y\mu_{90}>18 \mathrm{\,mag\,arcsec}^{-2}
\end{equation*}

The validity of colour star-galaxy selection will be discussed further in Section \ref{sec:VISCLASS} where we perform visual classifications on all sources in the D10 region using high resolution HST data.  

One potential caveat to using a colour-based star-galaxy separation is that strong emission lines from star-formation/AGN can potentially cause galaxies to have stellar-like colours at specific redshifts and thus be removed from our sample \citep[see][for the effect of strong emission line sources on NIR colours]{Atek11}. While we perform additional visual classifications in Section \ref{sec:VISCLASS} which aim to quantify these misidentifications, it is interesting to consider the types of sources and redshifts that may be affected. H$\alpha$ at 6568\AA\ is the most prominent line emission that may significantly affect the colours used in our star-galaxy separation. At $0.5<z<0.7$ H$\alpha$ falls within the Y-band and with sufficient line luminosity could alter our NIR colours. However, in using a (H-Ks)-(Y-J) colour selection, an increased Y-band flux would only redden the colour and such an object would still be selected as a galaxy. However, at $z\gtrsim0.85$ H$\alpha$ transitions to the J-band and would act to make that galaxy colours closer to that of stars. At $z\gtrsim0.96$ the [OIII] 5007\AA\ line, which is also associated with strong star-formation, also transitions to the Y-band and, for comparable EWs, the effect of emission lines of Y-J colour are negated. Hence, only strongly star-forming and AGN sources at $0.85<z<0.96$ may potentially have their colours significantly affected and be identified as stars in our star-galaxy separation. If resolved, these sources will be identified in our visual classifications. However, for unresolved sources with star-like colours, separation between stars and galaxies is problematic. We note that such galaxies would appear as roughly uniformly-blue-scattered points about the galaxy locus in Figure \ref{fig:StarGal} (as they are likely to have a range of line luminosity). However, there does not appear to be a large number of sources scattered below our star-galaxy separator line. This is unsurprising as the on-sky number density of sources like those discussed in \cite{Atek11} are extremely low and as such, we predict that the misclassification of such sources is unlikely to significantly affect our sample.

\begin{figure*}
\begin{center}
\includegraphics[scale=0.7]{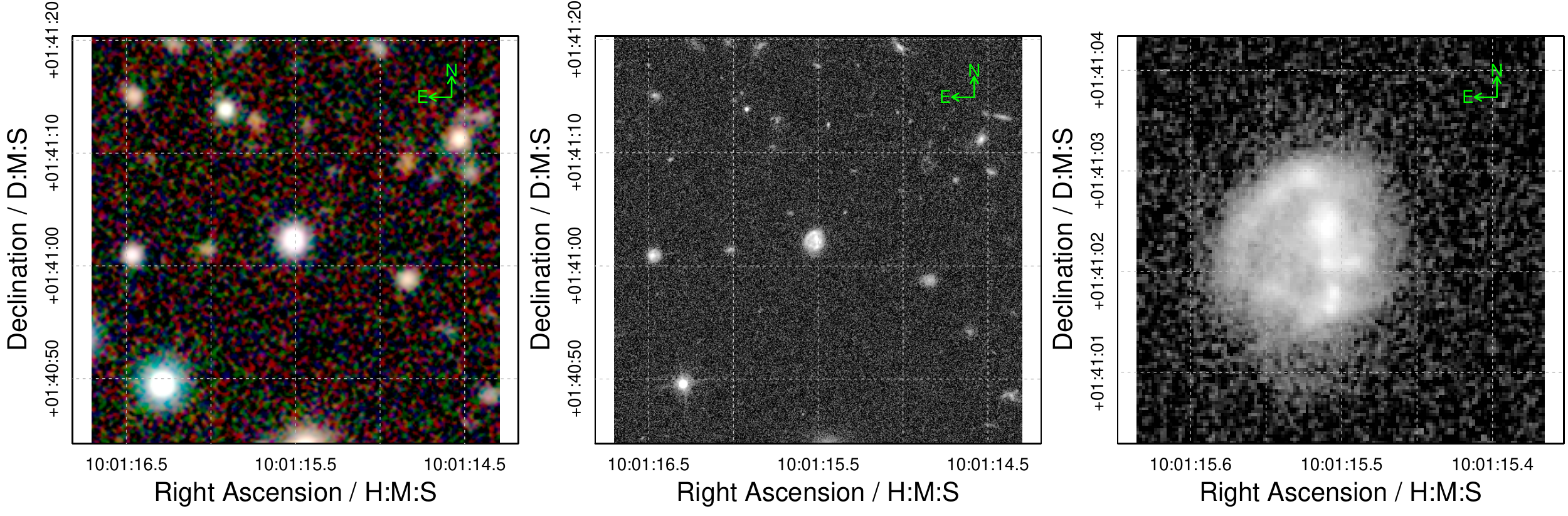}
\includegraphics[scale=0.7]{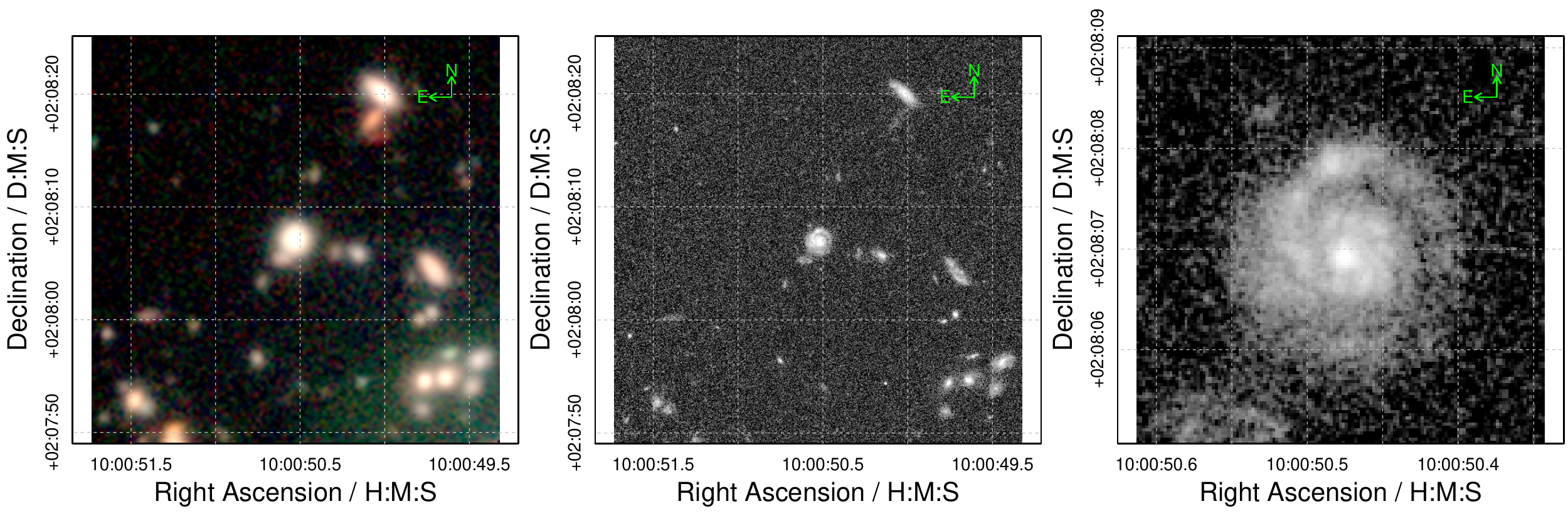}
\includegraphics[scale=0.7]{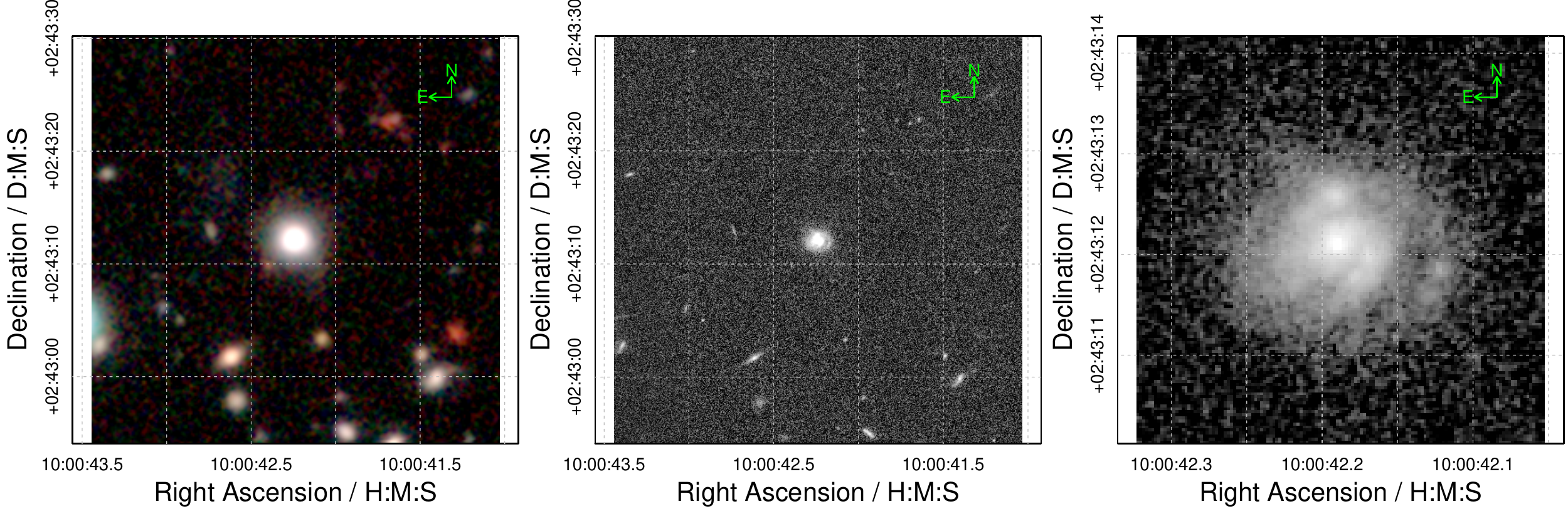}
\caption{Examples of how galaxies could potentially be visually mis-identified as stars when using NIR data alone for visual classifications. Left: UltraVISTA Y, J, Ks rgb images centred on an unresolved, spherical source in the UltraVISTA data. Middle: HST F814W image of the same region. Right: Zoom in of HST image at the position of the galaxy. In terms of visual classification, the central sources are largely indistinguishable from stars in the VISTA images alone (the blue source in the bottom left hand corner of the top row is a star). However, HST resolution imaging reveal that the central sources are all galaxies with clearly defined spiral structure. All rows are plotted over the same size regions. }
\label{fig:HSTcomp}
\end{center}
\end{figure*}

\subsection{Visual Classification}
\label{sec:VISCLASS}

\subsubsection{Initial HST Visual Classification in D10}

To test the validity of our colour-based star-galaxy separation and to remove any additional artefacts from our input catalogue we perform a visual classification. While we would ideally like to undertake visual classifications using high-resolution imaging (such as HST), such data are not available over all of the DEVILS area. However, we can use the the deep HST imaging in D10 (COSMOS) to identify regions of parameter space which require further visual classification and to compare visual classification to those obtained via the VISTA imaging alone. For example, are sources which are colour selected as galaxies but appear small, compact and spherical in the VISTA imaging truely galaxies or stars when viewed with HST? To this end we visually classify \textit{ALL} $\sim$30,000, Y$<$21.2\,mag sources (both stars and galaxies, prior to masking) in the D10 region using both a NIR three-colour image (Y, J, Ks) and single band HST F814W image. We generate postage stamps of all sources and visually classify them as either a star, galaxy or artefact/confused photometry/subregion of galaxy in both images separately (see Figure \ref{fig:HSTcomp}).

We find that our initial colour selections are extremely robust at identifying stars and galaxies down to Y$\sim$21.2\,mag, as only 1.4$\%$ of colour-selected stars appear to be galaxies in the HST imaging, and only 1.6$\%$ of colour-selected galaxies appear to be stars. We also find that the majority ($>95\%$) of sources classified as artefact/confused photometry/subregion of galaxy are in our masked regions, with the rest being sub-regions of a larger structure. However, a caveat is that this only applies for the magnitude range covered by the DEVILS sample (Y$<$21.2\,mag).

\begin{figure*}
\begin{center}
\includegraphics[scale=0.7]{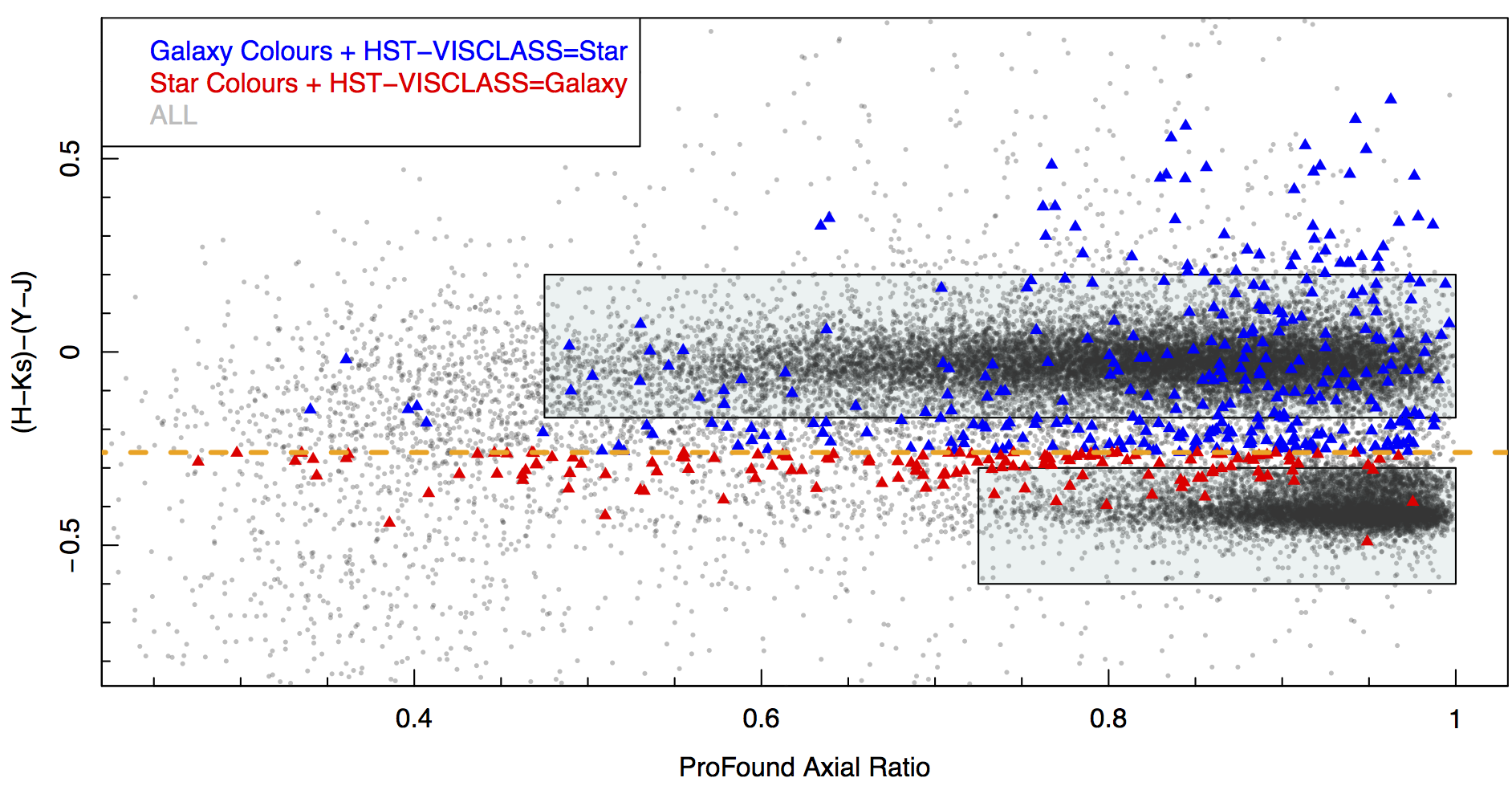}
\caption{Visual classifications using HST imaging in D10. Grey points display all Y$<$21.2\,mag sources in D10, with our initial star-galaxy separator displayed as the dashed orange line. Blue points display sources which were NIR colour selected as galaxies, but appear to be stars when visually inspected using HST resolution data. Red points display sources which were NIR colour selected as stars, but appear to be galaxies when visually inspected using HST.  Note that red and blue points are displayed with larger symbols, and no transparency to highlight their location. The grey coloured sample dominates in number over all coloured points in almost all regions of this figure. Unsurprisingly, the majority of mis-classified stars and galaxies sit close to the star-galaxy separation line. The two shaded boxes represent regions of this parameter space where we can robustly separate stars and galaxies using NIR colour and axial ratio. }
\label{fig:HSTVis}
\end{center}
\end{figure*}

Using these HST-based classifications, we can therefore define regions within our NIR colour space with which to perform further visual classifications in our other fields (where HST data are not available for the full region), and potential pitfalls in identifying galaxies as stars and vice versa when only VISTA imaging is available. Figure \ref{fig:HSTVis} displays the NIR-colour vs axial ratio used in our initial star-galaxy separation. Based on our colour selection and visual classifications, we colour-code sources which were not correctly defined using colour alone ($i.e.$ galaxies that have stellar colours and/or stars which have galaxy colours). Unsurprisingly, the majority of these misclassifications sit very close to the star-galaxy separator dividing line.  We define two regions of this parameter space where we can robustly isolate stars and galaxies (shaded regions in Figure \ref{fig:HSTVis}). There are a small number of red stars which cover the galaxies part of this parameter space (blue triangles which fall in the upper grey-shaded region). However, these will not bias our sample as they will simply be additional stars which we spectroscopically observe (and then remove). This will very marginally increase our total observation time. Given these sources can not be identified a-priori without HST-resolution data, and we do not have this data over all of the  DEVILS area, they remain within our target sample.       

For our final visual classifications we inspect all sources outside of these two regions in all three DEVILS fields. Note, that for our final input catalogues we repeat the visual classifications in D10 without the aid of HST data to be consistent across all regions. Thus, we visually inspect all sources which \textit{do not} meet one of the following criteria.
\\

\noindent Robust Galaxies: 
\begin{equation}
-0.17<(H-Ks)-(Y-J)<0.2 \\ \& \\
Axial Ratio>0.475 
\end{equation}

\noindent Robust Stars: 
\begin{equation}
-0.6<(H-Ks)-(Y-J)<-0.3 \\ \& \\
Axial Ratio>0.725 
\end{equation}

As noted above, D02 and D03 only have the VISTA imaging to perform our visual classifications (no contiguous HST data exists over the full region). As such, we explore some potential pitfalls in VISTA-based classifications using a comparison between HST and VISTA in D10. The most striking example of this is that almost all sources which are red in our three-colour images but appear spherical and unresolved, turn out to be galaxies when inspected in HST. Figure \ref{fig:HSTcomp} shows examples of such cases. Thus, when visually inspecting the NIR three-colour images in all fields, we take care to not misclassify galaxies as stars based on their unresolved nature. For the DEVILS input catalogue we wish for our sample to be highly complete, and will not be significantly affected by a low level of stellar contamination. Thus our visual classifications will be aimed at only removing robustly identified stars, artefacts and sub-regions instead of removing potentially ambiguous sources. In cases where a classification is unclear, we opt to retain the source in our sample.

\subsubsection{Final NIR Visual Classifications in All Fields}

Following the process outlined in the previous section, we select all sources that fall outside the two regions described in Figure \ref{fig:HSTVis} and are not in the masked regions described in Section \ref{sec:Mask} (leaving $\sim$9,000 sources for visual classification) and produce postage stamp NIR three-colour images. We then visually classify these sources as either stars, galaxies or artefacts/sub-regions of galaxies, and assign these flags to our master input catalogue. We find that of 8,945 sources visually inspected 1,065 change their initial NIR colour classification between star-galaxy (592 star to galaxy, and 473 galaxy to star). To be conservative in the process, we do not remove the small number of sources that are visually classified as stars from our sample, but only include additional sources which are visually classified as galaxies.

\begin{figure*}
\begin{center}
\includegraphics[scale=0.55]{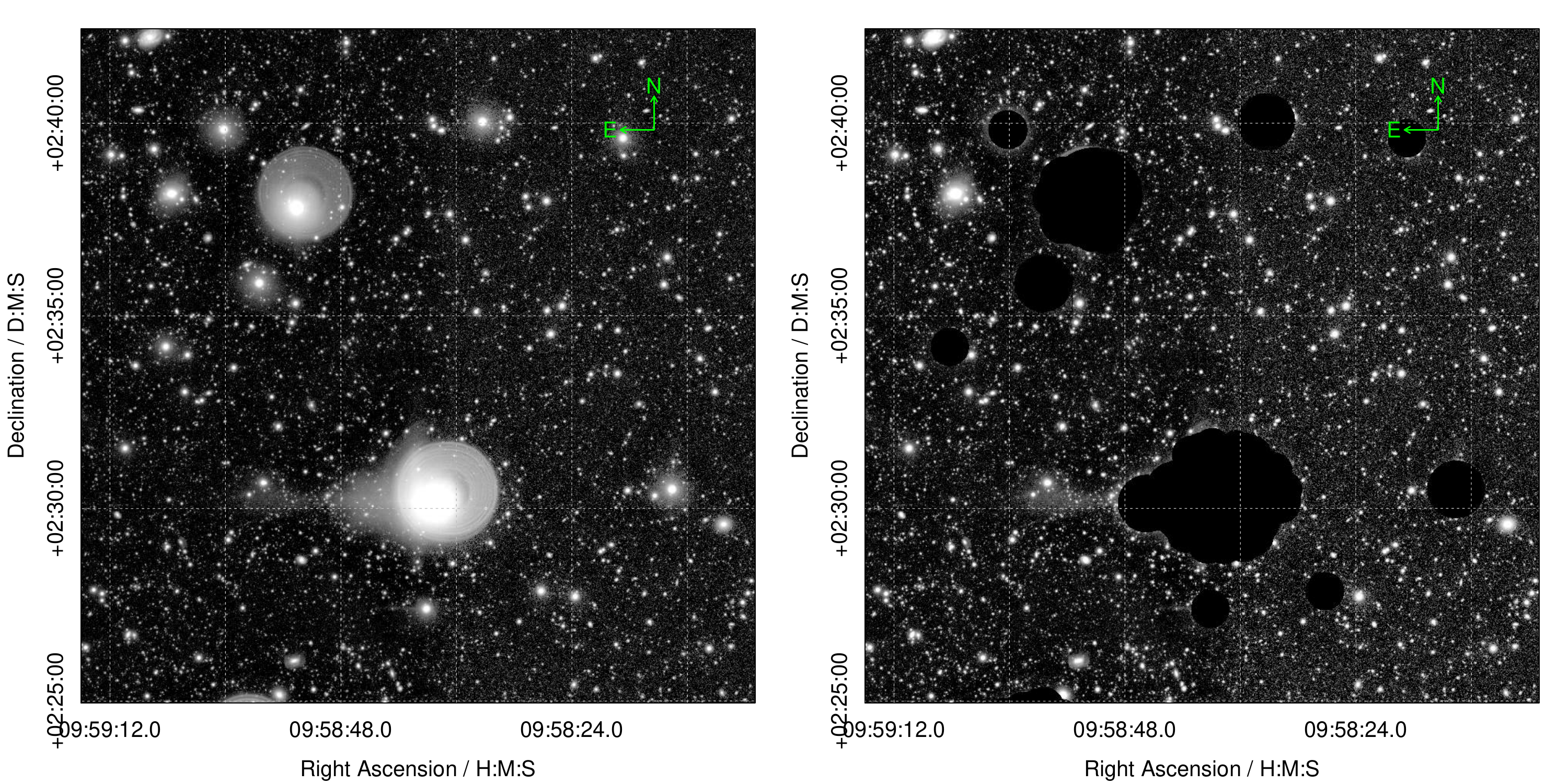}
\caption{Left: Example of bright haloes around stars in UltraVISTA Y-band image. These haloes can cause increased sky brightness, poor source detection and confused photometry, and thus must be masked. Right: The same region after masking.}
\label{fig:halo}
\end{center}
\end{figure*}

\begin{figure*}
\begin{center}
\includegraphics[scale=0.8]{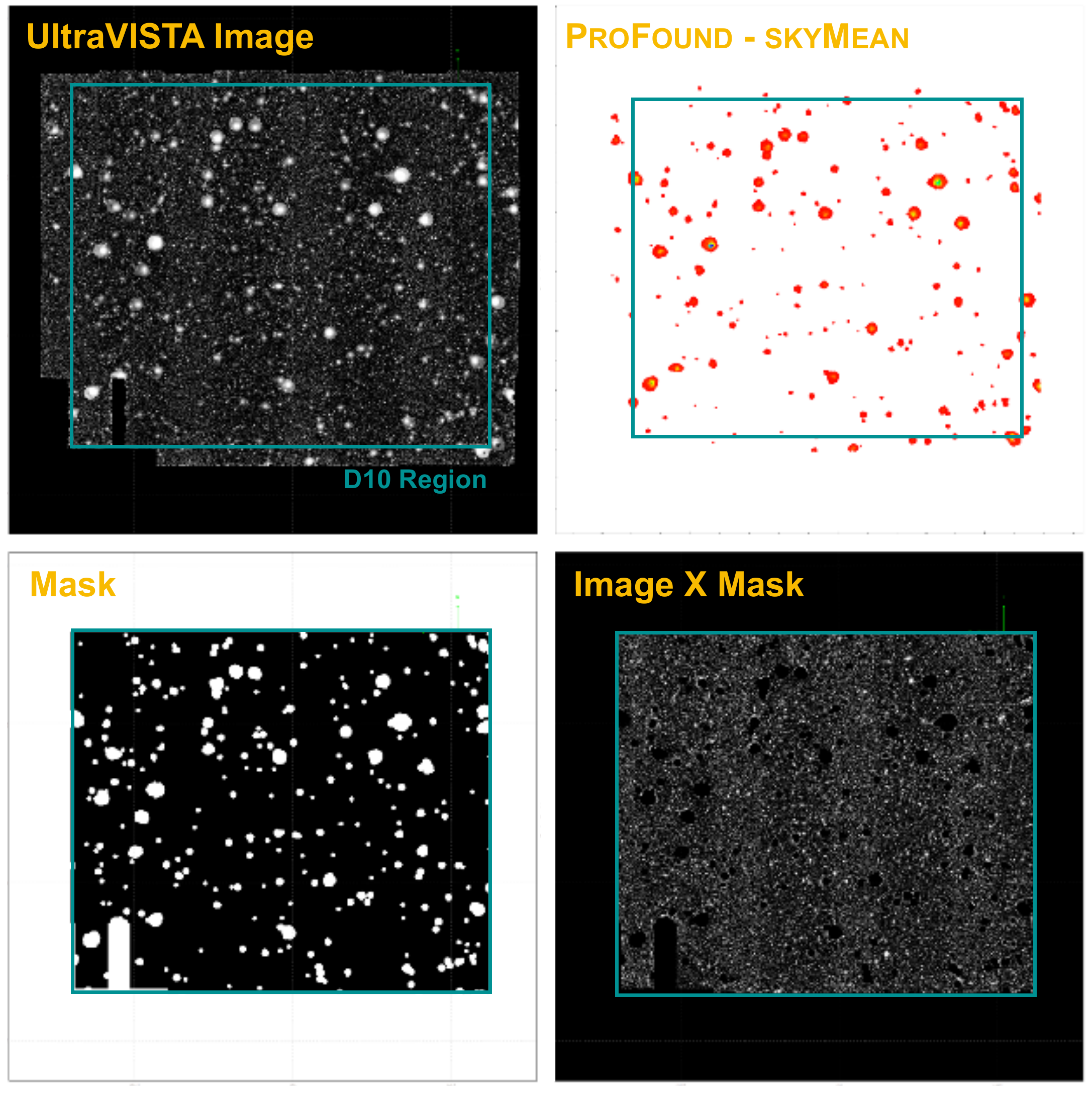}
\caption{Masking process in D10. Top left: UltraVISTA Y-band imaging in COSMOS, green box displays the D10 field. Top right: the distribution of high \textsc{ProFound - skyMean} values across the field. Bottom left: the binary mask generated by our bright halo/source identification (white = masked regions). Bottom right: the field with mask applied. }
\label{fig:Mask}
\end{center}
\end{figure*}

\subsection{Masking}
\label{sec:Mask}

Within the VISTA imaging used to select targets in our sample, instrumental effects in the optics produce ghosting and haloes around bright stars (as in all Cassegrain-style telescopes). These regions cause increased sky brightness, poor source detection and confused photometry (see Figure \ref{fig:halo}). As such, we mask the regions around bright stars in our input catalogue. In producing this mask, it is important to quantify the area of each field which is masked in order to perform any analyses using volume/area ($i.e.$ Section \ref{sec:NumberCounts}). 

The \textsc{ProFound} outputs provide useful parameters with which to identify bright halo regions. Here we use the segment \textsc{skyMean} value in our Y-band \textsc{ProFound} run, which gives the local sky background at each segment position. In regions of bright haloes, this sky value is elevated above the background level (see top right panel of Figure \ref{fig:Mask}). We find that to optimally mask regions in the VISTA data, we require different mask parameters for UltraVISTA and VIDEO due to the different depths of the data. For UltraVISTA we identify all segments where \textsc{skyMean}$>1.8\times$SD[\textsc{skyMean}] and mask a 250\,pixel (85$^{\prime\prime}$) radius region around the centre of the segment. In regions with large bright haloes, multiple masked 250\,pixel regions overlap to form a large masked region. This process also masks non-circular regions of scatter light, such as reflections.  While this process captures large bright haloes, it does miss some smaller haloes around stars. In a second step we also identify all stars with Y-band surface brightness, $Y\mu_{90}<20$\,mag\,arcsec$^{-2}$ , and $Y<13.25$\,mag. These sources are also masked with a 200\,pixel (68$^{\prime\prime}$) radius region. For VIDEO we repeat this process using a \textsc{skyMean}$>2.25\times$SD[\textsc{skyMean}] cut and 150/100 pixel regions respectively.  In an additional step, we find that the UltraVISTA data have some very large bright haloes which are offset radially from the bright star that produces them and are not fully encompassed by the above process. We identify these regions visually and apply an additional 500\,pixel (170$^{\prime\prime}$) radius mask to the halo. Note that all mask sizes and selections have been optimised by trial and improvement to remove artefacts in the data.

Finally, we mask out regions of the DEVILS fields which are close to the edge of a VIRCAM pointing ($i.e.$ the bottom left corner of the D10 field shown in the top left panel of Figure \ref{fig:Mask}) and regions of the VISTA imaging that sit outside of the DEVILS target region. Figure \ref{fig:Mask} displays the masking process in the D10 field. The top right panels shows the distribution of high \textsc{skyMean} values across the field, the bottom left panel shows the binary mask generated by our bright halo/source identification, and the bottom right displays the the field with mask applied. In Table \ref{tab:Mask} we detail the resultant masked areas. For how this affects the number of target sources see Table \ref{tab:summary}.

\begin{table} 
\begin{center}
\caption{Area of fields masks in bright star/halo masking.}
\label{tab:Mask}
\begin{tabular}{c c c c c }
Field & Total & Masked & Unmasked & Fraction  \\
 & (deg) & (deg) & (deg) & Masked \\
\hline
\hline

D02 & 3.040 & 0.214 & 2.825 & 0.076 \\
D03 & 1.501 & 0.086 & 1.415 & 0.061\\
D10 & 1.504 & 0.136 & 1.368 & 0.099\\

\end{tabular}
\end{center}
\end{table}

\begin{figure*}
\begin{center}
\includegraphics[scale=0.7]{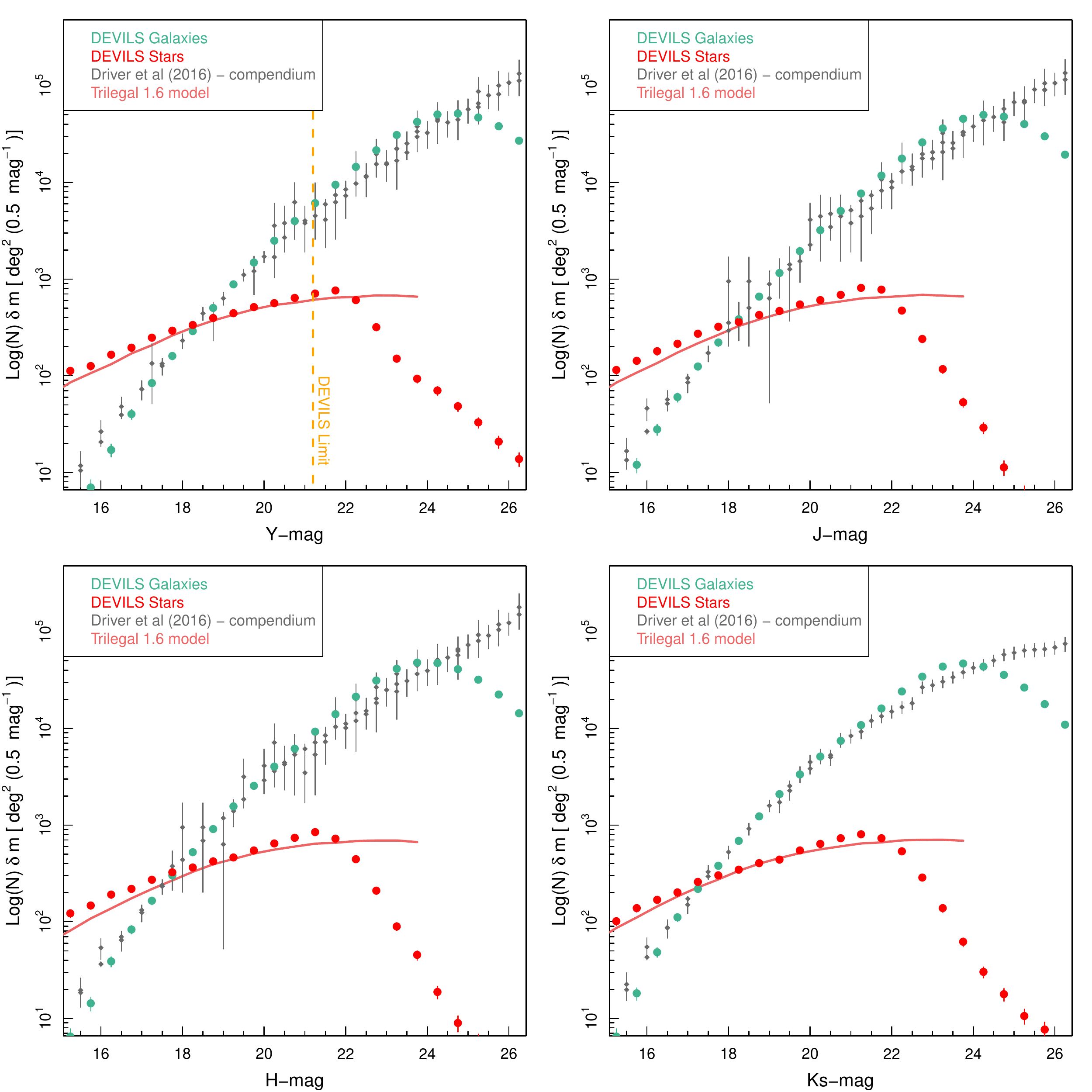}
\caption{Faint NIR number counts derived from our \textsc{ProFound} source detection and extraction in the DEVILS regions for galaxies (green large points) and stars (red large points). We over-plot the compendium of number counts from \citet{Driver16}, grey small points, and find consistency over the majority of the magnitude range in our sample. We find an increase in the number density of faint sources (mag$>$22) in comparison to \citet{Driver16}. We also over plot the TRILEGAL (see text) model stellar number counts as the red line.}
\label{fig:NumberCounts}
\end{center}
\end{figure*}

\begin{table*}

\begin{center}
\caption{Summary of generation of target catalogues. Col2: Total number of segments identified by \textsc{ProFound} in the region. Col 3:  Total number of \textsc{ProFound} with Y-mag$<$21.1. Col 4: Number of segments identified as potential galaxies and stars using the colour selection outlined in Sec. \ref{sec:StarGal}. Col 5: Number of colour-selected stars and galaxies in masked region. Col 6: Number of colour-selected stars and galaxies in unmasked region. Col 7: Number of sources visually classified as stars, galaxies and artefacts (total inspected = 8,945 - artefacts include erroneous sources, subregions of bright galaxies that have been split by \textsc{ProFound}, in these cases we ensure that the centre of the galaxy is still included in our sample, and sources which are very close to a bright source and are likely to have their photometry compromised). Col 8: existing number of sources with redshifts, total number if the total at Y-mag$<$21.1 in the region and sample is the number in our final sample.  Col 9: Final DEVILS target sample for AAT observations (colour-selected galaxies in un-masked regions which currently don't have a redshift, plus sources with stellar colours but visually appear to be galaxies). }
\label{tab:summary}
\begin{scriptsize}

\begin{tabular}{c c c c c c c c c}
Field & ProFound & Y < 21.2 & Colour Sel & Masked  & Unmasked & Vis-Class (Gals/   & Previous & Final \\
& Segments & Segments & (Gals/Stars) &(Gals/Stars) &  (Gals/Stars) &  Stars/Artefacts)  &   (total/sample) & Targets\\
\hline
Sec. \ref{sec:Fields} & Sec. \ref{sec:ProFound} & Sec. \ref{sec:ProFound} & Sec. \ref{sec:StarGal} & Sec. \ref{sec:Mask} & Sec. \ref{sec:Mask} & Sec. \ref{sec:VISCLASS} & Sec. \ref{sec:Prev} & Sec. \ref{sec:Input}  \\
(col 1) & (col 2) & (col 3) & (col 4) & (col 5)  & (col 6) & (col 7)  & (col 8) & (col 9) \\
\hline
\hline

D02 & 1,197,887 & 51,078 & 36,048/14,662 & 4,413/3,106 & 31,635/11,556  & 2,812/1,128/1,440 & 9,630/7838 & 21,602\\
D03 & 568,333 & 24, 585 & 17,052/7,350 & 2,209/1,601  & 14,843/5,749 & 1,230/456/637 & 4,713/3,830  & 10,040\\
D10 & 811,366 & 29,597 & 19,607/9,926 & 2,799/1,933 & 14,657/902  & 857/93/294 & 6,867/6,419 & 7,575\\
\hline
\textbf{Total} & \textbf{2,577,589} & \textbf{105,260}  & \textbf{72,707/31,938}  & \textbf{9421/6640}  & \textbf{61,135/18,207}   & \textbf{4,899/1,677/2,369}  & \textbf{21,210/18,087}  & \textbf{39,217} 

\end{tabular}
\end{scriptsize}

\end{center}
\end{table*}

\subsection{Faint NIR Number Counts}
\label{sec:NumberCounts}

As an additional, albeit coarse, verification of our \textsc{ProFound} photometry, zeropoint scaling and star-galaxy separation, we also calculate the faint NIR number counts for our Y, J, H and Ks photometry. Note that this does not include our visual classifications as this process is only performed on the DEVILS Y$<$21.2\,mag sample, and here we wish to investigate the number counts to much fainter magnitudes.  

We take all sources identified in our \textsc{ProFound} catalogues and apply our star-galaxy separation discussed in the previous section, see Figure \ref{fig:StarGalall}. We bin in $\delta m=0.5$\,mag and use a total DEVILS area of 6.045\,deg$^2$. In order to estimate errors we assume our sources are predominately from $0.2<z<1.0$ and calculate a cosmic variance error of 5.4\% using equation 4 from \cite{Driver10}, we linearly combine this with the Poisson error in each bin \citep[as in][]{Driver16}. Figure \ref{fig:NumberCounts} displays the \textsc{ProFound}-derived number counts in all four VISTA bands. For comparison we over-plot the compendium of data from \cite{Driver16} collated from GAMA, G10/COSMOS, and various HST programs  \cite[see][for details]{Driver16}. For the Driver et al points, we include their random, fitting, zeropoint and cosmic variance errors. The DEVILS number counts are consistent with the Driver et al results in the $17<$mag$<22$ regime, which covers the bulk of our spectroscopic target sample. Most notably, we see an excess in the faint number counts at mag$>22$, suggesting an additional contribution from faint galaxies which could significantly contribute to the EBL \citep[$i.e.$ see][]{Driver16}. This is potentially due to the increased flux being detected by \textsc{ProFound} over \textsc{SExtractor} for faint galaxies, as described in Section \ref{sec:photComp}, the potential lack of normalisation consistency between the diverse array of data compiled by \cite{Driver16} - $i.e.$ we find an excess in the number counts at the point where the \cite{Driver16} relations transition between GAMA/G10-COSMOS measurements and those derived from HST, and/or inadequate separation of stars and galaxies at these faint magnitudes (see below). We do also see a small decrement in all of the number counts at the very bright end (mag$<17$) which is potentially due to the fragmentation of bright sources in the \textsc{ProFound} runs (which will be amended in our visual classifications) or the choice of these deep fields avoiding bright nearby structures. However, our final sample only contains a very small number of sources in this magnitude range as previous spectroscopic surveys have already targeted this population in the DEVILS regions. 

As a crude validation of our star-galaxy separation, we also over-plot the number counts for stars in the DEVILS regions in comparison with the models from the TRIdimensional modeL of thE GALaxy \citep[TRILEGAL\footnote{\url{http://stev.oapd.inaf.it/cgi-bin/trilegal}}, $e.g.$][]{Girardi05}. To simulate our DEVILS stellar number counts with TRILEGAL, we supply the position and size of each of the DEVILS regions and combine to take into account the varying stellar density as a function of field position. In all bands the model stellar number densities are comparable to our star counts, most notably in the Y-band where we perform our target selection and at 17.5$<$Y$<$21.2\,mag, where the majority of our targets lie. We do note that our stellar number counts are somewhat steeper than those found in the deep HST data form the Great Observatories Origins Deep Survey (GOODS) South field \citep{Windhorst11}. They obtain a slope of $\sim$0.045 dex/mag at NIR wavelengths for 17$<$NIR-bands$<$25\,mag, in comparison to our $\sim$0.1\,dex/mag at  for 17$<$NIR-bands$<$21.5\,mag. However, considering Figure 11 of \cite{Windhorst11} we find that at NIR wavelengths, the flattening of the stellar number counts occurs prominently at $>$20\,mag and there are large error bars at $<$19\,mag. Over-plotting our number counts on those from Figure 11 of \cite{Windhorst11}, we find them to be completely consistent within errors in the 17.5$<$NIR-bands$<$21.2\,mag regime. 

We note that the cut off in our stellar counts at mag$>$22 is largely due to our Y$\mu_{90}<24.5$\,mag\,arcsec$^{-2}$ cut in star-galaxy separation. At fainter surface-brightnesses it becomes difficult to distinguish between stars and galaxies in the VISTA data alone. However, we note that if we simply use our total source number counts (star+galaxies) and subtract the TRILEGAL model in each magnitude bin, it does not fully account for the discrepancy in the faint galaxy number counts in comparison to \cite{Driver16} noted previously. We do not go into a detailed analysis of the separation of stars and galaxies, and their number counts at these faint magnitudes which are well below the DEVILS limit ($>$22\,mag), as a detailed description of the faint number counts from DEVILS and their implication will be presented in Koushan et al. in prep.

\section{Final Input Catalogues}
\label{sec:Input}

\subsection{Spectroscopic Target List}
\label{sec:Prev}

Following all of the analysis in Section \ref{sec:Targets}, we produce a final input catalogue of targets for AAT spectroscopy using the following identification and selection criteria:
 \\
 \\
$\bullet$ Source was identified and segment parameters defined by \textsc{ProFound} using a stacked Y, J, H, Ks VISTA image (Section \ref{sec:ProFound})\\
\\
$\bullet$ Source \textit{total} photometry measured by \textsc{ProFound} after zeropoint scaling and galactic extinction correction with Y$<21.2$\,mag (Section \ref{sec:ProFound} and \ref{sec:2MASS})\\
\\
$\bullet$ Source \textit{colour} photometry measured by \textsc{ProFound} has (H-Ks)-(Y-J)$>-0.26$ and Y$\mu_{90}<24.5$\,mag\,arcsec$^{-2}$ (Section \ref{sec:StarGal})\\
\\
$\bullet$ Source Y-band surface brightness measured by \textsc{ProFound} has Y$\mu_{90}>18$\,mag\,arcsec$^{-2}$ (Section \ref{sec:StarGal})\\
\\
$\bullet$ Source does not fall in a masked region of the image (Section \ref{sec:Mask})\\
\\
$\bullet$ If visually inspected, the source is classified as a galaxy (Section \ref{sec:VISCLASS})\\

This leaves an initial sample of 57,304 DEVILS targets. Selected columns from the final DEVILS input catalogues (including all Y$<$21.2\,mag sources) are made public via the AAO data central archive\footnote{\url{datacentral.aao.gov.au/services/query/}}, and are described in Table \ref{tab:inputCat}.

\begin{table*}

\begin{center}
\caption{DEVILS input catalogue columns and description. Full catalogue is made publicly available at \url{https://datacentral.aao.gov.au/services/query/}}
\label{tab:inputCat}
\begin{scriptsize}

\begin{tabular}{l l l l }

Column Name & Column Descriptor & UCD & Units \\
\hline
\hline

CATAID & Unique DEVILS survey identifier & meta.id & none \\
FIELD & DEVILS Field & meta.code.member & none \\
VIDEOID & VIDEO ID of 2$^{\prime\prime}$ matched source & meta.id.cross & none \\
UVISTAID & UltraVISTA ID of 2$^{\prime\prime}$ matched source & meta.id.cross & none \\
COSMOS2013ID & COMSOS2013 ID of 2$^{\prime\prime}$ matched source in COSMOS2015 Catalogue & meta.id.cross & none\\
G10CATAID & G10COSMOS ID of 2$^{\prime\prime}$ matched source from \citep{Andrews17} & meta.id.cross & none\\
RA & Right Ascension of source in VISTA Y-band & pos.eq.ra;em.IR.Y & deg\\
DEC & Declination of source in VISTA Y-band & pos.eq.dec;em.IR.Y & deg\\
YMAG & VISTA Y-band magnitude & em.IR.Y & mag\\
YMAGERR & VISTA Y-band magnitude error & stat.error;em.IR.Y & mag\\

STARCLASS & Star-galaxy separation flag based on NIR colours and surface brightness & meta.code.class & none \\
 & (0=Galaxy, 1=Star) & & \\
 
MASK\_FLAG & Mask flag (0=unmasked, >0=masked) & meta.code.class & none\\

VISCLASS & Visual classification flag base on VISTA YJKs rgb images (0=galaxy, 1=star,  & meta.code.class & none \\
  & 2=artefact, 3=shredded region of bright galaxy, 4=near bright source  & & \\
  & with confused photometry, NA=not classified) & & \\
  
AXRAT & Axial ratio of source defined from ProFound segments in VISTA Y-band & phys.size.axisRatio;em.IR.Y & none\\
SB90 & Average surface brightness to R90 in VISTA Y-band & phot.flux.density.sb;em.IR.Y & mag\,arcsec$^{-2}$\\
NIRCOL & VISTA (H-Ks)-(Y-J) colour used in star-galaxy separation  & phot.color & none \\

\end{tabular}
\end{scriptsize}

\end{center}
\end{table*}

\subsubsection{Existing Spectra}
\label{sec:existingSpec}

Each of the three DEVILS fields was specifically targeted for the high number of previously spectroscopically targeted sources. We compile a robust list of these sources in each region and set the priority (P) of these objects to 1 (not to be observed). In this previous spectroscopic sample we include, in D02 and D03: \\
\\
$\bullet$ Sources from the OzDES catalogue of confirmed redshifts. This includes a compilation of targeted OzDES observations, SDSS, GAMA, SNLS, DEEP2, 2dFGRS, PanSTARRS-AAOmega, VVDS, VIPERS, VUDS and ongoing targeted observations in ECDFS carried out by the LADUMA team (Wu et al, in prep) - for further details of these observations and catalogue, see \citep{Childress17}. \\
\\
$\bullet$  In total, there are 9,630 previous redshifts in D02 and 4,713 previous redshifts in D03. Of these 7,838 and 3,830 respectively are in our final target sample. These are assigned a priority of P=1 (meaning not targetable).   \\

and in D10:\\
\\
$\bullet$ Sources from VVDS \citep{LeFevre14} ZFLAGS=3 and ZFLAGS=4.\\
\\
$\bullet$ Sources from zCOSMOS \citep{Lilly07} with Z\_CC>2 and Z\_CC<6, or Z\_CC>12 and Z\_CC<16, or Z\_CC>22 and Z\_CC<26. This includes spectroscopically confirmed primary targets, AGN and secondary targets.\\
\\
$\bullet$ Sources from VUDS \citep{LeFevre13} quality flag>2.\\
\\
$\bullet$ Sources from hCOSMOS \citep{Damjanov18} with confirmed redshifts.\\
\\
$\bullet$ In total there are 6,867 previous redshifts in D10. Of these 6,419 are in our final sample and are set to P=1. \\
\\
      
These spectroscopic observations and their composition will be described in detail in the first DEVILS data release paper.

\subsection{Calibration Sources}

\subsubsection{Sky Fibres}

In order to select sky fibre positions, we can also utilise \textsc{ProFound}. One of the \textsc{ProFound} outputs is an aggressively dilated object mask (\textsc{objects\_redo}), which masks all pixels which could potentially contain source flux. In order to space potential sky fibre positions uniformly within our fields, we consider each sub-region processed through \textsc{ProFound} (see Section \ref{sec:ProFound}) independently. We select all unmasked pixels in the \textsc{objects\_redo} map that are classed as a sky pixel and do not border an object pixel. We then exclude all sky pixels that are within 20$^{\prime\prime}$ of a Y$<21.5$\,mag object centre or within 1.5$^{\prime}$ of a bright Y-mag$<15$ source.  We then randomly select 10 potential sky fibre positions in each subregion and ensure that no two positions are within 1$^{\prime}$ of each other. Lastly, we visually inspect all sky fibre positions to ensure that they do not contain source flux. This provides a total of 520, 241, and 420 potential sky fibre positions in D02, D03 and D10 respectively which are passed to our fibre assignment algorithm. In each observation we observe 25 blank sky positions.        

\subsubsection{Flux Standards}

We select flux standards in a similar manner to GAMA and OzDES. In both D02 and D03, we use identical flux calibration standards as OzDES, who select all F-class stars at $16.6<$r$<18.4$\,mag \citep{Yuan15}. Observing identical flux calibration stars to OzDES allows for a much more robust accuracy of the flux calibration. For D10 we use a similar selection to GAMA. Spectroscopic standards are selected from SDSS using (from the SDSS DR14 catalogue) fibermag\_r $>$ 16.9 and 16.6 $<$ psfmag\_r $<$ 18.4, and classified as either \textsc{SPECTROPHOTO\_STD} or \textsc{REDDEN\_STD}. We observe three spectroscopic standards in each observation.

\subsubsection{Guide Stars}

For guide stars we select all sources at 13.7$<$R1Mag$<$14.4 sources from the USNO-B guide-star catalogue and exclude sources with proper motions of >15\,mas\,yr$^{-1}$. We then perform a 2$^{\prime\prime}$ match to our \textsc{ProFound} catalogues and exclude all sources which would not be colour selected as stars. We then use the \textsc{ProfFound} RAcen and Deccen values as our guide positions to be consistent with our target astrometry ($i.e.$ we use the VISTA positions of USNO-B selected guide stars). All potential guide stars are visually inspected to ensure they are isolated, single, unsaturated stars. In total we obtain $\sim100$ potential guide stars per deg$^{2}$, of which 7-8 are selected in each pointing.

\subsection{Tiling and Fibre Assignment}

Targets are assigned priorities based on their Y-band magnitude, with fainter sources having higher priority:  Y>21\,mag=P7, 20<Y<21\,mag=P6, 19<Y<20\,mag=P5,   Y<19\,mag=P4. This allows objects which are likely to require more repeat observations to be preferentially targeted early in the survey. We then also produce a set of bad weather priorities which inverse these to preferentially target bright sources. Targets are assigned to fibres using the greedy tiling algorithm outlined in \cite{Robotham10} and used extensively in GAMA \citep{Driver11, Baldry12, Liske15} and for the Sydney-AAO Multi-object Integral-field spectrograph (SAMI) Galaxy Survey \citep{Bryant15}. This adds additional weights to priorities based on close on-sky clustering to allow complex regions with high levels of potential fibre collisions to be  preferentially targeted. In the majority of cases the 400 2dF fibres are allocated as follows: $\sim$360 targets, 25 sky, 3 standard stars, 8 guides (remaining fibres broken or unusable).

\section{Survey Strategy, Data Reduction and 2017B Observations}

\subsection{Nightly Observations}
DEVILS targets were observed with the 2dF+AAOmega system on the AAT with program ID A/2017B/011. Fibre flat observations were taken with the Quartz\_75\_A, 75W lamp and arc observations with the FeAr\_1, FeAr\_2, CuAr\_1, CuAr\_2, CuHe\_1, CuNe\_1 lamps. Data are typically observed in a 6\,sec flat, 45\,sec arc, 2$\times$1800\,sec sequence (modulo changes to weather and exact tile assignment at the start/end of the night). $\sim30$ dark frames are taken each run and 10 bias frames are observed each day.       

\subsection{Redshift Feedback Exposure Times}
\label{sec:redFeed}

Within the DEVILS survey we do not use fixed total integration times based on prior information regarding each source. This is due to the fact that it is difficult to make a robust prediction for the exposure time required to obtain a redshift based on observed-frame colour and magnitude alone (this will be explored in detail in further work). The required exposure time is a complex function of spectral type (emission/absorption), line strength and redshift ($i.e.$ where key spectral feature fall within your observing window). If we wish to obtain a $>95\%$ completeness survey with no bias to spectral type using fixed exposure times, we would by necessity have to over expose for many emission-line sources where these lines fall in easily observable, high transmission regions of the spectral window. To overcome this, instead we adopt a new observing strategy which uses a short-timescale redshift-success feedback loop to maximise our survey efficiency. In this strategy we observe each target on a 1h timescale and check for a secure redshift. If a redshift is confirmed then the target is removed from our input catalogue; if not, the source is prioritised for a repeat observation on the following night. Multiple repeat observations are then combined prior to the redshift checks. In this manner sources are only observed for the minimum (rounded up to the nearest hour) exposure time required to obtain a secure redshift \citep[see][for our primary exploration of this method for the G15Deep sample]{Kafle18}. 

Given CCDs with no read-noise, fast reconfiguration times and availability of targets, this method would become maximally efficient with very short exposure times (i.e. with 15min exposures one would never expose for $>15$\,min longer than required to obtain a redshift). However, in practice each sub-exposure must be easily sky-noise dominated (as not to stack large contributions of read-noise in each sub exposure), and long reconfiguration times and short exposures would lead to large overheads. For AAOmega at the AAT 1800sec sub-exposures are sufficiently sky-noise dominated and reconfiguration times of the 2dF fibre positioner are $\sim$45-50min for our relatively-complex configurations (reconfigurations are undertaken while observing as 2dF has two observing plates). The combination of these two time constraints necessitates minimum target exposure times of $\sim$1\,h. Note that this process is also proposed for a number of the 4m Multi-Object Spectrograph Telescope \citep[4MOST,][]{deJong14} consortium surveys, the Taipan survey \citep{daCunha17} and potential surveys undertaken with the Mauna Kea Spectroscopic Explorer telescope \citep[MSE,][]{McConnachie16}. Notably, 4MOST will suffer less from read-noise (newer CCDs) and reconfiguration time  \citep[4MOST's Australian European Southern Observatory Positioner, AESOP, has reconfigure times of $<$1min;][]{Haynes16}  and thus will be able to explore shorter exposure time feedback loops.

 \begin{figure*}
\begin{center}
\includegraphics[scale=0.905]{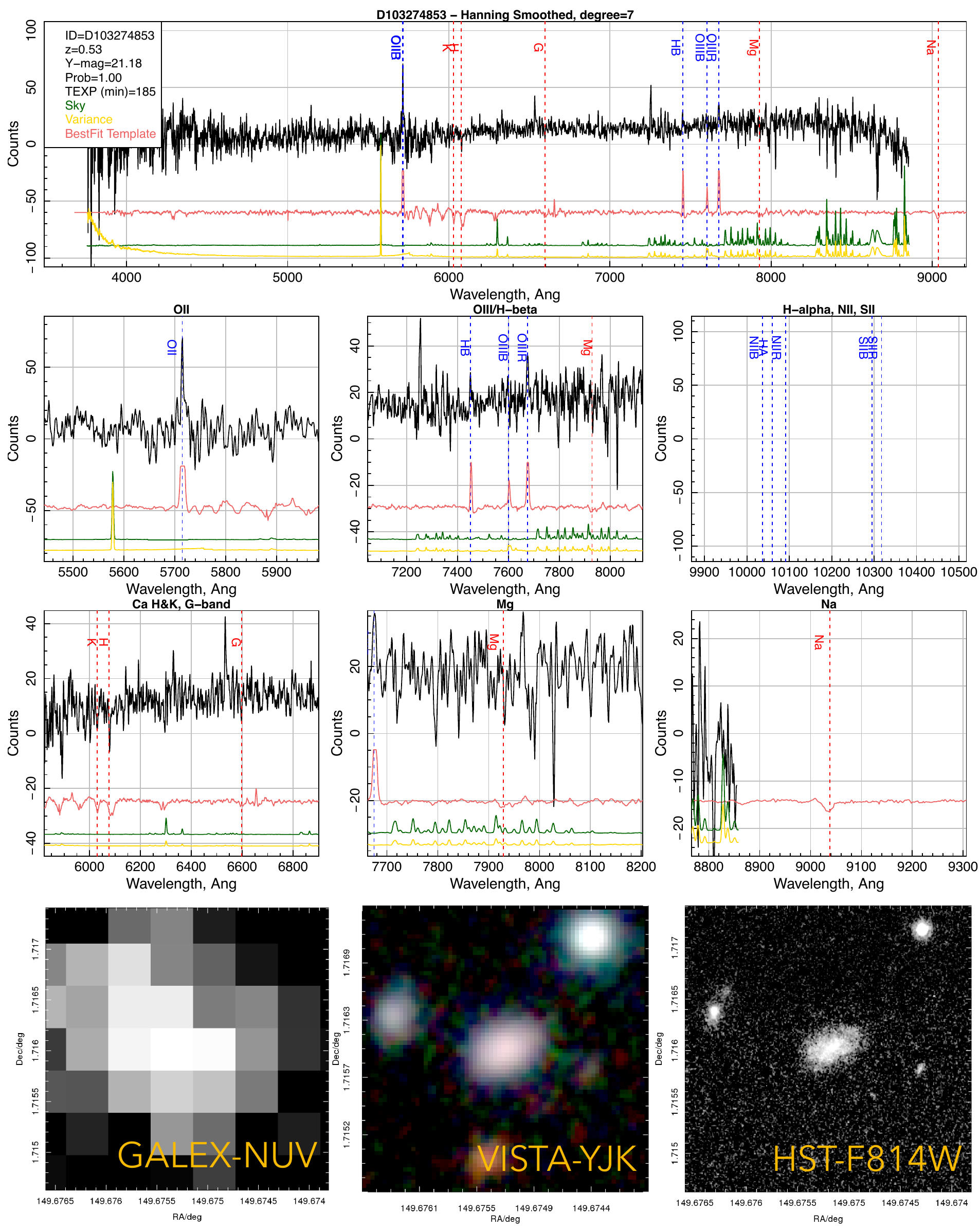}

\caption{Example spectrum of source D103274853 in the D10 region. The top row shows the full spectrum, middle rows key regions of emission (top) and absorption (bottom) line, and bottom row GALEX, VISTA and HST imaging for this source. Note H$\alpha$ and Na fall outside of the spectral range for this source. The galaxy has faintly detected OII, H$\beta$ and OIII emission lines and Ca H\&K absorption feature. The target had its redshift correctly identified by \textsc{Autoz} \citep{Baldry14}  in just over 3h integration with a probability of 1 (very secure). Key source parameters are given in the legend, but rounded to 2\,decimal places. for ease of plotting. Over-plotted are the variance and sky spectrum from the stacked AAT observation, as well as the best-fit template from \textsc{Autoz}.}
\label{fig:Example}
\end{center}
\end{figure*}       

\subsection{Data reduction using the DEVILS Tool for Analysis and Redshifting (TAZ)}

The difficulty of observing in the mode discussed above is that we require our data to be reduced, redshifted, target catalogues updated and new target configurations produced on a very short timescales. The survey becomes maximally efficient if this process occurs on a $\sim$12\,h timescale ($i.e.$ redshift information from a night's observations are used to target for the following night). 

In order to perform this short timescale feedback, we developed the DEVILS Tool for Analysis and Redshifting (TAZ). Our pipeline allows data to be synced from the telescope to the DEVILS archive nightly, TAZ then reduces this data using the 2dFDR software package \citep[see][]{Croom04,AAO15} in a bespoke highly-parallelisable fashion, spectra are extracted, repeat observations stacked, stacked spectra redshifted using \textsc{Autoz}  \citep{Baldry14}, target catalogues updated and tiling files produced for the next night's observing. 

Full details of our TAZ reduction will  be presented in the first DEVILS data release paper. However, we simply note here that the pipeline takes $\sim4-5$\,h to run from the raw data being added to the database, to new fibre configuration files being uploaded to the telescope for the next night's observing with no human intervention. All raw and reduced data, extracted spectra, meta data, catalogues, logs and diagnostic plots are written to the DEVILS database which exist in identical clones in both Perth and on the AAO Data Central system.

\subsection{December 2017 \& January 2018 observations}
\label{sec:Obser}

\begin{figure*}
\begin{center}
\includegraphics[scale=0.58]{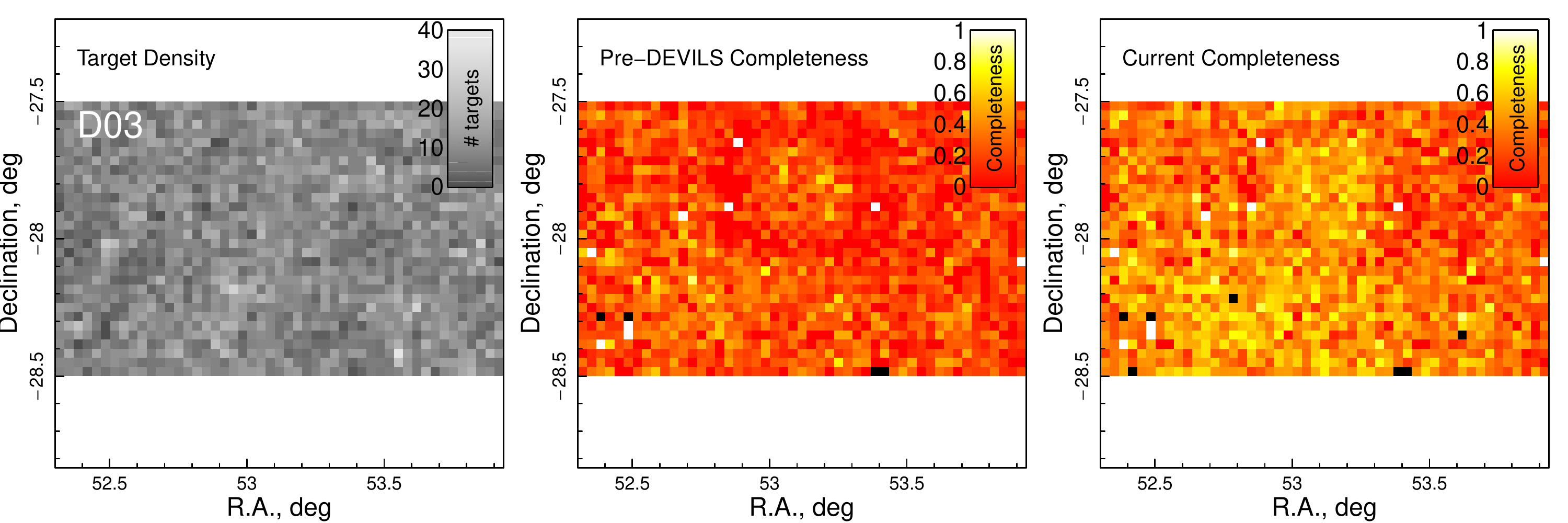}
\includegraphics[scale=0.58]{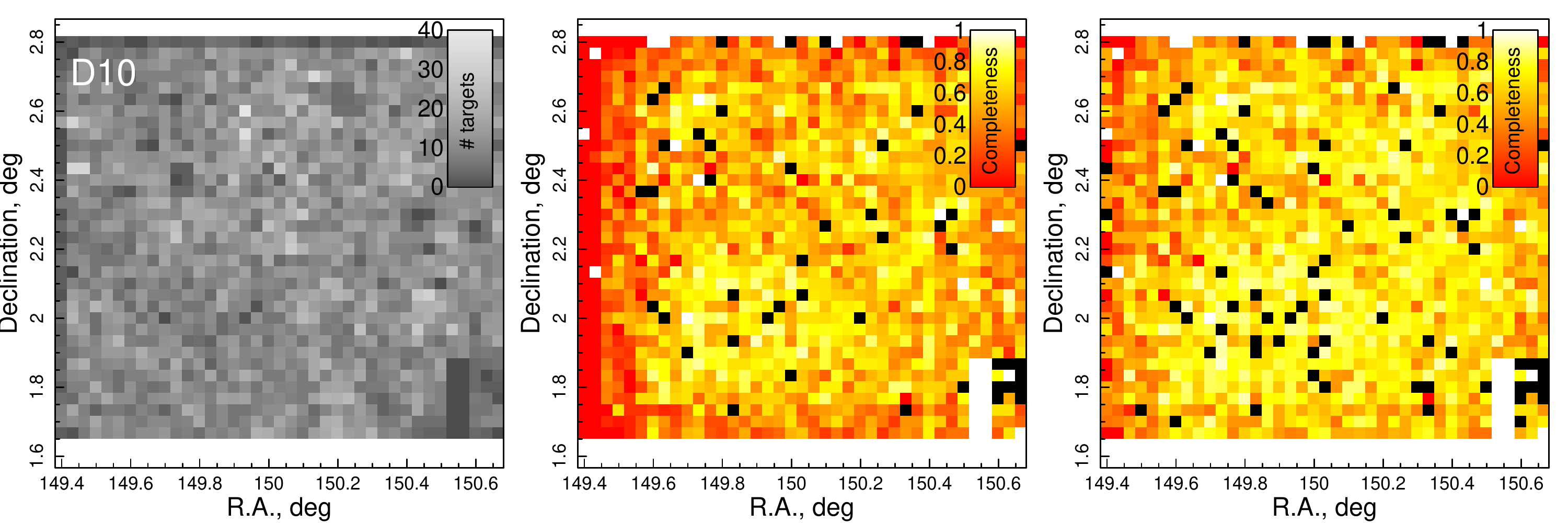}

\caption{The pre-DEVILS and post-2017B observations completeness to Y$<$21.2\,mag targets in D03 (top) and D10 (bottom) - gridded in 2$\times$2\,arcmin bins. Left: Y$<$21.2\,mag source density, middle: the pre-DEVILS redshift completeness, right: the current redshift completeness. Bins with $>95\%$ completeness are shown in black.}
\label{fig:comp}
\end{center}
\end{figure*}

DEVILS observations began in December 2017 with 19\,nights scheduled in the 2017B semester. While the detailed description of these observations will be outlined in our first data release paper, here we describe the primary results of the first observations and survey progress.  

We typically observed 6 fibre configurations per night (3 on half nights) targeting $\sim2100$ sources on a full night. As we prioritised faint (Y$>$21.0\,mag) targets which require the longest exposure times, we observed 8,856 unique sources over 80 different configurations, and obtained 4,353 secure redshifts. We lost $\sim33\%$ of observing time to bad weather and a site evacuation due to forest fires. Our current observed-targets-to-redshift-success rate is only $\sim50\%$ as we have prioritised the faintest sources in our sample, and many targets have not yet been observed to the integration time required to obtain a redshift. The currently confirmed redshifts and exposure times are completely consistent with predicted survey progress based on target magnitude and colour.    

An example spectrum of a faint source and a selection of corresponding multi-band images can be seen in Figure \ref{fig:Example}. This source sits at the faint end of our sample (Y=21.18\,mag), and took the combination of 3$\times \sim$1\,h exposures to obtain a secure redshift. As part of the TAZ pipeline we produce similar diagnostic plots for all spectra in the DEVILS sample. As a series of other examples we include similar plots for a faint (Y=21.12\,mag) absorption line system, bright (Y=19.25\,mag) emission line system and bright (Y=19.38\,mag) absorption line system in Appendix \ref{sec:appendix}.  

To highlight our current survey progress, Figure \ref{fig:comp} displays the target density (left), pre-DEVILS completeness to Y$<$21.2\,mag (middle) and currently completeness including DEVILS redshifts (right) in the D03 and D10 fields gridded in 2\,arcmin bins - D02 was not observed in 2017B. Within D03 we have already significantly increased the completeness over the entire region, while in D10 we have made significant progress in filling in the non-uniformity in completeness produced (predominantly) by the zCOSMOS footprint. DEVILS has already increased the median completeness to our Y$<$21.2\,mag limit in 2$\times$2\,arcmin bins has increased from 25\% to 38\% in D03 and 54\% to 67\% in D10 (for mean completeness: 26\% to 39\% in D03 and 54\% to 64\% in D10).

At the time DEVILS was observing the D03 field, the Dark Energy Survey \citep{Flaugher05} was also discovering transient sources in this field using DECam on the CTIO 4m telescope. As part of our fibre exchange with OzDES, a number of 2dF fibres were allocated to these transients in the D03 region during the December and January runs. 22 transients were observed, obtaining redshifts for 18, and classifying 6 of them as supernovae (these will be presented in two upcoming ATEL publications)

In addition to the transients, 55 Active Galactic Nuclei (AGN) from the OzDES reverberation mapping program were also observed during the January run. This program has been running since 2013, and the additional observations undertaken by DEVILS increases the number of points that can be used and therefore increases the likelihood of a successful AGN-lag measurement.

\subsection{Planned Public Data Releases}
\label{sec:DR}

The DEVILS survey has a number of planned public data releases both in terms of AAT spectroscopy and associated multi-wavelength imaging with consistent processing, and derived properties. These data will be hosted and made public via AAO Data Central\footnote{\url{https://datacentral.aao.gov.au}} once the core science of the project has been completed. It is currently expected that our initial data release containing spectra from the D10 region will occur in 2020 with subsequent full data release in 2022. However, this is subject to change based on time allocation and completion of core science projects. The preliminary (DR0) DEVILS input catalogue and a cutout service for imaging in the D10 region (covering imaging data from X-ray to far-IR wavelengths) is currently available through this service.      

\section{Summary}

The Deep Extragalactic Legacy Survey (DEVILS) is a spectroscopic campaign at the AAT aimed at bridging the near and distant Universe by producing the highest completeness survey of galaxies and groups at intermediate redshifts ($0.3<z<1.0$). The key science aims of the survey are to measure the late time evolution of the high mass end of the halo mass function (as predicted by $\Lambda$CDM) and the effect of environment in regulating galaxy evolution of the last 8 billion years. A summary overview of our key science goals is presented in Section \ref{sec:KeyScience}

Using the \textsc{ProFound} source finding code and UltraVISTA/VIDEO imaging, we have selected a sample of 57,000 Y$<$21.2\,mag galaxies over $\sim$6\,deg$^{2}$ in three well-studied deep extragalactic fields (COSMOS,XMM-LSS, and ECDFS, see Section \ref{sec:Targets} for our photometric selection). Within this process we perform robust star-galaxy separation (Section \ref{sec:StarGal}), masking of artefact regions (Section \ref{sec:Mask}), and visual classifications (Section \ref{sec:VISCLASS}). We form a final target list of sources which do not currently have a secure redshift in our sample, and discuss our AAT observing strategy for obtaining redshifts for these sources. As part of this process, we have developed a nightly redshift-feedback mechanism to maximise survey efficiency, allowing sources to be observed for the minimum time required to obtain a redshift (Section \ref{sec:redFeed}). DEVILS observations began in late 2017 (Section \ref{sec:Obser}) and we have currently obtained 4,353 new redshifts.  Strategies such as these will be essential for the next generation of large spectroscopic surveys, $i.e.$ the Wide Area VISTA Extragalactic Survey \citep[WAVES,][]{Driver16c}. DEVILS observations will continue until 2021-22, with planned data releases via AAO Data Central.

\section*{Acknowledgements}

DEVILS is an Australian project based around a spectroscopic campaign using the Anglo-Australian Telescope. The DEVILS input catalogue is generated from data taken as part of the ESO VISTA-VIDEO \citep{Jarvis13} and UltraVISTA \citep{McCracken12} surveys. DEVILS is part funded via Discovery Programs by the Australian Research Council and the participating institutions. The DEVILS website is \url{https://devilsurvey.org}. The DEVILS data are hosted and provided by AAO Data Central (\url{datacentral.aao.gov.au}). Parts of this research were conducted by the Australian Research Council Centre of Excellence for All Sky Astrophysics in 3 Dimensions (ASTRO 3D), through project number CE170100013.

\appendix 
\section{Other Examples of 2017B Spectra}
\label{sec:appendix}

\begin{figure*}
\begin{center}
\includegraphics[scale=0.42]{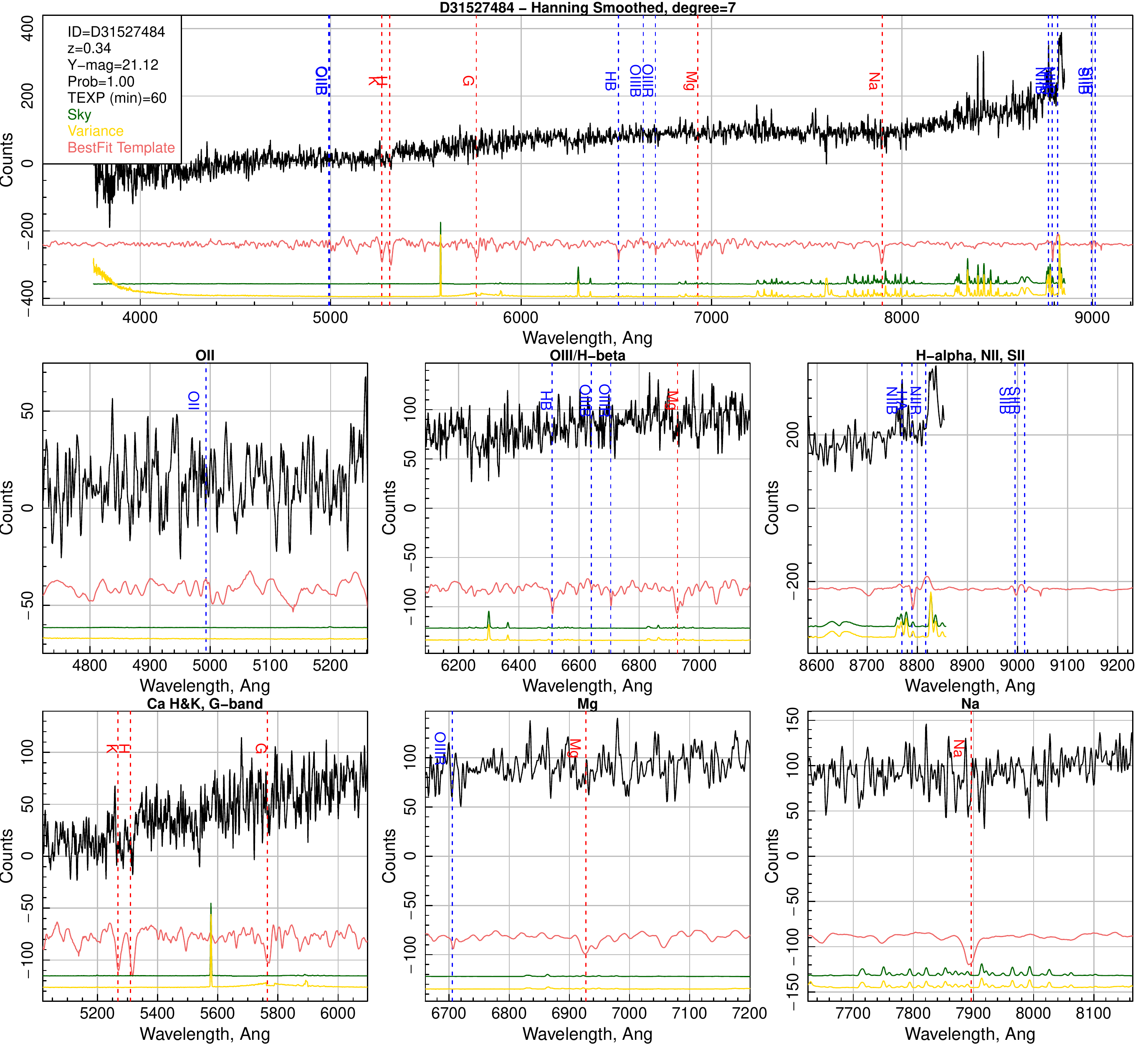}
\caption{The same as Figure \ref{fig:Example} but for a faint (Y=21.12\,mag) absorption line system at $z\sim0.34$ }
\label{fig:faintRed}
\end{center}
\end{figure*}

\begin{figure*}
\begin{center}
\includegraphics[scale=0.42]{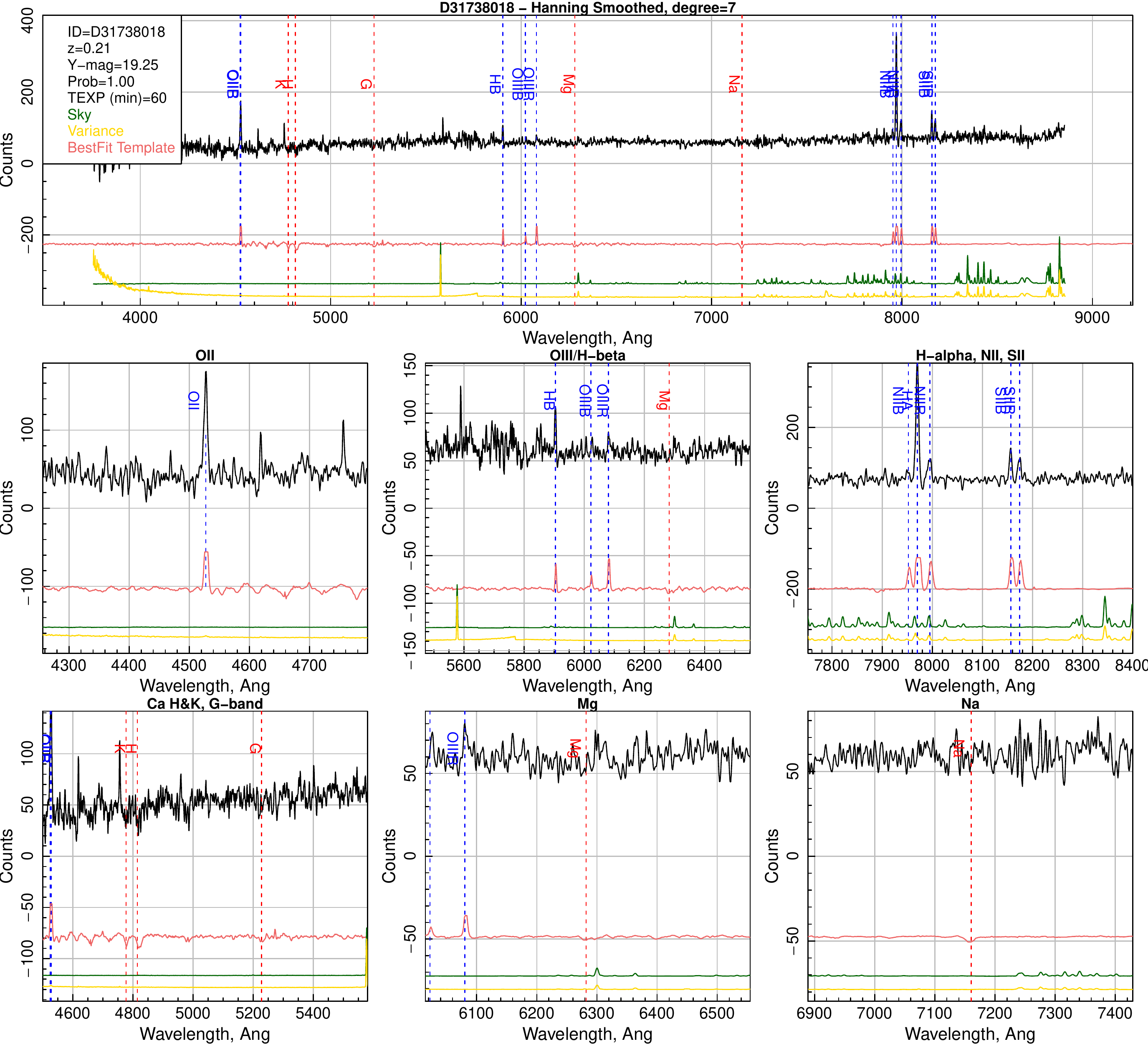}
\caption{The same as Figure \ref{fig:Example} but for a bright (Y=19.25\,mag) emission line system at $z\sim0.21$ }
\label{fig:brightBlue}
\end{center}
\end{figure*}

\begin{figure*}
\begin{center}
\includegraphics[scale=0.42]{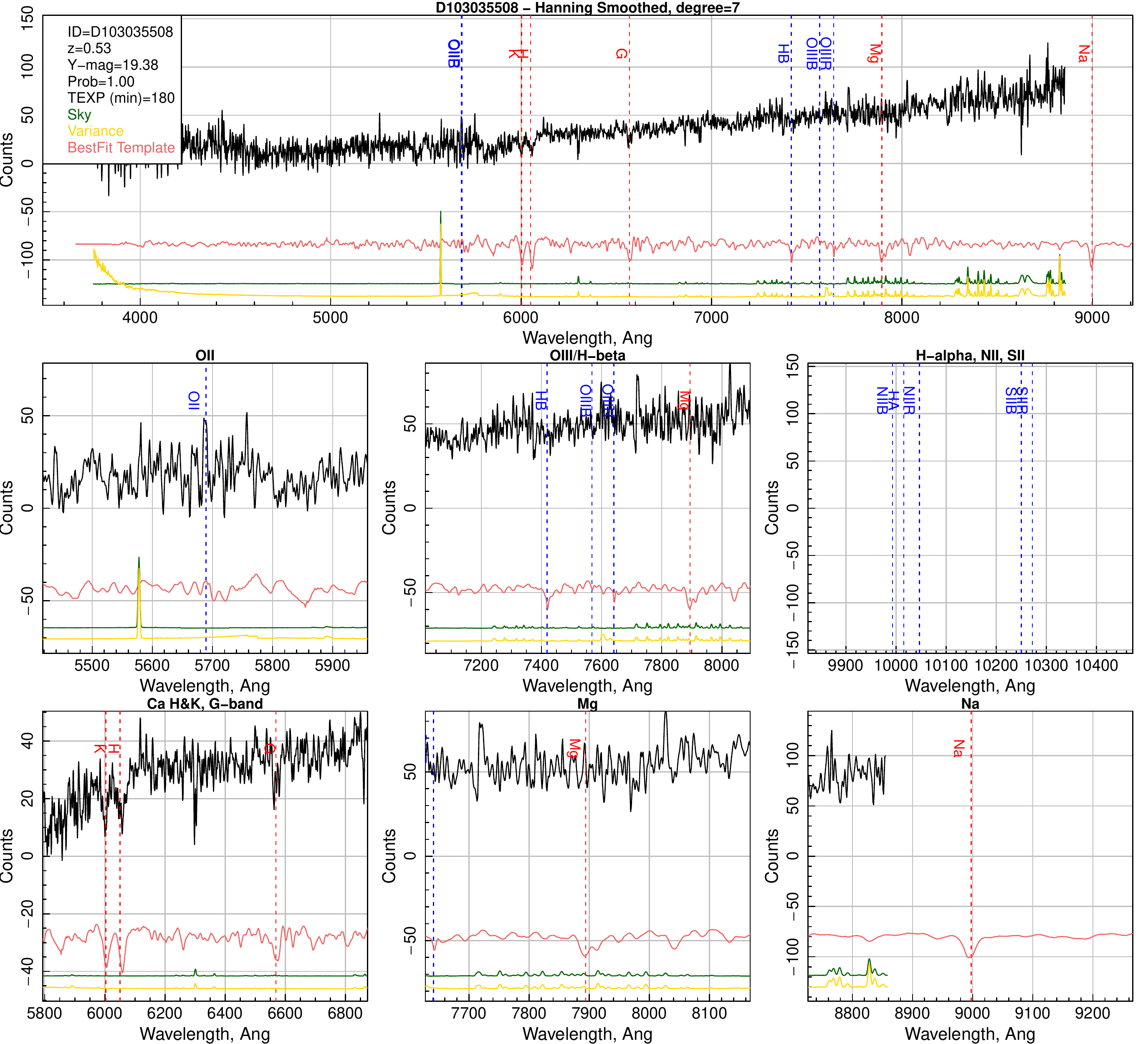}
\caption{The same as Figure \ref{fig:Example} but for a bright (Y=19.38\,mag) absorption line system at $z\sim0.53$ }
\label{fig:brightRed}
\end{center}
\end{figure*}

This appendix provides further examples of the spectra obtained in our 2017B observations for different galaxy type. We include faint (Y=21.12\,mag) absorption line system at $z\sim0.34$ \ref{fig:faintRed}, a bright (Y=19.25\,mag) emission line system at $z\sim0.21$ \ref{fig:brightBlue} and a bright (Y=19.38\,mag) absorption line system at $z\sim0.53$ \ref{fig:brightRed}

%\appendix 
 %\section{Addition affiliations}
\onecolumn
\noindent \textit{$^{7}$ Astrophysics Research Institute, Liverpool John Moores University, IC2, Liverpool Science Park, 146 Brownlow Hill, Liverpool, L35RF, UK}\\
\textit{$^{8}$ Leiden Observatory, Leiden University, Niels Bohrweg 2, NL-2333 CA Leiden, the Netherlands}\\
\textit{$^{9}$ National Centre for Nuclear Research, Astrophysics Division, PO Box 447, PL-90-950 Lodz, Poland}\\
\textit{$^{10}$ HH Wills Physics Laboratory, University of Bristol, Tyndall Avenue, Bristol, BS8 1TL, UK}\\
\textit{$^{11}$ School of Physics and Astronomy, Monash University, Clayton, VIC 3800, Australia}\\
\textit{$^{12}$ ESA/ESTEC SCI-S, Keplerlaan 1, 2201 AZ Noordwijk, The Netherlands}\\
\textit{$^{13}$ Department of Physics and Astronomy, 102 Natural Science Building, University of Louisville, Louisville KY 40292, USA}\\
\textit{$^{14}$ Oxford Astrophysics, Department of Physics, Keble Road, Oxford OX1 3RH, UK}\\
\textit{$^{15}$ Department of Physics, University of the Western Cape, Bellville 7535, South Africa}\\
\textit{$^{16}$ ASTRON, the Netherlands Institute for Radio Astronomy, Postbus 2, 7990 AA, Dwingeloo, The Netherlands}\\
\textit{$^{17}$ Vanderbilt University, Department of Physics and Astronomy, Nashville, TN 37240, USA}\\
\textit{$^{18}$ Centre for Astrophysics and Supercomputing, Swinburne University of Technology, Hawthorn 3122, Australia}\\
\textit{$^{19}$ School of Earth \& Space Exploration, Arizona State University, P.O. Box 871404, Tempe, AZ 85287-1404, USA}

\bsp	% typesetting comment
\label{lastpage}
\end{document}